\documentclass{scrartcl}

\usepackage[a4paper, total={6.5in, 8in}]{geometry}

\usepackage{amsmath,amsthm}
\usepackage{bm}
\usepackage{amsfonts}
\usepackage[square,numbers]{natbib}
\usepackage{placeins}
\usepackage{dsfont}
\usepackage{amssymb}
\usepackage[plain,noend]{algorithm2e}
\usepackage{url}
\usepackage{graphicx,epsf} 
\DeclareGraphicsExtensions{.png,.pdf,.jpg}
\usepackage{enumerate}
\usepackage{booktabs}
\usepackage{amssymb}
\usepackage{eucal}
\usepackage{dsfont}
\usepackage{comment}
\usepackage{placeins}
\usepackage{bm}
\usepackage[utf8]{inputenc}
\usepackage{natbib}
\newcommand{\bch}{}
\newcommand{\ech}{}

\date{}
\newtheorem{lemma}{Lemma}
\newtheorem{proposition}{Proposition}

\title{Multiple hypothesis screening using mixtures of non-local distributions with applications to genomic studies\\
}

\author{Francesco Denti\footnote{francesco.denti@unicatt.it}\\
\textit{Università Cattolica del Sacro Cuore, Milan}\\
Stefano Peluso\\ 
\textit{University of Milan-Bicocca}\\
Michele Guindani\\ 
\textit{University of California, Los Angeles}\\
Antonietta Mira\\
\textit{Universit\`a della Svizzera italiana and University of Insubria}
}

\begin{document}


\maketitle

\begin{abstract}
The analysis of large-scale datasets, especially in biomedical contexts, frequently involves a principled screening of multiple hypotheses. The celebrated two-group model jointly models the distribution of the test statistics with mixtures of two competing densities, the null and the alternative distributions. We investigate the use of weighted densities and, in particular, non-local densities as \emph{working} alternative distributions, to enforce separation from the null and thus refine the screening procedure. We show how these weighted alternatives improve various operating characteristics, such as the Bayesian False Discovery rate, of the resulting tests for a fixed mixture proportion with respect to a local, unweighted likelihood approach. Parametric and nonparametric model specifications are proposed, along with efficient samplers for posterior inference. By means of a simulation study, we exhibit how our \bch model compares with \ech both well-established and state-of-the-art alternatives in terms of various operating characteristics. Finally, to illustrate the versatility of our method, we conduct three differential expression analyses with publicly-available datasets from genomic studies of heterogeneous nature.
\end{abstract}

\section{Introduction}
Multiple hypothesis tests are often needed in the statistical analysis of large biomedical datasets to screen whether $N$ appropriately defined test statistics $\bm{z}=\{z_i\}_{i=1}^N$ are realizations from a given \emph{null} model or not. For instance, screening procedures are pivotal for detecting differentially regulated genes associated with disease occurrences.\citep{VantWout2003} In this context, mixture models represent a flexible statistical tool, widely employed to cast the hypothesis testing problem in terms of selection and estimation of competing models. In this direction, Efron\cite{Efron2004} proposed a two-group model to select and estimate an empirical null distribution and the corresponding alternative. Mixture models have also been proposed for distributions of $p$-values.\citep{Pounds2003} In a Bayesian framework, Do et al\cite{Do2005} employed Dirichlet process mixture models of Gaussian densities to describe null and alternative components. Martin and Todkar\cite{Martin2012} developed a likelihood-based analysis of the two-group model, with a semiparametric specification of the non-null density. Muralidharan\cite{Muralidharan2012} proposed an empirical Bayes hierarchical mixture model to simultaneously estimate the effect size and the local or tail-area false discovery rate for each test statistic. 

Arguably, the objective is to identify relevant cases generated from the alternative model, but the amount of separation between competing mixture components can crucially affect the performance of the tests.
To our knowledge, the available approaches pose no control on the possible detrimental overlap between the null and the non-null distribution.
Here, we propose a likelihood-based analysis of the two-group model, where the non-null distribution is explicitly chosen to improve the discriminating power of the testing procedure. More specifically, we first define a class of weighted densities obtained by rescaling a density function via an appropriately defined weight function.\citep{Rao1985} The class includes many known distributions as special cases, like the skew Normal\citep{Azzalini1985, Ohagan1976} and the non-local densities proposed by Johnson and Rossell.\cite{Johnson2010} Then, we suggest using non-local likelihood functions as \emph{working} alternative distributions to enforce improved separation from the null model. The term \emph{working} highlights that \bch these distributions are not chosen to represent the actual distributions of the data under the alternative hypothesis\ech, but only to improve the screening of the hypotheses. Thus, this modification of the two-group model, which results from the incorporation of available prior knowledge about the support of the data at the level of the likelihood, gives us direct control over the amount of separation between the two distributions.

The paper is structured as follows. First, we introduce the concept of weighted densities and develop an easily interpretable parametric Bayesian two-group model in Section \ref{modeling}. A Bayesian nonparametric extension is also proposed. In Section \ref{theory} we prove how the use of a non-local likelihood leads to increased power and AUC, and lower Bayesian False Discovery and Bayesian False Omission rates with respect to a non-weighted likelihood approach. 
We employ a computationally efficient collapsed Gibbs sampler for estimating both the parametric and nonparametric specifications of the model. To conduct posterior inference, in Section \ref{postinf} we discuss the adopted post-processing of the results and provide an estimate of the local false discovery rate (\emph{lfdr}\cite{Efron2004, Efron2007}), which is -- additionally -- constrained in $\left[0,1\right)$, a natural requirement nevertheless often violated in the literature. We compare our methodology against established alternatives on simulated scenarios in Section \ref{Sec::simu} and on benchmark gene and proteomic expression datasets in Section \ref{Sec::app}. Section \ref{concl} discusses some potential extensions and conclusions.

\section{Non-local likelihood two-group model}
\label{modeling}
\subsection{Weighted densities and non-local distributions}
\label{Section21}

Let $X$ be a random variable with support $\mathcal{S}_X$ and probability density function $\pi\left(x;\eta\right)$. Let   $w\left(x;\xi\right)$ be a non-negative function with parameters $\xi$, such that $\mathbb{E}_{\pi}\left[w\left(X;\xi\right)\right]<\infty$. Then, a  (proper) \textit{weighted density} function is defined by rescaling $\pi\left(x;\eta\right)$ via the weight function over $\mathcal{S}_X$, i.e.
\begin{equation}
	\pi_{w}(x; \xi, \eta) = 
	\frac{w(x; \xi)}{\mathbb{E}_{\pi(x;\eta)}\left[ w(X; \xi)\right]}\pi(x;\eta). 
	\label{DEF:WD}
\end{equation}
Weighted densities of the form  \eqref{DEF:WD} have been  previously introduced by Rao,\cite{Rao1985} who provides a formalization as an adjustment to enhance  density specification when knowledge about the data generating mechanism is available. In the context of robust Bayesian analysis, they have been discussed in Bayarri and Berger\cite{Bayarri1998} and, more recently, in Ruggeri et al. \cite{Ruggeri2019}

Many well-known distributions can be expressed as weighted densities characterized by specific weight functions. Trivially, a truncation of the random variable $X$ on $\left[a,b\right]\ \in\mathcal{S}_X$ can be obtained by setting $w(x;\xi)=\mathbb{I}_{\{x\in \left[a,b\right]\} }$ with $\xi = (a,b)$. More complex truncations are obtainable by considering the sum of indicator functions on disjoint sets. A more elaborated example is the skew Normal distribution, which is defined by weighting a Gaussian density via a Gaussian cumulative distribution function (c.d.f). One can also show that multivariate repulsive distributions belong to this family. For example, define $A=\{(s, j): s=1, \ldots, k ; j<s\}$ and let $g: \mathbb{R}^{+} \rightarrow [0, M]$ be a strictly monotone differentiable function, with $g(0)=0$, $g(x)>0$ for all $x>0$ and $M<\infty.$ Then, 
with $w(x) = \min _{\{(s, j) \in A\}} g\left(||x_{s}- x_{j}||_2\right)$ we obtain the repulsive distribution of Petralia et al.\cite{Petralia2012aaa}

This paper considers another type of weighted densities: non-local densities. Non-local priors have been introduced by Johnson and Rossell \cite{Johnson2010}: these priors balance the convergence rates of the Bayes factor under the null and alternative hypotheses as the number of samples increases. Here, we recast their use as working alternative densities in a likelihood-based approach to multiple testing. A density $\pi_{NL}\left(x\right)$ is a \emph{non-local density} on $\mathcal{S}^0_{X}\subset\mathcal{S}_X$  if, for every $\varepsilon >0$, there is a $\zeta>0$ such that $\pi_{NL}( x ) < \varepsilon \: \text{ for all } x \in \mathcal{S}_X$ for which $\inf _ { x_0 \in  \mathcal{S}^0_{X}} \left| x - x_0 \right| < \zeta$.\\ 
Hence, non-local densities assign a negligible amount of probability to the subspace $\mathcal{S}^0_{X}$. Following the Bayesian literature, we will refer to a density that does not satisfy the previous definition as a \emph{local density}.\\ 
A non-local density can be operatively defined by rescaling a local one. For example, if we consider a univariate Normal distribution, the weighted density obtained by assuming  $w(x;x_0,k)=\left(x-x_0\right)^{2k}$ defines the so-called moment (MOM) distribution, whereas $w(x;x_0,\xi)=\exp\{\sqrt2-\tau\nu/\left(x-x_0\right)^2\}$ with $\xi=\left(\nu,\tau\right)$ defines the exponential-moment density (eMOM).\citep{Rossell2013}  More generally, a non-local distribution around $x_0$ is obtained by imposing that $w\left(x;\xi\right)\rightarrow 0 $ as $x\rightarrow x_0$, regardless of the form of $\pi(x;\eta)$.\citep{Rossell2017a} We exploit this behavior by employing a non-local density  to identify  significant observations beyond a region of irrelevance.

\subsection{Non-local likelihood and two-group model}
\label{subsection23}
We focus on multiple tests of $N$ hypotheses. Let $z_i$ denote a standardized test statistic, $i=1, \ldots, N$, and let $H^{(i)}_0: z_i \sim f_0$ be the $i$-th null hypothesis. This setting is typical, for example, of large-scale screening in genomics. Here, the objective is to quickly identify a few targets of interest, e.g.,  genes that are differentially expressed across conditions. Alternative hypotheses do not typically represent a well-determined belief about the true distribution of the statistics. Still, their purpose is to help reach a conclusion about the evidence against the null. Thus, any specific distributional assumption for the alternative hypothesis, say $H^{(i)}_1: z_i \sim f_1$, can be seen as a \emph{working} alternative distribution used to detect differentially expressed genes. In other words, the choice of $f_1$ should be made to improve the operating characteristics of the model.

Under the assumption of exchangeable hypotheses, one could describe the hypothesis testing problem using a two-group model mixture formulation \citep{Efron2004} by assuming 
\begin{equation}
	z_i | \rho, f_0,f_1 \overset{i.i.d.}{\sim}f(z_i)= (1-\rho)f_0(z_i)+\rho f_1(z_i)
	\label{e1}	
\end{equation}
where $i=1,\dots,N$ and $\rho \in  \left(0,1\right)$ denotes the mixture weight. More specifically, let $\phi(z;\mu,\sigma^2)$ denote a Normal density with mean $\mu$ and variance $\sigma^2$.\bch A natural choice for $f_0$ would be the \emph{theoretical null} $\phi(z;0,1)$. However, a standard Gaussian distribution may be unrealistic, especially in a multiple hypothesis testing setting. Indeed, Efron \cite{Efron2007, Efron2007_cor} notes that the failed model assumptions, unobserved covariates, and correlation of measurements across and within statistical units can make the null distribution effectively wider or narrower than N(0,1). Hence, following Efron's paradigm, we propose to estimate an \emph{empirical null} distribution, which should capture departures from the theoretical null, but still be ``close'' to a standard Gaussian, with mean and variance estimated from the data.\ech Therefore, we model $f_0$ as a Normal distribution $\phi(z;\mu_0,\sigma^2_0)$, 
with Normal-Inverse Gamma prior concentrated around $\left(0,1\right)$ for $\left(\mu_0,\sigma^2_0\right)$. \\
In contrast, we model $f_1$ with a non-local distribution 
of the form $\pi_W\left(z;\xi,\eta\right)\propto w(z;\xi)\pi\left(z;\eta\right)$,  where $w(z;\xi)$ is a weight function that induces small (zero) mass around (at) the origin, in order  to enforce separation from $f_0$. As for the local density $\pi\left(z;\eta\right)$ we first propose a bi-modal mixture of two Normals,
$$
\pi\left(z;\tilde\alpha,\{\mu_j,\sigma^2_j\}_{j=1}^2\right)=
(1-\tilde\alpha)\phi(z;\mu_1,\sigma^2_1)+\tilde\alpha \phi(z;\mu_2,\sigma^2_2),
$$
with $\tilde\alpha\in(0,1)$. In most cases, $\mu_1$ and $\mu_2$  have opposite signs, to capture the behavior of the tails. To this extent, we assume $\mu_1$ and $\mu_2$ to be constrained on the negative and positive semi-axis, respectively. For example, in the analysis of a genomic dataset,  it may be of interest to identify under- and over-expressed groups of observations.\\
Let $\tilde{\bm{\theta}}=\left(\rho,\tilde\alpha,\{\mu_j,\sigma^2_j\}^2_{j=0},\xi\right)$ and $\tilde{\bm{\theta}}_1$ be the sub-vector of parameters that pertain to the non-null distribution. Then, model \eqref{e1} can be re-written as
\begin{equation}
	\begin{aligned}
		z_i|\tilde{\bm{\theta} }\, 
		{\sim}&\,  		\left(1-\rho\right) \phi(z_i;\mu_0,\sigma^2_0) + 
		\rho
		\frac{w(z_i;\xi)}{
			\tilde{\mathcal{ K}} (\tilde{\bm{\theta}}_1)
		}\left[(1-\tilde\alpha)\,\phi(z_i;\mu_1,\sigma^2_1)+\tilde\alpha\,\phi(z_i;\mu_2,\sigma^2_2)\right]
		, \label{Nollik}
	\end{aligned}
\end{equation}
where $\tilde{\mathcal{ K}}(\cdot)$ is the normalizing constant of the non-null distribution. 
For computational convenience, we reparameterize $f_1$ in model \eqref{Nollik} as a mixture of weighted kernels: 
\begin{equation}
	\begin{aligned}
	&f_1(z_i|\{\mu_j,\sigma^2_j\}_{j=1}^2,\alpha,\xi)=
	(1-\alpha)\frac{w(z_i;\xi)\phi(z_i;\mu_1,\sigma^2_1)}{
		{\mathcal{ K}_1}
	}+\alpha\frac{w(z_i;\xi)\phi(z_i;\mu_2,\sigma^2_2)}{{\mathcal{ K}_2}
	}
	, 
	\end{aligned}	
	\label{Nollik_rewrite}
\end{equation}
with $\alpha=\tilde{\alpha}\mathcal{ K}_2/\tilde{\mathcal{K}}$ and $\mathcal{ K}_j=\mathbb{E}_{ \phi \left(z; \mu_j,\sigma^2_j\right)}\left[w(Z;\xi)\right]$ for $j=1,2$. 
In Section 1.1 of the Supplementary Material we show how this equivalence holds in the general case of mixtures with $J$ components.
\bch
To provide a visual example, in the Section 3.1 of the Supplementary Material we report a figure that illustrates how the weight function influences a priori the alternative and marginal densities, and the corresponding relevance probability. Finally, in Section \ref{theory}, we will show how the induced separation between the two competing densities improves the operating characteristics of the weighted model.
\ech 
\subsubsection{Model augmentation with latent membership labels}

It is useful to introduce the latent allocation variables $(\lambda_i, \gamma_i)$, $i=1, \ldots, N$, that explicitly identify the mixture components each observation is sampled from: 
\begin{equation}
	z_i|\bm \theta  \stackrel{i.i.d.}{\sim} \begin{cases}
		\phi(z_i;\mu_0, \sigma^2_0) \; &\text{if} \quad  \lambda_{i}=0,\gamma_{i}=0,\\
		w(z_i;\xi)\phi(z_i;\mu_1, \sigma^2_1)/\mathcal{K}_{1}
		\; &\text{if} \quad \lambda_{i}=1, \gamma_{i}=1,\\
		w(z_i;\xi)\phi(z_i;\mu_2, \sigma^2_2)/\mathcal{K}_{2}
		\; &\text{if} \quad \lambda_{i}=1, \gamma_{i}=2,
	\end{cases}
	\label{Nollik2}
\end{equation}
where $\bm{\theta}=\left(\{\mu_j,\sigma^2_j\}^2_{j=0},\xi,\bm{\Gamma},\bm{\Lambda}\right)$, with $\bm{\Lambda}=\left(\lambda_1,\ldots,\lambda_N\right)$ and $\bm{\Gamma}=\left(\gamma_1,\ldots,\gamma_N\right)$. 
Note that $\gamma_i$ is enough to identify in which of the three cases the $i$-th item is located, therefore $\lambda_i$ has the only scope of improving model interpretability. We refer to  the distribution induced by \eqref{Nollik2} as a non-local likelihood (Nollik). We complete model specification as follows:
\begin{equation}
	\begin{aligned}
		z_i|\bm{\theta}
		&\overset{i.i.d.}{\sim} Nollik\left(\cdot|\lambda_i, \gamma_i, \{\mu_j,\sigma^2_j\}^2_{j=0},\xi\right),\\
		\gamma_i | \lambda_i,\alpha &\overset{i.i.d.}{\sim} Cat\left( 1-\lambda_i, \alpha\lambda_i, (1-\alpha)\lambda_i\right),\quad 
		\lambda_i|\rho \overset{i.i.d.}{\sim} Bern(\rho),  \quad \rho \sim Beta(a_\rho,b_\rho),\\
		\mu_j|\sigma^2_j &\sim TN\left(m_j,\sigma^2_j/\kappa_j,\mathcal{M}_j\right), \quad
		\sigma^2_j \sim IG(a_j,b_j), \quad j=1,2,\\  
		\left(\mu_0,\sigma^2_0\right) &\sim NIG\left(m_0,\kappa_0,a_0,b_0\right), \quad \alpha \sim Beta(a_\alpha,b_\alpha), \quad \xi \sim Q
		\label{MOD}
	\end{aligned}
\end{equation}
where $\gamma_i\in \{0,1,2\}$, $Cat(\bm p)$ indicates a categorical distribution with support on \{0,1,2\} and probability vector $\bm p$, $NIG$ a Normal-Inverse Gamma, and $TN$ a truncated normal distribution, with $\mathcal{M}_1=\mathbb{R}^-$,
$\mathcal{M}_2=\mathbb{R}^+$ being the truncation regions. Finally, $Q$ is the distribution of the parameters in the weight function.
Interpretability of the parameters in model \eqref{MOD} is straightforward.  In addition, posterior simulation can be easily performed via Gibbs Sampling. For further details, see Section 2.1 of the Supplementary Material.  

\subsection{A Bayesian Nonparametric extension}
In the proposed setup, the distribution under the alternative is a \emph{working} alternative aimed at improving the screening between relevant and irrelevant tests. From a hypothesis testing perspective, one should only require $f_1$ to be longer-tailed than $f_0$, with the non-null $z_i$’s tending to occur far away from the origin.\citep{Efron2007} However, the assumption of a specific parametric form under the alternative hypothesis can be too restrictive, and it may not be able to capture multi-modality or heavy-tailed behavior. Hence,  to reflect the desired flexibility and lack of knowledge about $f_1$,
we can extend \eqref{Nollik_rewrite} to a \emph{Dirichlet Process Mixture Model} (DPMM) with non-local mixing kernels. The DPMM is defined as $$\tilde{f}(z) = \int \varphi(z;\vartheta)G(d\vartheta), \: G\sim DP(a,H), $$ where $\varphi(z;\vartheta)$ denotes a generic kernel density parameterized by $\vartheta$ and $DP$ indicates the Dirichlet Process with concentration parameter $a$ and base measure $H$.\citep{Ferguson1973} It is well known that the realizations of a DP are almost surely discrete, $G=\sum_{j=1}^{+\infty}{\omega}_j \delta_{x_j}$  where $x_j \sim H$ and according to the stick-breaking representation \citep{Sethuraman1994} ${\bm{\omega}}=\{{\omega}_j\}_{j\geq 1}\sim SB(a)$, i.e. ${\omega}_j=u_j\prod_{l=1}^{j-1}(1-u_l)$, $u_l\sim Beta(1,a)$ for $l\geq 1$.\\
Through the stick-breaking representation, we obtain a broad class of densities that favor realizations away from the origin as 
\begin{equation}
	f_1(z_i|\tilde{\boldsymbol{\theta}}_1^{DP})=
	\sum_{j\geq 1}\omega_j \frac{w(z_i;\xi)\phi_j(z_i;\mu_j,\sigma^2_j)}{
		{\mathcal{ K}_j}
	}, \label{Nollik_rewriteBNP}
\end{equation}
where $\tilde{\bm{\theta}}_1^{DP}=\left(\{ \omega_j\}_{j\geq 1},\{\mu_j,\sigma^2_j\}_{j\geq 1},\xi\right)$ and $\mathcal{ K}_j=\mathbb{E}_{ \phi \left(z; \mu_j,\sigma^2_j\right)}\left[w(Z;\xi)\right]$ for $j\geq 1.$
We remark that, similarly to the parametric case, model \eqref{Nollik_rewriteBNP} can be expressed as $f_1=w(z,\xi)\pi(z,\eta)/\tilde{\mathcal{K}} $, i.e., a non-local distortion of a nonparametric local density.

Despite the similar nomenclature, the proposed model is essentially different from the \emph{weighted DP} of Sun et al\cite{Sun2018} (see also \cite{Zellner1986, Dunson2007}), where the authors employ a  Dependent DP \citep{MacEachern2000} in a regression framework to allow the error terms of observations with similar predictors' values to be characterized by similar distributions.\\
An alternative approach may assume a non-local distribution for the base measure of the prior process. However, without an appropriate choice of the concentration parameter $a$, such a prior choice does not prevent the resulting mixture from assigning non-negligible mass to regions around the origin.\citep{Denti2020a} 

Once again, we introduce latent allocation variables that assign every observation $z_i$ to either the null ($\lambda_i=0, \: \gamma_i=0$) distribution or one of the countable components of the alternative density weighted DP density ($\lambda_i=1,\: \gamma_i=l, l\geq1$).
We collect the new parameters in  $\bm{\theta}^{DP}=\left(\{\mu_j,\sigma^2_j\}^{+\infty}_{j=0},\xi,\bm{\Gamma},\bm{\Lambda}\right)$, so we can write
\begin{equation}
	\begin{aligned}
		z_i|{\bm{\theta}}^{DP}\stackrel{i.i.d.}{\sim} \begin{cases}
			\phi(z_i;\mu_0, \sigma^2_0) \; &\text{if} \quad  \lambda_{i}=0,\gamma_{i}=0,\\
			\frac{w(z_i;\xi)}{
				\mathcal{K}_{l}
			} \phi(z_i;\mu_l,\sigma^2_l) \; &\text{if} \quad  \lambda_{i}=1,\gamma_{i}=l, \: \forall l\geq 1.\\
		\end{cases}
		\label{NollikBNP2}
	\end{aligned}
\end{equation}
In the following,  we refer to the two-group mixture \eqref{NollikBNP2} between the empirical null and the nonparametric alternative as a Bayesian nonparametric non-local likelihood (BNP-Nollik) model. In summary, our Bayesian nonparametric extension can be represented as  
\begin{equation}
	\begin{aligned}
		z_i|\bm{\theta}
		&\overset{i.i.d.}{\sim} BNP\text{-}Nollik\left(\cdot|\lambda_i, \gamma_i, \{\mu_j,\sigma^2_j\}^{+\infty}_{j=0},\xi\right)\\
		\pi\left(\gamma_i=l|\lambda_i,\bm{\omega}\right) &{=} \lambda_i\cdot \omega_l  + (1-\lambda_{i})\cdot \delta_{0}(l),\quad \forall l\geq 0, \quad \:\:\:\:\:\:
		\lambda_i|\rho \overset{i.i.d.}{\sim} Bern(\rho), 
		\\
		\rho &\sim Beta(a_\rho,b_\rho),\:\:\:\:\:\:
		\bm{\omega} \sim SB(a), \:\:\:\:\:\: \xi \sim Q,\\
		\left(\mu_0,\sigma^2_0\right) &\sim NIG\left(m_0,\kappa_0,a_0,b_0\right),\quad 
		\left(\mu_j,\sigma^2_j\right) \sim G = NIG\left(m_G,\kappa_G,a_G,b_G\right),\\
	\end{aligned}  
	\label{MODBNP}
\end{equation}
where we assume $\omega_0=0$ and $m_G,\kappa_G,a_G,b_G$ denote the hyperparameters of the Normal-Inverse Gamma distribution adopted as DP base measure for the alternative distribution. 
Lastly, a Gamma distribution can be adopted as a prior for the concentration parameter $a$.

\section{Properties of non-local two-group model}
\label{theory}
\begin{figure}[t!]
	\centering
	\includegraphics[width=.8\linewidth]{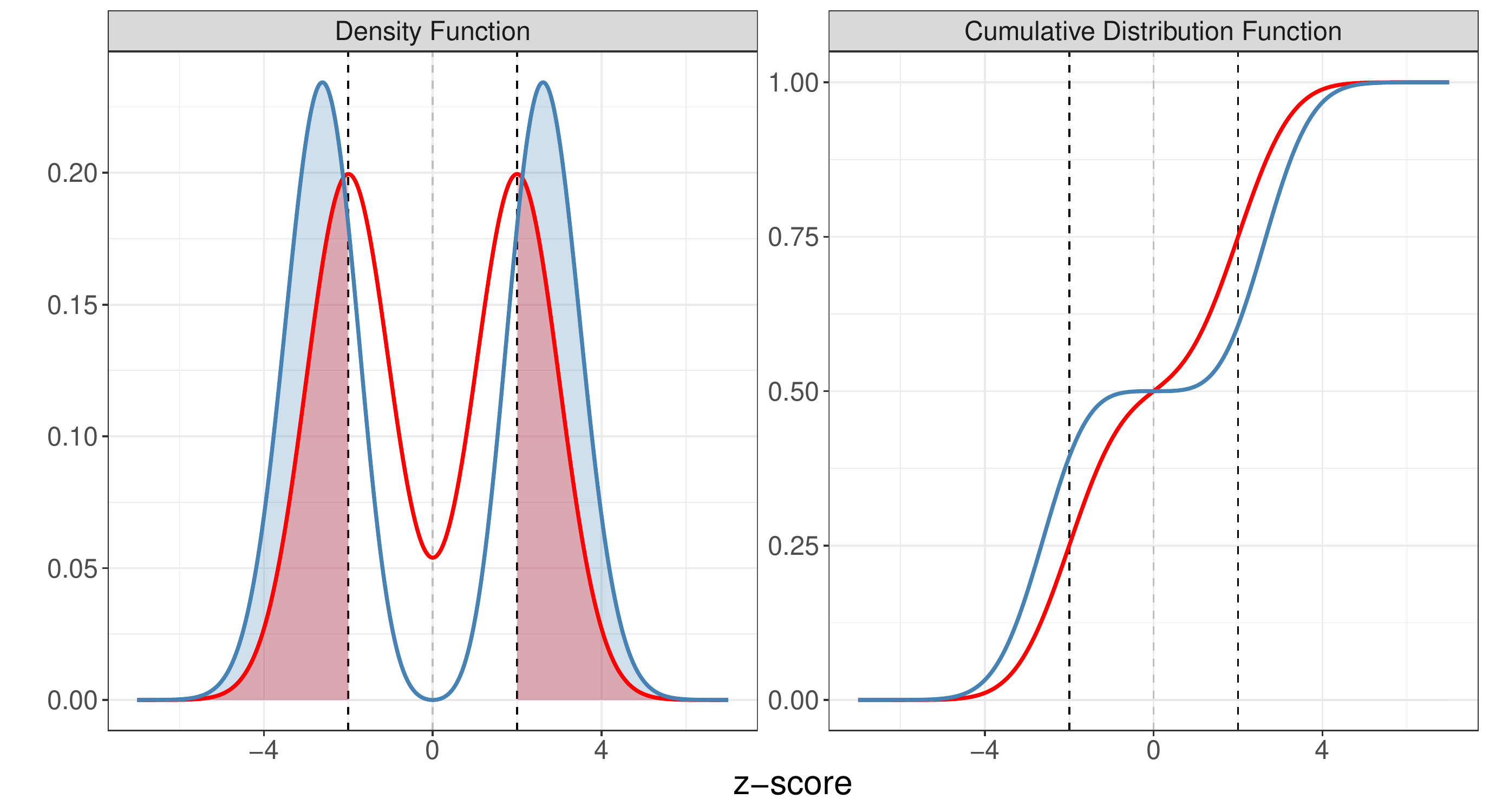}
	\caption{Comparison between local (red) and non-local (blue) distributions. The acceptance region $\mathcal{A}=\left[-2,2\right]$ is highlighted by vertical dashed lines. The left panel compares the density functions, the right one the c.d.f.'s. The colored areas in the left panel represent the power $\mathbb{P}\left[Z\notin\mathcal{A}|H_1\right]$.}
	\label{fig:explain}
\end{figure}

To simplify notation, we denote with $f_1(z)=\pi(z;\eta)$ a local density for the alternative distribution and with  $$f_1^{NL}(z)=\pi_{NL}(z;\xi,\eta)=\frac{w(z;\xi)}{\mathcal{K}}\pi(z;\eta)$$ 
its weighted distortion as in \eqref{DEF:WD}, where $w(z;\xi)$ is a non-local weight function and $\mathcal{K}=\int_{-\infty}^{\infty} w(s;\xi)\, \pi(s;\eta)\, ds$ is the normalizing constant. \\
The screening process determines the specification of an interval $\mathcal{A}=\left[\underline{z},\bar{z}\right]$ (i.e., the acceptance region) outside of which the $z$-scores are flagged as relevant, and the corresponding null hypotheses are rejected. Let $\mathcal{R}=\mathbb{R}/\mathcal{A}$ denote the rejection region. Without loss of generality, we assume $\bar{z}>0$ and $\underline{z}<0$. 
Following Efron,\cite{Efron2004,Efron2007} given an acceptance region  $\mathcal{A}$, we define the \emph{Bayesian False Discovery Rate} as 
\begin{equation}
	\begin{aligned}
	FDR(\mathcal{A}) =& \mathbb{P}\left[H_0|Z\notin\mathcal{A} \right] = \frac{\mathbb{P}\left[Z\notin\mathcal{A}|H_0\right](1-\rho)}{\mathbb{P}\left[Z\notin\mathcal{A}\right]  }
	=\frac{(1-\rho)\int_{\mathcal{R}}f_0(z)dz}{\int_{\mathcal{R}}(1-\rho)f_0(z_i)+\rho f_1(z_i)dz}.
		\end{aligned}
\end{equation}
Analogously, we can also define the \emph{Bayesian False Omission Rate} $FOR(\mathcal{A}) = \mathbb{P}\left[H_1|Z\in\mathcal{A} \right]$, and the \emph{power} (sensitivity) $1-\beta(\mathcal{A}) = \mathbb{P}\left[Z\notin\mathcal{A}|H_1\right]$, where with $\beta$ we indicate the type II error probability. 
Similar quantities can be defined when the assumed alternative distribution is non-local, i.e., for $f_1^{NL}(z)$. We denote them with $FDR^{NL},
FOR^{NL}$ and $1-\beta^{NL}$, respectively. We will show that modeling the unknown alternative with a non-local density improves these operating characteristics given a fixed mixing proportion $\rho$. 
With this in mind, we compute the differences 
\begin{equation}
	\begin{aligned}
		\Delta FDR(\mathcal{A})&=FDR(\mathcal{A})-FDR^{NL}(\mathcal{A}),\\
		\Delta FOR(\mathcal{A})&=FOR(\mathcal{A})-FOR^{NL}(\mathcal{A}),\\
		\Delta \beta(\mathcal{A})&=\beta(\mathcal{A})-\beta^{NL}(\mathcal{A}),
	\end{aligned}
\end{equation}
to provide a direct assessment of the relative performances in the unweighted and weighted versions. In Section 1.2 of the Supplementary Material, we show that all these differences simplify into the comparison of the discrepancies between the c.d.f.'s of the local and non-local distribution $\Delta F_1(z)=F_1(z)-F_1^{NL}(z)$ evaluated at the extremes of the acceptance region, implying that: 
\begin{equation}
	\begin{aligned}
	\Delta F_1(\bar{z}) &\geq \Delta F_1(\underline{z})
	\;\;\Rightarrow\;\;
	\Delta FDR\geq0,\;\; \Delta FOR\geq0,\;\; \Delta \beta\geq0. \label{eq:cond}
	\end{aligned}
\end{equation}
Thus, a sufficient condition for ensuring improved Bayesian FDR, Bayesian FOR, and power of the non-local weighted alternative is that the weighted c.d.f. is lower than its unweighted counterpart in $\bar{z}$ (so that $\Delta F_1(\bar{z})>0$), and higher in $\underline{z}$ (so that $\Delta F_1(\bar{z})<0$). This also implies that the screening procedure has a higher ROC curve and a higher AUC index (refer to Section 1.2 of the Supplementary Material for more details).\\
To provide a visual intuition, we display a simple example in Figure \ref{fig:explain}. Given an acceptance region $\mathcal{A}=\left[-2,2\right]$, delimited by vertical dashed lines, we depict the local and non-local densities in red and blue, respectively. With similar colors we highlight the areas representing the power $\mathbb{P}\left[Z\notin\mathcal{A}|H_1\right]$. The non-local weight pushes the density mass away from the origin, resulting in sharper increments in the corresponding c.d.f. distant from zero.

To state a formal result, we need to postulate some reasonable regularity assumptions on the behavior of the weight function, additionally to the ones introduced in Section \ref{Section21}. We start by recalling that, generally, a weight function $w\equiv w(z;\xi)$ is non-local w.r.t. $z_0$ if (i) $\lim_{z \rightarrow z_0} w(z;\xi)=0$.\citep{Rossell2017a} In the hypothesis testing setting we consider here,  $z_0=0$ represents the only interesting point where to induce vanishing mass. Thus, we require (ii) $w$ to be weakly monotone decreasing (increasing) on the negative (positive) semi-axis. With no additional information about how to weight the support of $z$, we require (iii) $w(-z;\xi) = w(z;\xi) \; \forall z$ i.e., $w$ is an even function. The effect of the weight function has to vanish far away from the origin: an essential requirement is (iv) $ w(z;\xi)\pi(z;\eta) = \mathcal{O}(\pi(z;\eta))$ as $z \rightarrow \pm \infty$, i.e., the non-local density shows the same or a faster asymptotic decay than the corresponding local density.
This is always the case for bounded weights. If $w(\cdot) $ satisfies the conditions (i)-(iv), we refer to it as a \emph{proper} weight function. We can then prove the following propositions.
\begin{proposition}
	Consider a null hypothesis $H_0:z\sim f_0$ characterized by an acceptance region $\mathcal{A}$.
	Let $w(z;\xi)$ be a proper weight function, $f_1$ a symmetric local density and $f_1^{NL}=\frac{w(z;\xi)}{\mathcal{K}}f_1$ its non-local distortion.
	Then, within the framework of the two-group model \eqref{e1} assuming a fixed mixing proportion $\rho$, modeling the alternative distribution with $f_1^{NL}$ rather than with $f_1$ ensures lower Bayesian FDR and Bayesian FOR, and higher power and AUC. 
\end{proposition}
The symmetry of the alternative distribution $f_1$ seems a reasonable assumption for two-tailed tests. The same result holds for one-tailed tests. If the symmetry hypothesis is removed, it is more difficult to derive a result that holds in general.
However, with the introduction of a few alternative assumptions, we can prove the following:
\begin{proposition}\label{prop2}
	Consider a null hypothesis $H_0:z\sim f_0$ characterized by an acceptance region $\mathcal{A}$.
	Let $w(z;\xi)$ be a proper weight function, $f_1$ a local density and $f_1^{NL}=\frac{w(z;\xi)}{\mathcal{K}}f_1$ its non-local distortion. Define $\mathcal{S}=\{z:w(z;\xi)\leq\mathcal{K}\}$, and assume that $\mathcal{S} \subseteq \mathcal{A}$.
	Then, within the framework of the two-group model \eqref{e1} assuming a fixed mixing proportion $\rho$, modeling the alternative distribution with $f_1^{NL}$ rather than with $f_1$ ensures lower Bayesian FDR and Bayesian FOR, and higher power and AUC. 
\end{proposition}
We remark that these are general properties that hold every time a two-tailed test is adopted. Given an acceptance region, a two-group model with alternative non-local density and weight function satisfying (i)-(iv) has higher power, lower Bayesian FDR, and lower FOR than the corresponding local version. 
In Sections 1.3 and 1.4 of the Supplementary Material, we report the proofs of both propositions, concluding with an example in Section 1.5. Last, \bch in Section 3.3 of the Supplementary Material,\ech we discuss another advantage of the non-local specification: its robustness to prior misspecification. Specifically, the two-group model is sensitive to the choice of the distribution for $\rho$, which directly controls the overlap between $f_0$ and $f_1$ in the absence of other constraints. With the help of a simulation study, we show how the non-local specification helps control the number of false positives and provides more reliable estimates of the posterior probability of relevance.

\section{Posterior Inference} 
\label{postinf}
The posterior distributions $\pi\left(\bm{\theta}|\bm{z}\right)$ for models \eqref{MOD} and \eqref{MODBNP} are not analytically tractable and we need to rely on Gibbs sampling schemes for posterior inference. For the parametric model, the full conditional distributions for $\xi$ and $(\mu_j,\sigma^2_j)$ $j=1,2$ require a Metropolis step. We adopt an adaptive Metropolis to improve the acceptance rate, as in Roberts and Rosenthal.\cite{Roberts2009} For the BNP Nollik model, we use the truncated representation of Ishwaran and James,\cite{Ishwaran2001a} where the infinite sum in \eqref{Nollik_rewriteBNP} is substituted with a sufficiently large number of mixture components $J$. The conditional specification allows faster computations than samplers based on P\'olya Urn schemes.\bch The samplers for both model specifications, along with comparisons in terms of the computational costs, are reported in Section 2 of the Supplementary Material.\ech

We recover $\mathcal{A}$ by thresholding the probability of selecting the alternative distribution. 
In the two-group model, this is equivalent to thresholding the $lfdr$, defined as 
$lfdr(z) = (1-\rho)f_0(z)/f(z)$. Thus, the \emph{acceptance region} $\mathcal{A}$ is
\begin{equation} 
	\mathcal{A}=\bigg\{z\in \mathbb{R}: lfdr(z)\geq \nu^* \bigg\}=\bigg\{z\in \mathbb{R}: 
	\frac{\rho f_1(z)}{f(z)}\leq \nu \bigg\},
	\label{accregion}
\end{equation}
where $\nu=1-\nu^*$, $\nu\in (0,1)$. \\
The fully Bayesian specification of our model allows the estimation of the parameters and functions thereof and the quantification of the uncertainty of the estimates. In particular, we are interested in the posterior probability of $H_0^{(i)}$ being rejected given by $P_{1}(z_i)=\mathbb{P}\left(\lambda_i=1|z_i\right)$, i.e., the probability of $z_i$ being flagged as relevant.
Once the MCMC sample is collected, we estimate $P_{1}(z_i)$ evaluating the ergodic mean $\hat{P}_{1}(z_i)=\sum_{t=1}^{T}{\lambda_{it}}/T$, where $T$ is the total number of iterations and $\lambda_{it}$ is the value of the chain for the parameter $\lambda_i$ at the $t$-th MCMC step. For any $z \in \mathbb{R}$, we estimate the posterior probability of relevance $P_1(z)$  by interpolating the estimates at the observed $z_i'$s. Alternatively, we can first estimate the densities $\hat{f}_0$ and $\hat{f}_1$ and consequently compute $\widehat{lfdr}(z)$ as defined in \eqref{accregion}.  The function $\hat{P}_1(z)$ is then obtained as $\hat{P}_{1}(z)=1-\widehat{lfdr}(z)$. Our Bayesian model naturally constrains the range of both $lfdr(z)$ and $P_1(z)$ in $\left[0,1\right)$, and enforces \bch$\mathbb{P} \left[ H_1|z=0 \right]=0$,meaning that a statistic value $z=0$ implies irrelevance almost surely.\ech  
Based on the computed estimate, the hypothesis test is conducted by thresholding the function $P_1(z)$ and  deriving the corresponding critical values $\left(\underline{z},\bar{z}\right)$ on the 
$z$-scores domain. We choose a threshold $\nu$ that controls, at a given level $\alpha$, the Bayesian FDR (BFDR) defined in Newton et al\cite{Newton2004}:
\begin{equation}
	\mathrm{BFDR}(\nu)=E(\mathrm{FDR} | \mathrm{Y})=
	\frac{ \sum_{i=1}^{N} \left(1-P(z_i)\right)\mathbb{I}_{\{ P(z_i)>\nu \} }}{ \sum_{i=1}^{N}\mathbb{I}_{\{ P(z_i)>\nu \} }}.
	\label{BFDR}
\end{equation}
For a specified level of $\alpha$, we obtain the threshold as the minimum $\nu$ for which $\mathrm{BFDR}(\nu)<\alpha$. 

\section{Simulation Study}
\label{Sec::simu}

For the following applications, we will focus on three specific weight functions, one improper ($w_0$), and two proper and bounded in $\left[0,1\right]$:
\begin{equation}
	\begin{aligned}
	w_0(z;k)=z^{2k},\quad\quad
	w_{1}(z ; \xi,k)=1-e^{ -\left(\frac{z}{\xi}\right)^{2 k}},\text{ and} \quad\quad
	w_{2}(z ; \xi,k)=e^{ -\left(\frac{z}{\xi}\right)^{-2 k}},
	\label{univ::weight}
		\end{aligned}
\end{equation}
characterized by different behaviors in the way they converge to zero. For example, the weight functions $w_0$ and $w_2$ have a similar structure to the MOM and eMOM weight, respectively. However, the latter presents a sharper decay than $w_1$, comparable to the iMOM distribution, leading to large areas of low density for the same values of $k$ and $\xi$. It is interesting to compare the two proper weight functions $w_1$ and $w_2$ in terms of their behavior around the origin. Figure 7 in the Supplementary Material shows the shape of the two weight functions for different values of $k,\xi\in\{1,2,3,4\}$. We can appreciate the different effects that the two parameters induce on the chosen functions: $\xi$ affects the functions globally, imposing a milder growth as the parameter increases. In contrast, $k$ affects the function only in a neighborhood of the origin. Therefore, the two parameters are crucial in modeling the decay of the non-local weights and tuning the amount of separation between the null and the alternative distributions. In the following, we will \bch set $w_0(z) = z^2$ and fix $k=2$ in $w_1$ and $w_2,$ \ech since in our experiments the resulting power $2k$ provides a reduction of the weight in a reasonably large neighborhood of the origin sufficient to enforce the required separation. \bch In Section 3.2 of the Supplementary Material, we report an additional simulation study that showcases the robustness of our model to several choices of the parameters $\xi$ and $k$\ech.

Once the weight functions are chosen, we can discuss the specification of the hyperprior parameters for both the parametric and nonparametric model specifications of the \emph{working} alternative density $f_1$ in our model.\\
In the parametric case, we first assume $\xi\sim IG(a_\xi,b_\xi)$, setting $a_\xi=20$ and $b_\xi=57$. This choice, a priori, ensures $\mathbb{E}\left[\xi\right]=3$, while the $\mathbb{V}\left[\xi\right]=0.5$. As Figure 7 in the Supplementary Material shows, $\xi\approx 3$ enforces very low weight on the interval $\left[-1,1\right]$ when combined with $k=2$. For the mixture proportion $\alpha$, we set $a_\rho=1$ and $b_\rho=9$, based on the assumption that only a small fraction of the observations is relevant. Moreover, we have no prior information about the proportions of the bi-modal mixture that models $f_1$ in Equation \eqref{Nollik}. Thus, we adopt an Uniform prior imposing $a_\alpha=b_\alpha=1$. Regarding the NIG specification for the parameters $\{\mu_j,\sigma^2_j\}_{j=1}^2$ of the alternative local distribution in \eqref{MOD}, we set $\kappa_j=1$, $a_j=2$, $b_j=5$. This implies, a prior, that $\mathbb{E}\left[\sigma_j^2\right]\approx 1.67$ and $Var\left[\sigma^2_j\right]=6.25$. 
This way, we are fairly uninformative while preventing the values of the variances from assuming indefinitely large values. This choice helps prevent the estimation of extremely flat posteriors that would jeopardize the classification of the relevant observations into the under-expressed and over-expressed sets. Moreover, we adopt $m_1=-3$ and $m_2=3$.
For the parameters $\left(\mu_0,\sigma^2_0\right)$ of $f_0$ we need to specify a NIG that places most of the mass around $(0,1)$. Therefore, we set $a_0=b_0=10$ to induce a density for $\sigma^2_0$ peaked around 1. We finally set $\kappa_0=100$ and $m_0=0$, so that $\mathbb{V}\left[\mu_0|\sigma^2_0\right]=\sigma^2_0/100$.\\
In the nonparametric case, we truncate the stick-breaking process at $J=30$. We then set the concentration parameter $a$ equal to 1 and we choose a $NIG(0,0.01,3, 1)$ as the base measure $G$ for the DP. These values are selected so that $\mathbb{E}\left[\mu_l|\sigma^2_l\right]=100\sigma_l^2$  and $\mathbb{E}\left[
\sigma^2_l\right]=1/2$.\citep{Rodriguez2008} All the other specifications are equal to the parametric case.

We test the performance of our model on 50 datasets generated under four scenarios, \bch adopting all the weight specifications listed in \eqref{univ::weight} under the parametric model. In addition, we also estimate the nonparametric model with the $w_1$ weight function\ech. Each simulated dataset contains 1,000 observations: 90\% of the sample is drawn from $f_0$, the remaining 10\% from $f_1$. The data generating mechanisms for the 4 scenarios are assumed as follows: (S1) $ z_i \sim 0.90 \mathcal{N}\left(0,1.5\right)+ 0.05 \mathcal{N}\left(5,1\right) + 0.05 \mathcal{N}\left(-5,1\right)$; (S2) $z_i \sim 0.90 \mathcal{N}\left(0,0.25\right)+ 0.05 \mathcal{N}\left(3,1.5\right) + 0.05 \mathcal{N}\left(-3,1.5\right)$; (S3) each $z_i \sim \mathcal{N}\left(\gamma_i,1\right)$, where $\gamma_i$ is sampled from the mixture $0.90\delta_0+0.1\mathcal{N}\left(-3,1\right)$. This scenario was previously proposed in Efron \cite{Efron2008}; (S4) $z_i\sim \mathcal{N}\left(\gamma_i,1\right)$, where $\gamma_i$ is sampled from the mixture: $0.90\delta_0+0.10 \left( 0.5 \mathcal{U}_{\left[-4,-2\right]}+0.5\mathcal{U}_{\left[2,4\right]}\right).$ This scenario is similar to the one proposed by Muralidharan.\cite{Muralidharan2012}

\bch We run the MCMC for 35,000 iterations, discarding the first 10,000 as burn-in period. We then thin the chains every 5 iterations to reduce autocorrelation, retaining a total of 5,000 simulations.\ech Visual inspection of the traceplots reveals good mixing, and the convergence of the chains was also assessed using standard MCMC  diagnostics.\citep{oro22547} In each simulation scenario, we compute the estimate $\widehat{lfdr}(z)$. In the nonparametric case, we evaluate the posterior densities $f_0$ and $f_1$ on a grid of points at each Markov chain iteration and then consider 
their point-wise averages. More specifically, given $T$ MCMC steps, we have
\begin{equation}
	\begin{aligned}
		\hat{f}^{BNP}_0(z) = \frac1T\sum_{t=1}^T \phi(z;\mu_{0,t},\sigma^2_{0,t}),\quad \quad \hat{f}^{BNP}_1(z) = \frac1T\sum_{t=1}^T\sum_{j=1}^J\omega_j\phi(z;\mu_{j,t},\sigma^2_{j,t}). 
	\end{aligned}
\end{equation}
We flag the relevant hypotheses by thresholding the posterior probability of the alternative with a value that controls the BFDR \eqref{BFDR} at a 5\% level.

\begin{figure}[ht!]
    \centering
    \includegraphics[width=.9 \linewidth]{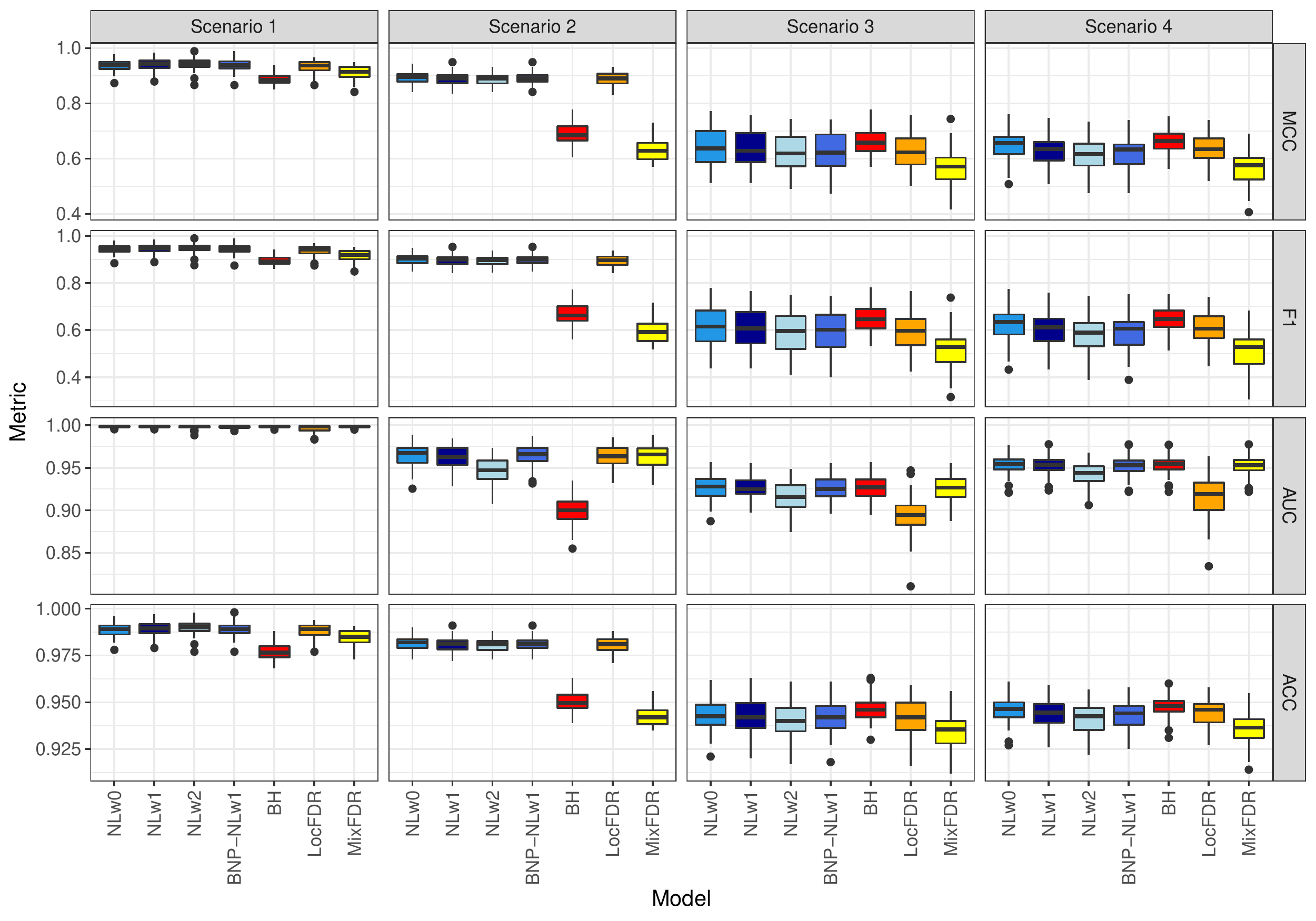}
    \caption{Boxplot of the different performance scores (rows) obtained over 50 replications under four scenarios (colums) by the Nollik models (\texttt{NLw0-2}) with different weights, the BNP-Nollik model (\texttt{BNP-NLw1}), the \texttt{BH} procedure, the \texttt{LocFDR}, and the \texttt{MixFDR}.}
    \label{fig:sim1}
\end{figure}

\begin{figure}[ht!]
    \centering
    \includegraphics[width=.9 \linewidth]{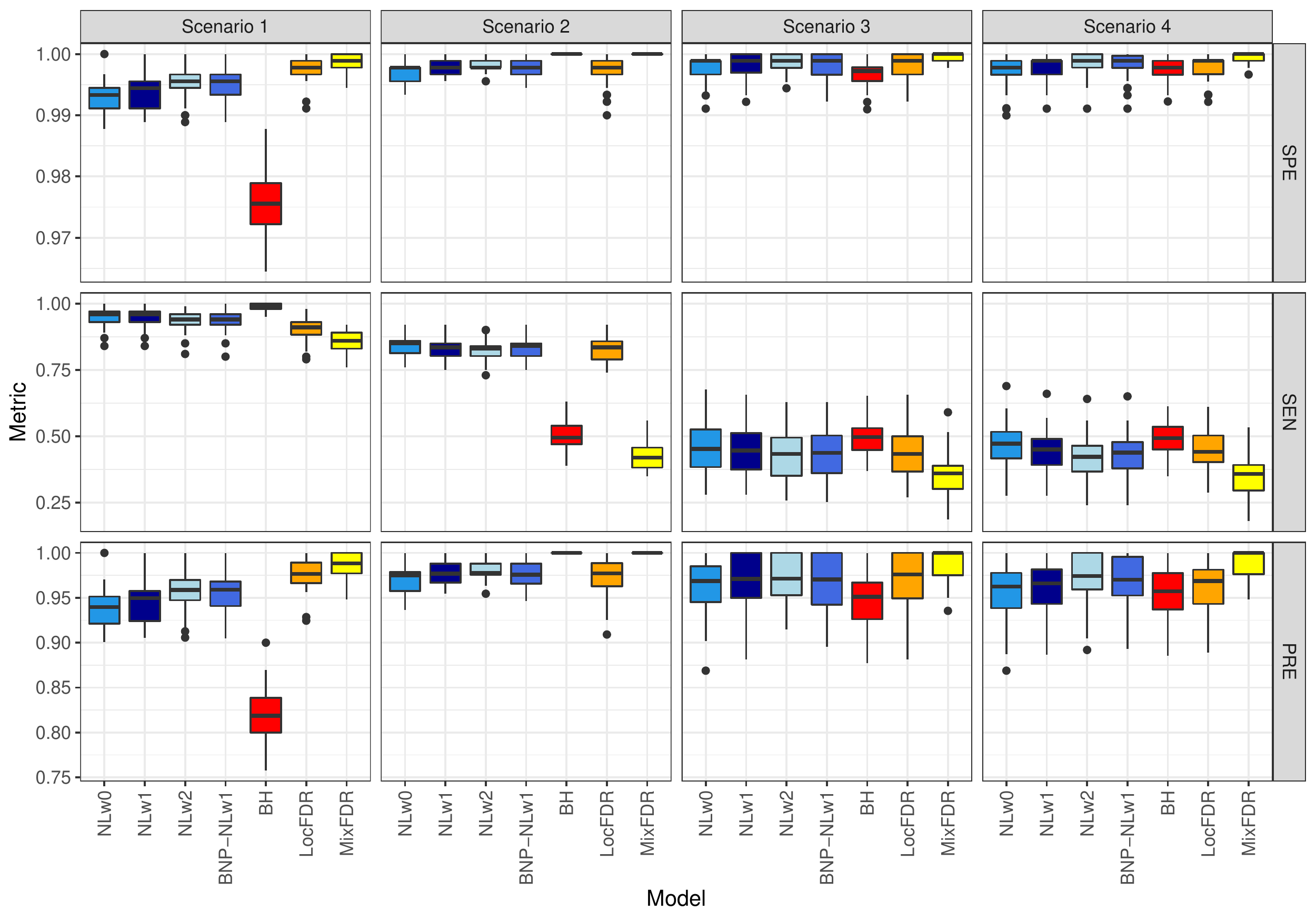}
    \caption{Boxplot of the different screening rates (rows) obtained over 50 replications under four scenarios (colums) by the Nollik models (\texttt{NLw0-2}) with different weights, the BNP-Nollik model (\texttt{BNP-NLw1}), the \texttt{BH} procedure, the \texttt{LocFDR}, and the \texttt{MixFDR}.}
    \label{fig:sim2}
\end{figure}

We compare the results obtained by our method with the \texttt{MixFDR} model, \citep{Muralidharan2012} the \texttt{LocFDR} model, \citep{Efron2004} and the Benjamini-Hochberg procedure (\texttt{BH}).\citep{Benjamini1995}
For the first two competitors, we threshold $\widehat{lfdr}$ at 0.20, as suggested by the authors. We threshold the \texttt{BH} adjusted p-values at 0.05. To quantify the relative performance of the models, we compute several indices describing the operating characteristics of the procedures. More precisely, we calculate the accuracy (ACC), specificity (SPE), sensitivity (SEN), precision (PRE), and Area Under the Curve (AUC) of the different methods. Moreover, we compare Matthew's Correlation Coefficients (MCC) and the $F_1$ scores, defined as 
\begin{equation*}
	\begin{aligned}
	\mathrm{MCC}=\frac{T P \times T N-F P \times F N}{\sqrt{(T P+F P)(T P+F N)(T N+F P)(T N+F N)}}, 
	\quad \quad F_{1}=\frac{2}{\text { SEN }^{-1}+\text { PRE }^{-1}},
	\end{aligned}
\end{equation*}
where FP and FN denote the number of false positives and false negatives, respectively. Similarly, TP and TN denote the number of true positives and true negatives. \bch For all indices, we report the boxplots of their distributions across the 50 Monte Carlo replications in Figures \ref{fig:sim1} and \ref{fig:sim2}. Figure \ref{fig:sim1} contains indexes that summarize the overall classification performance (AUC, ACC, MCC, $F_1$), whereas Figure \ref{fig:sim2} showcases the different screening rates (PRE, SEN, SPE)\ech. 


\bch
All the Nollik procedures lead to similar results, underlying the robustness of our proposal to different weight choices. We point out that an unbounded weight function, like $w_0$, is expected to over-inflate the mass far away from the origin, and therefore it is not optimal for density estimation. Nonetheless, it appears to function correctly as \emph{working} density for estimating the posterior probability of rejection in all tests. It is also interesting to note that, despite we introduced the nonparametric specification to reflect a potential lack of knowledge regarding the true shape of the alternative distribution $f_1$, the parametric and nonparametric methods return similar results in all the different cases considered here. Again, this behavior suggests that, albeit the parametric specification is not optimal for density estimation, in many cases it may be sufficient to obtain competitive performances.

Overall, the simulations suggest that the Nollik models obtain very good results across all the scenarios for almost every score. Most importantly, all the scores obtained by the Nollik models are always in line, if not better, with the ones obtained by the other well-established methods. We do not observe the same robustness across scenarios in competitors. For example, \texttt{BH} presents low precision and specificity in Scenario 1, where the actual null distribution does not coincide with the theoretical one, i.e., $N(0,1)$. The \texttt{MixFDR} always presents the highest specificity, but its results deteriorate in terms of sensitivity, especially in Scenario 2.
Nonetheless, we note that the \texttt{LocFDR} procedure always provides comparable performances to those of the Nollik. The main advantage to the Nollik framework resides in the fully Bayesian approach, which provides straightforward uncertainty quantification and posterior inference. However, the improved performance of our modeling framework comes at a price. As it fully relies on MCMC, Nollik is computationally more expensive than any of the competitors we considered. For example, the average time taken by different Nollik specifications to run ranged between 10 and 30 minutes to be completed (with the nonparametric specification being the most expensive), while the competitors provided results in a matter of seconds. For a more detailed description of the computational cost of the algorithm, see Section 2.3 of the Supplementary Material.\ech

\section{Differential Gene Expression Case Studies}
\label{Sec::app}
We apply our model to three different gene expression datasets using the weight function $w_1$ in Equation \eqref{univ::weight}. Results obtained with $w_2$ are similar; a summary can be found in Section 3.4.2 of the Supplementary Material. We compare the results with Efron's \texttt{LocFDR}, the \texttt{MixFDR}, and the \texttt{BH} procedures.
\bch In the first two applications, we run 70,000 MCMC iterations, and, after discarding the first 20,000 as a burn-in period, we thin the remaining chain every 10 iterations. In the third case, we increase the burn-in to 50,000 iterations and thin the chain every 20 steps to reduce autocorrelation. \ech 
We adopt the hyperparameter configuration detailed in Section \ref{Sec::simu}. On the one hand, we will show how our model can capture the overall data distribution leading to similar results as Efron's \texttt{LocFDR}, while also allowing for uncertainty quantification in a coherent, fully Bayesian framework. On the other hand, the mixture formulation allows for more flexible modeling of the irregularities in the empirical null distribution, such as lepto- or platykurtosis (see Figures 10 and 11 in the Supplementary Material). These characteristics are mostly ignored by the \texttt{BH} procedure, leading to potential loss of relevant (abundance of irrelevant) genes in the case of leptokurtosis (platykurtosis) of $f_0$. \bch In Section 3.4.3 of the Supplementary Material, we also report plots that summarize the discoveries obtained by the various methods on the different datasets and their pairwise agreements.\ech

\subsection{HIV Microarray Data}
A benchmark example of gene expression case study is the \texttt{HIV} microarray matrix.\citep{VantWout2003,Efron2007} The dataset is publicly available in the R package \texttt{locfdr}. The experiment goal is to compare the gene expression values of 7,680 microarray genes of $4$ HIV negative subjects with $4$ HIV positive patients. Microarray data are continuous, therefore for each gene we compute the corresponding t-statistics to test the difference of expression among the two groups. We transform the data using the c.d.f. of a Student's t-distribution with $6$ degrees of freedom. Efron's \texttt{LocFDR} (thresholded at 0.2) flags 160 genes as relevant, the \texttt{MixFDR} flags 64 genes, while the \texttt{BH} (thresholded at 0.05) only 18. 
Nollik, in its parametric version, estimates a proportion of relevant hypotheses of $\hat{\rho}=0.079\: (s.d.\:0.011)$, whereas the estimated proportion of the over-expressed genes among the flagged ones is $\hat{\alpha}=0.121\: (s.d.\:0.050)$. The parameter $\hat{\xi}$ of the weight function is estimated as $2.062 \:(s.d.\:0.306)$. \bch The empirical null is characterized by $\hat{\mu}_0 = -0.108\:(s.d.\:0.012)$ and $\hat{\sigma}^2_0 = 0.557\:(s.d.\:0.023)$, which suggests the potential presence of correlation across tests. \ech We control for a BFDR level of $5\%$, corresponding to a threshold on $\hat{P}_1(z)$ equal to 0.840. This leads to 143 genes flagged as relevant. From the left panel of Figure \ref{fig:1} we can see how the functions $\hat{P}_1(z)$ for our method and Efron's \texttt{LocFDR} are very similar, \bch while the \texttt{MixFDR} results in the most conservative method, with the lowest $\hat{P}_1(z)$ on the entire domain.\ech \\

\subsection{Microbiome Abundance table: the Torondel dataset}
\begin{figure*}[t]
	\centering
	\includegraphics[width=.49\textwidth]{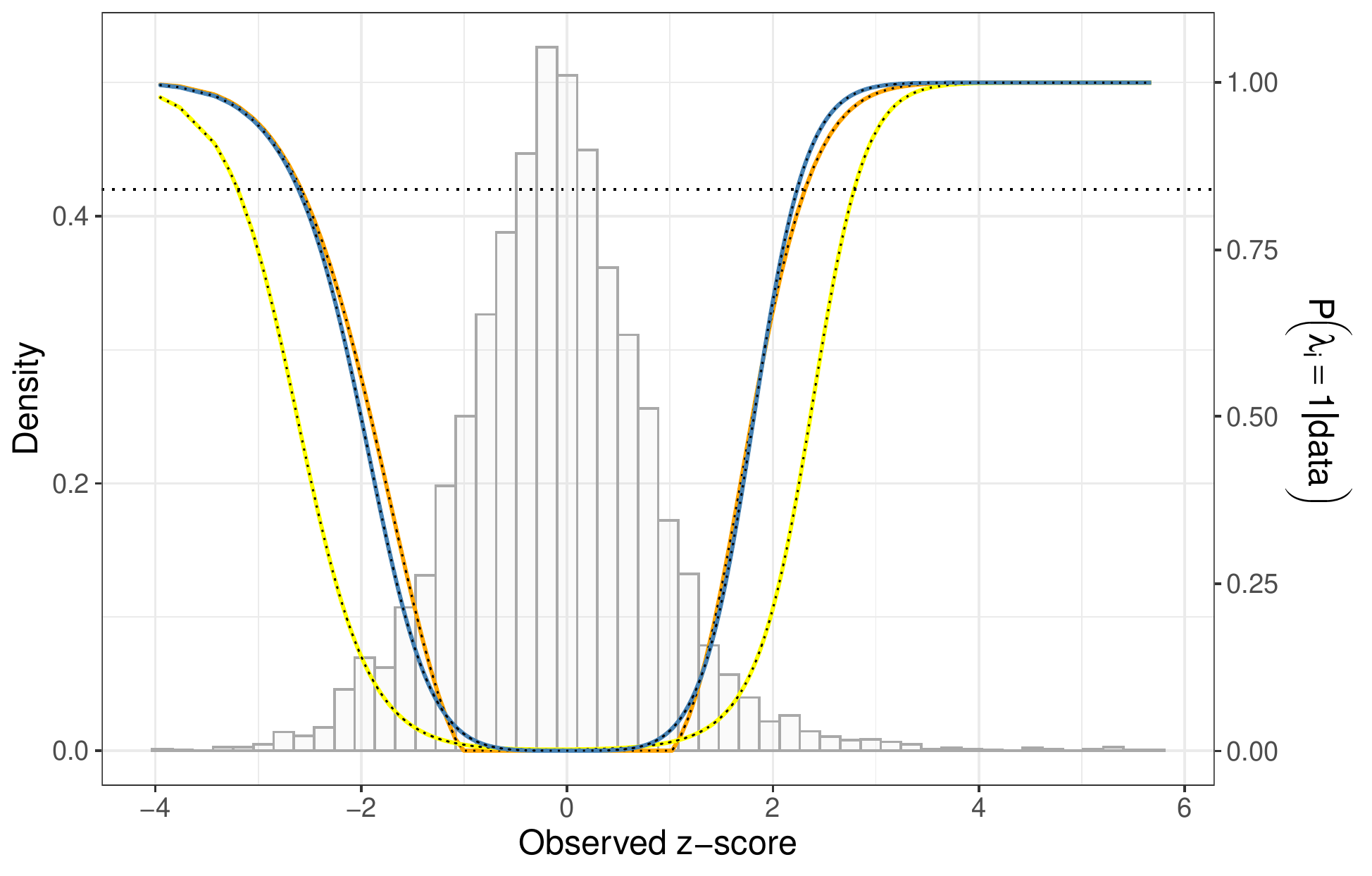}
	\includegraphics[width=.49\textwidth]{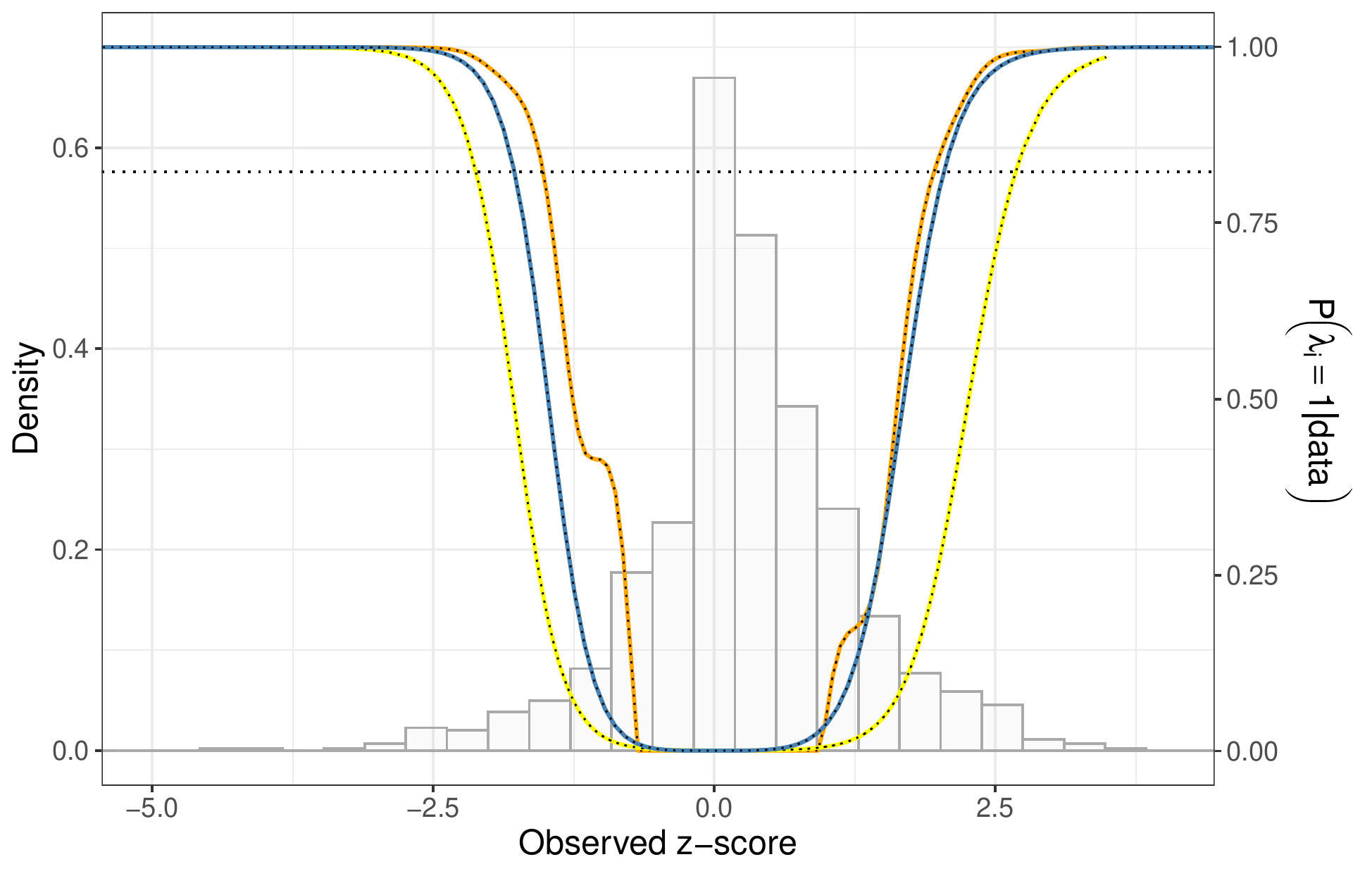}
	\caption{\texttt{HIV} (left) and \texttt{Torondel} (right) datasets. Histograms of the data with function $P_1(z)$ superimposed computed \bch via the \texttt{LocFDR} (orange), the \texttt{MixFDR} (yellow) and Nollik (blue). The horizontal dotted lines represent the estimated threshold controlling for a BFDR of 5\%.\ech}
	\label{fig:1}
\end{figure*}

Many models and software have been developed by bioinformaticians to address the challenges that count data from sequencing studies raise for investigating differential expression (e.g., \emph{edgeR} and \emph{baySEQ} \citep{Hardcastle2010,Robinson2007}). Among these models, Love et al\cite{Love2014} have proposed \emph{Deseq2}, a method for differential analysis based on Negative Binomial regression. To conduct multiple hypothesis testing, \emph{Deseq2} thresholds the \texttt{BH} adjusted p-values computed from estimated Wald statistics. 
Here, we apply this method to the \texttt{Torondel} dataset,\citep{Torondel2016} available from the R library \texttt{microbiomeSeq}.
The abundance table comprises the frequencies of 8883 taxa found in 81 pit latrines: 29 from Tanzania, 52 from Vietnam. Let $x_{ij}$ denote the frequency for taxon $i$ in the pit latrine $j$. We first filter out all the taxa having variance of the relative counts $r_{ij}=x_{ij}/\sum_{j\geq1} x_{ij}$ lower than $10^{-7}$. The inclusion of this extremely sparse taxa might distort the analysis producing a high number of negligible test statistics which may mislead the estimation of $f_0$. \bch After these preprocessing steps, we are left with 1,204 taxa.\ech
The Wald statistics are known to be asymptotically Normal in the number of samples. A preliminary data analysis shows that the assumption is reasonable for the rescaled Wald statistics. The \texttt{BH} procedures flags only 1 taxon as relevant, the \texttt{MixFDR} flags 39 taxa, while the \texttt{LocFDR} 107.
To better address irregularities in the tails of the data, we employ the BNPNollik. 
We obtain a proportion of relevant taxa equal to $\hat{\rho}=0.139 \: (s.d.\:0.023)$, while $\hat{\xi}=2.913 \: (s.d.\:0.689)$. Controlling for a BFDR at level 0.05 induces a threshold at 0.822, with 91 taxa marked as relevant, as reported in the right panel of Figure \ref{fig:1}, \bch in between the discoveries provided by \texttt{MixFDR} and \texttt{LocFDR}.\ech \\

\subsection{Grouped Proteomic Data: the Ubiquitin-protein interactors dataset} 
In numerous studies, the case group may be composed of $J$ different subsets reflecting specific experimental conditions (e.g., stages of a disease, drug dosages, etc.), while the control group remains the same. At one extreme, a separate analysis for each subgroup would result in a potential loss of statistical power. At the other extreme, pooling all the data together is not optimal, since test statistics are not independent across subgroups. In other words, we need to capture commonalities across subgroups induced by the comparisons with a shared control group.\bch To conduct a unified analysis, we extend our approach into a hierarchical model where we jointly estimate different Nollik distributions, one for each experimental condition. In this specification, the dependence across groups induced by the shared control set is captured via a common relevant proportion $\rho$. \ech Let $z_{ij}$ be the test statistics relative to hypothesis $i$ in the $j$-th subgroup. 
Model \eqref{e1} becomes 
\begin{equation}
	z_{ij} | \rho, f_{0,j},f_{1,j} \sim f_j(z_{ij})= (1-\rho)f_{0,j}(z_{ij})+\rho f_{1,j}(z_{ij}) 
	\label{eH}	
\end{equation}
where $f_{0,j}$ and $f_{1,j}$ are subgroup-specific null and alternative distribution, respectively. Within each subgroup, the model is \eqref{Nollik}-\eqref{Nollik2}, with $\boldsymbol{\theta}_j$ being its specific set of parameters. At the same time, $\rho$ is the same across all the subgroups. The advantages of this model specification are threefold. First, the efficiency and interpretability of the Nollik model are unaltered. Second, the parameters $\rho$ captures the commonality structure, allowing for the borrow of information across subgroups. Third, this model allows the estimation of $P_{1,j}(z)$ functions specific for each group, capturing the differences in the various proportions of relevant hypothesis across conditions. 

\begin{figure*}[t]
	\centering
	\includegraphics[width=.9\linewidth]{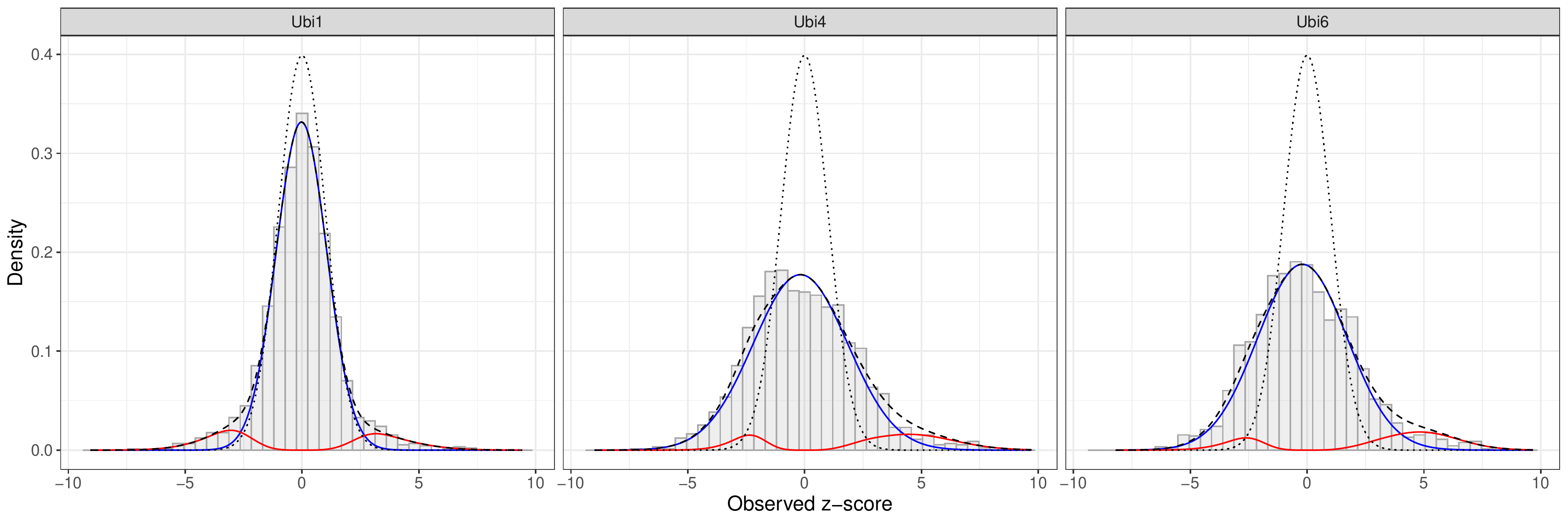}
	\caption{\texttt{Ubiquitin-protein interactors} dataset. 
	Estimated null (black, dashed), alternative (red), and overall (blue) densities for the of the three subgroups considered (z-scores shown as histograms). The dotted lines denote the $\phi(0,1)$ densities. Weight function: $w_1$.}
	\label{fig:3}
\end{figure*}

We analyze a mass spectrometry proteomic data for differential protein expression, freely available in the R package \texttt{DEP}. The proteins are grouped into three sub-groups, reflecting the different intensities of Label-Free Quantification (LFQ) of the mass spectrometry used to preprocess the data: \texttt{Ubi1}, \texttt{Ubi4}, and \texttt{Ubi6}. See Zhang et al\cite{Zhang2017} for additional details on the data. 
We follow the data analysis pipeline indicated by Zhang et al,\cite{DEP} and obtain 1899 values of proteomic expressions by evaluating the contrasts of the three experimental conditions with the common control group. We evaluate the differential expressions with \emph{Limma}, an Empirical Bayes procedure that produces \emph{moderate t-statistics}, computed as $d/(s+s_0)$, where $d$ is the difference in the sample means, $s$ is the pooled standard deviation and $s_0$ is a small constant, added to avoid divisions by an extremely small
variance estimate.\citep{Limma}\\  
The estimated overall proportion of relevant tests is $\hat{\rho}= 0.101\: (s.d.\:0.014)$. Figure \ref{fig:3} shows the data and the estimated densities stratified by condition. The subgroup-specific models lead to the estimation of different parameters and numbers of relevant proteomic expressions, as summarized in Table \ref{tab:my_label}, \bch where we also report the discoveries obtained by the competitors when applied to each subgroup independently. Here, we observe that \ech the platykurtic shape of the histograms \texttt{Ubi4} and \texttt{Ubi6} lead the \texttt{BH} procedure to likely overestimate the number of relevant proteins.

\begin{table*}[t!]
	\centering
	\begin{tabular}{l ccccccc}
		\toprule
		&\multicolumn{3}{c}{Posterior estimates} & \multicolumn{4}{c}{\# Relevant proteins}\\
		\cmidrule(lr){2-4} \cmidrule(lr){5-8}
		&$\hat{\xi}_j$ &  $\hat{\alpha}_j$ & Threshold  & Nollik & \texttt{MixFDR} & \texttt{LocFDR} & \texttt{BH}  \\
		\midrule
		\texttt{Ubi1} & $2.656\:(s.d.\:0.405)$ &$0.447 \: (s.d.\:0.249)$ &  0.815 & 107 & 61 & 92 & 132\\
		\texttt{Ubi4} & $2.202\:(s.d.\:0.508)$ &$0.348 \: (s.d.\:0.157)$ &  0.903 & 17  & 7  & 13 & 466\\
		\texttt{Ubi6} & $2.366\:(s.d.\:0.539)$ &$0.387 \: (s.d.\:0.121)$ &  0.882 & 31  & 10 & 19 & 457\\
		\bottomrule     
	\end{tabular}
	\caption{\texttt{Ubiquitin-protein interactors}. Estimates and numbers of relevant proteins according to different methodologies stratified by subgroup (\texttt{Ubi1}, \texttt{Ubi4}, \texttt{Ubi6}).}
	\label{tab:my_label}
\end{table*}

\section{Discussion}
\label{concl}
In this article, we have proposed a weighted 
alternative density for multiple hypothesis testing, which leverages on non-local distributions. We have shown how a non-local alternative likelihood can be used as a convenient working density for hypotheses' screening, as it increases the separation from the null distribution. In particular, the trimodal structure of the proposed parametric model, with parameters appropriately tuned for the screening of a large number of hypotheses, allows the segmentation of the z-scores into under-expressed, null, and over-expressed once they are assigned to the Normal distributions centered in $\mu_1<0$, $\mu_0$, and $\mu_2>0$, respectively. \bch In summary, the results we observed in the various simulation studies suggest that Nollik is particularly suited for the following scenarios: in case of an asymmetric behavior of over- and under-expressed test statistics, under potential hyperprior misspecification for the proportion of the relevant statistics, and when the overlap of the two competing distributions hinders the estimation of the empirical null.\\
An acknowledged limit of our approach stands in its computational cost. Albeit efficient, our implementation fully relies on MCMC, which makes it more costly than every competitor we considered. One solution to this problem would be the development of a Variational Bayes approximation\cite{Blei2017} of the Nollik posterior, which would scale up its applicability to massive collections of hypotheses.\\ Methodologically, \ech
the simple yet flexible structure of the Nollik models paves the way for relevant extensions. First, the weight functions can be readily generalized to accommodate multivariate data, following Johnson and Rossell.\cite{Johnson2010} Given a $d$-dimensional vector $\bm{z}$, we can define the quantity 
$Q ( \bm{z} ) = \frac{\left( \bm{z} - \bm{z}_0 \right)'  \Sigma^{ - 1 } \left( \bm{z} - \bm{z}_{ 0 } \right)}{n \xi \sigma^{ 2 }}$,
where $\Sigma$ is a positive definite matrix and $\sigma^2$ and $\xi$ are scalars, and then extend the weight functions to
$
w_1\left(\bm{z};\xi\right) = 1 -\exp\left[-Q ( \bm{z} ) ^{k}\right]$ and $w_2\left(\bm{z};\xi\right) = \exp\left[-Q ( \bm{z} ) ^{-k}\right]$.
This multivariate likelihood can be useful, for example, in spatial settings, where hypotheses are typically associated with clusters. Second, a covariate-adjusted framework can be naturally addressed without increasing the complexity of the model. Let $\bm{X}$ be a dataset of $p$ dimensional measurements and denote as $\bm{X}_i=\left(X_{i1},\ldots,X_{ip}\right)$ the vector of values specific for individual $i$. Then we can introduce the dependence via $\bm{\Lambda}$, specifying $\lambda_i \sim Bern\left(p_i\right), \: p_i = g(\bm{X}_i,\bm{\eta})\: \forall i$. This formulation has two main advantages: (i) the tractability of the MCMC is not altered, being $\bm{\Lambda}$ separated from the other parameters in the hierarchical structure; (ii) the covariates directly affect parameters driving allocation to latent classes. The function $g(\bm{X}_i,\bm{\eta})$ can be assumed as the usual logistic or probit link, for which efficient samplers are readily available.\citep{Polson2013,Durante2019}

\section{Software and data availability statement}
\label{sec5}

Software in the form of \texttt{R} and \texttt{C++} code is available at the Github repository \url{https://github.com/Fradenti/Nollik_2GM}. The datasets that support the findings of this study are openly available in the aforementioned \texttt{R} packages.

\clearpage
\begin{center}
    {\Huge \textbf{Supplementary Material}}
\end{center}
\setcounter{section}{0}
\setcounter{figure}{0}
\setcounter{table}{0}

\section{Theoretical Results}
\label{theoS}
\subsection{Equivalence between weighted mixture and mixture of weighted kernels}
Consider the following weighted alternative distribution
\begin{equation}
	f^W_1(z|\boldsymbol{\theta},\xi)= \frac{w(z,\xi)}{\tilde{\mathcal{K}}}f_1(z,\boldsymbol{\theta}), 
	\label{mix1}
\end{equation}
with $f_1$ a generic local distribution and $\tilde{\mathcal{K}}\equiv\tilde{\mathcal{K}}(\boldsymbol{\theta},\xi) = \int w(z,\xi)f_1(z,\boldsymbol{\theta}) dz$. Let us suppose that $f_1$ can be expressed as a mixture distribution $f_1(z|\boldsymbol{p},\boldsymbol{\theta}) = \sum_{j=1}^J p_j \phi_j(z,\theta_j)$, with weights $\boldsymbol{p}=(p_1,\ldots,p_J)$ and atoms $\boldsymbol{\theta}=\left(\theta_1,\ldots,\theta_J\right)$. Then, conditionally on $\boldsymbol{\theta}$ and $\xi$, we can rewrite 
\eqref{mix1} as a mixture of weighted kernels. Let $\mathcal{K}_j\equiv\mathcal{K}_j(\theta_j,\xi)= \int w(z,\xi) \phi(z,\theta_j)dz$. Then,
\begin{equation}
	f^W_1(z|\boldsymbol{\theta},\xi)= \frac{w(z,\xi)}{\tilde{\mathcal{K}}}\sum_{j=1}^J p_j \phi_j(z,\theta_j)
	=\sum_{j=1}^J \pi_j \phi^{W}_j(z,\theta_j), 
	\label{mix2}
\end{equation}
where $\pi_j=p_j\mathcal{K}_j/\tilde{\mathcal{K}}$ is the new mixing proportion and $\phi^{W}_j$ the corresponding weighted kernel. We underline that $\boldsymbol{\pi}=\left(\pi_1,\dots,\pi_J\right)$ is a vector of proper mixing weights, since $\tilde{\mathcal{K}}=\sum_{j=1}^J p_j\mathcal{K}_j$. The same reparameterization also holds for mixtures with an unbounded number of components due to the dominated convergence theorem. Going from a weighted mixture to a mixture of weighted kernels is computationally convenient because it allows decomposing the general normalizing constant $\tilde{\mathcal{K}}$ into $J$ different component-wise contributions $\mathcal{K}_j$. This way, we can break down the normalizing constant from the mixture weights, and one can compute the component-wise contributions in parallel.

\subsection{$FDR$, $FOR$, and Power as function of the Acceptance Region } 
Let $\mathcal{A}=\left(\underline{z},\bar{z}\right)$ be a given acceptance region. Recall that the Bayesian False Discovery Rate, Bayesian False Omission Rate, and power (sensitivity) are defined as $FDR(\mathcal{A}) = \mathbb{P}\left[H_0|Z\notin\mathcal{A} \right]$, $    FOR(\mathcal{A}) = \mathbb{P}\left[H_1|Z\in\mathcal{A} \right]$, and $1-\beta(\mathcal{A}) = 1-\mathbb{P}\left[Z\in\mathcal{A}|H_1\right]$, respectively.  Let also $\mathbb{P}\left[H_0\right]=(1-\rho)$, $F(z)=(1-\rho) F_0(z)+\rho F_1(z)$ and  $F^{NL}(z)=(1-\rho) F_0(z)+\rho F^{NL}_1(z)$. 
Then,
\begin{align*}
	FDR(\mathcal{A})-FDR^{NL}(\mathcal{A})\geq 0 \iff\\ \frac{\mathbb{P}\left[Z\notin\mathcal{A}|H_0\right](1-\rho)}{\mathbb{P}\left[Z\notin\mathcal{A}\right]  }-
	\frac{\mathbb{P}^{NL}\left[Z\notin\mathcal{A}|H_0\right](1-\rho)}{\mathbb{P}^{NL}\left[Z\notin\mathcal{A}\right]  } \geq 0 \iff\\
	\frac{\left[F_0(\underline{z})+1-F_0(\bar{z})\right](1-\rho)}{ F(\underline{z})+1-F(\bar{z})  }-
	\frac{\left[F_0(\underline{z})+1-F_0(\bar{z})\right](1-\rho)}{ F^{NL}(\underline{z})+1-F^{NL}(\bar{z})  }  \geq 0  \iff\\
	F^{NL}(\underline{z})+1-F^{NL}(\bar{z}) - (F(\underline{z})+1-F(\bar{z}))  \geq 0 \iff\\
	(1-\rho) F_0(\underline{z})+\rho F^{NL}_1(\underline{z})+1-(1-\rho) F_0(\bar{z}) - \rho F^{NL}_1(\bar{z}) - \\(1-\rho) F_0(\underline{z})-\rho F_1(\underline{z}) -1 +(1-\rho) F_0(\bar{z})+\rho F_1(\bar{z})  \geq 0 \iff \\
	F_1(\bar{z}) -  F^{NL}_1(\bar{z})  + F^{NL}_1(\underline{z})- F_1(\underline{z})  \geq 0.
\end{align*}

Similarly, we have
\begin{align*}
	FOR(\mathcal{A})-FOR^{NL}(\mathcal{A})&\geq 0 \iff\\ \frac{\mathbb{P}\left[Z\in\mathcal{A}|H_1\right](\rho)}{\mathbb{P}\left[Z\in\mathcal{A}\right]  }-
	\frac{\mathbb{P}^{NL}\left[Z\in\mathcal{A}|H_1\right](\rho)}{\mathbb{P}^{NL}\left[Z\in\mathcal{A}\right]  } &\geq 0 \iff\\
	\frac{F_1(\bar{z})-F_1(\underline{z})}{F(\bar{z})-F(\underline{z})} - \frac{F^{NL}_1(\bar{z})-F^{NL}_1(\underline{z})}{F^{NL}(\bar{z})-F^{NL}(\underline{z})} &\geq 0 \iff \\
	1-\frac{F_0(\bar{z})-F_0(\underline{z})}{F(\bar{z})-F(\underline{z})} - 1 + \frac{F_0(\bar{z})-F_0(\underline{z})}{F^{NL}(\bar{z})-F^{NL}(\underline{z})} &\geq 0 \iff \\
	F(\bar{z})-F(\underline{z})- F^{NL}(\bar{z})+F^{NL}(\underline{z}) &\geq 0 \iff \\
	F_1(\bar{z}) -  F^{NL}_1(\bar{z})  + F^{NL}_1(\underline{z})- F_1(\underline{z}) & \geq 0,
\end{align*}
and
\begin{align*}
	\beta(\mathcal{A})-\beta^{NL}(\mathcal{A})&\geq 0 \iff\\ \mathbb{P}\left[Z\in\mathcal{A}|H_1\right]-
	\mathbb{P}^{NL}\left[Z\in\mathcal{A}|H_1\right] &\geq 0 \iff\\
	F(\bar{z})-F(\underline{z})- F^{NL}(\bar{z})+F^{NL}(\underline{z}) &\geq 0 \iff \\
	F_1(\bar{z}) -  F^{NL}_1(\bar{z})  + F^{NL}_1(\underline{z})- F_1(\underline{z}) & \geq 0.
\end{align*}

So we showed that the differences in $FDR$, $FOR$ and $\beta$ between the unweighted and weighted case all simplify into the expression \begin{equation}
	F_1(\bar{z}) - F_1(\underline{z}) - ( F^{NL}_1(\bar{z})  - F^{NL}_1(\underline{z})) \geq 0,
	\label{diffe}
\end{equation} 
which is the difference in the areas under the densities $f_1$ and $f_1^{NL}$ computed over $\mathcal{A}$.
Notice that the True Negative Rate (or Specificity), $TNR(\mathcal{A})=\mathbb{P}\left[Z\in \mathcal{A}|H_0 \right]$, and the False Positive Rate $FPR(\mathcal{A})=\mathbb{P}\left[Z\notin \mathcal{A}|H_0 \right]$
are unaltered by our weighted distortion. \\
Now, suppose that \eqref{diffe} holds. Then, given the acceptance region $\mathcal{A}$, we have $\beta^{NL}(\mathcal{A})>\beta(\mathcal{A})$ (i.e., the non-local test has higher sensitivity or $TPR$) for the same level of $FPR(\mathcal{A})=1-TNR(\mathcal{A})$.
Therefore, the $ROC$ curve, generally defined as $(1-TNR(\mathcal{A}),TPR(\mathcal{A})),$ is uniformly higher in the weighted case. This also implies that $AUC^{NL}>AUC.$\\

\subsection{Proof of Proposition 1}

First, we report a Lemma proved in \cite{Dharmadhikari1983} that will be useful for the proof of Proposition 1.
\begin{lemma}
	Suppose now that $X$ and $Y$ are nonnegative random variables with $F(0) = G(0)$. Suppose  also that $F$ and $G$ have  densities $f$ and $g$ on $(0,+\infty)$ with  respect to Lebesgue measure. 
	Either of the following conditions imply that $F$ stochastically dominates by $G$, i.e., $\forall x, F(x) \leqslant G(x)$:\\
	a. The density $g$ crosses $f$ only once and from above.\\
	b. For all $t \in(F(0), 1)$,
	$\frac{\mathrm{d}}{\mathrm{d} t}\left\{F^{-1}(t)-G^{-1}(t)\right\} \geqslant 0$
	or, equivalently $f\left[F^{-1}(t)\right] \leqslant g\left[G^{-1}(t)\right]$.
\end{lemma}

The proof of Proposition 1 follows.

\begin{proof} 
	For conciseness, in this proof we will drop the subscript 1 from $f_1$ and $F_1$, since the alternative distribution is the only one of interest. Consider a generic random variable $Z$, characterized by a local density $f(z)$ symmetric in 0 and its weighted, non-local version $Z_{NL}$, with $f_{NL}(z)$. Denote with $\mathcal{K}$ the normalizing constant of the non-local density, $\mathcal{K}=\int w(z;\xi)f(z) dz$.
	Let $Z^T$ and $Z^T_{NL}$ indicate the truncations on the positive semi-axis of the r.v.s $Z$ and $Z_{NL}$, respectively. Thanks to the symmetry of the distributions, we can state that $f^T(z) = 2 f(z)\mathbb{I}_{\left[0,+\infty\right)}$ and let $F^T(z)=2F(z)-1$ for $z>0$ be its c.d.f. The same can be said about $f^T_{NL}(z)$. Applying the Lemma from in \cite{Dharmadhikari1983} -- reported above -- we want to conclude that $Z^T_{NL}$ stochastically dominates (I order) $Z^T$ $\forall z \left[0,+\infty\right)$, meaning that $F^T(z) \geq F^T_{NL}(z)$ and $F^T(z) > F^T_{NL}(z)$ for at least one $z$.\\
	
	To verify condition (a) of the Lemma, we need to study the sign of $\Delta(z)= f^T(z)-f^T_{NL}(z) = (1-w(z;\xi)/\mathcal{K})f^T(z)$. The function $g(z)=1-w(z,\xi)/\mathcal{K}$ is monotone decreasing, given the monotonicity of $w(z,\xi)$ on the positive semi-axis. 
	Temporarily suppose that the weight function $w$ is bounded from above, for all $z$, by some constant $K\geq 0$. It is clear that $\mathcal{K} \leq K$. Therefore, $g(z)$ is 1 in zero and for $z \rightarrow +\infty$ it tends to $1-K/\mathcal{K}\leq 0$, admitting a unique root $z^*$. This is true even if $w$ is unbounded, case that is recovered letting $K\rightarrow +\infty$.\\
	
	On the one hand, if the weight function is continuous then the monotonicity of $g$ and the positivity of $f^T$ imply that $\Delta(z)=g(z)\cdot f^T(z)$ has only one zero, occurring in $z^*$ as well. It follows that $\lim_{z\rightarrow z^*+}\Delta(z) = 0^-$ and $\lim_{z\rightarrow z^*-}\Delta(z)=0^+$, so we can conclude that $f^T$ crosses $f^T_{NL}$ just once and from above, thus the condition is satisfied. \\
	
	On the other hand, if the weight function exhibits any discontinuity point (e.g., $w(z,a)=1-\mathds{1}_{\{-a,a\}}$) the existence (and uniqueness) of the root $z^*$ is not guaranteed. However, we can redefine it as $z^*= \sup\{z\geq 0: g(z)>0\}$, i.e., the point where the function $g(z)$ (and therefore $\Delta(z)$) changes sign. Given the monotonicity of $g(z)$, $z^*$ is unique. Then, for the result to hold it is sufficient that $\lim_{z\rightarrow z^*+}\Delta(z) = a_2 < 0$ and $\lim_{z\rightarrow z^*+}\Delta(z) = a_1 >0$, which is again true given the sign of $g(z)$. 
	We also notice that the previous argument holds true even in the cases where the equation $g(z)=0$ admits multiple solutions.
	Therefore, on the positive semiaxis, $F^{T}_{NL}$ stochastically dominates $F^{T}$.\\
	
	The previous arguments imply that, $\forall z \geq 0$, $F^T(z) \geq F^T_{NL}(z) \iff 2F(z)-1 \geq 2F_{NL}(z)-1 \iff F(z) \geq F_{NL}(z)$, showing that, on the positive semi-axis, the c.d.f. of $Z$ is always greater that its weighted counterpart. 
	Exploiting the symmetry of the densities of the two random variables, the converse holds on the negative semi-axis, and then the result follows.\\
\end{proof}

\subsection{Proof of Proposition 2}
\begin{proof}
	Consider the function $g(z)=1-w(z,\xi)/\mathcal{K}$.  Let us denote with $\pm z^*= \pm \sup\{z \geq 0: g(z)>0\}$. If $w$ is proper, then $w$ is an even function, monotone on each semi-axes: therefore we can compute analytically $\pm z^*=\pm w^{-1}\left(\mathcal{K}\right)$. 
	Now let us define the function $H(z)=F_1(z)-F^{NL}_1(z)=\int_{-\infty}^{z}g(x)f_1(x)dx.$ Notice that Equation \eqref{diffe} can be rephrased as $H(\bar{z})-H(\underline{z})$.\\
	
	In general, $\lim\limits_{z \rightarrow -\infty} H(z)=0^-$ and $\lim\limits_{z \rightarrow +\infty} H(z)=0^+$. We can study the sign of the derivative $H'(z)=g(z)f_1(z)$. We observe that $H$ starts negative, decreases until its point of global minimum $-z^*$, then increases until its point of global maximum $z^*$ and 
	finally decreases towards zero from above as $z\rightarrow +\infty$. Let us call $\hat{z}$ the point where $H\left(\hat{z}\right)=0$. It must be that $\hat{z} \in \left[-z^*,z^*\right]$. Because $g$ is not requested to be strictly monotone, there could be cases where we can recover an interval $\hat{\mathcal{I}}$ such that $\forall z \in \hat{\mathcal{I}}=\left[\hat{z}_1,\hat{z}_2\right]$, $H(z)=0$. Evidently, $\hat{\mathcal{I}}$ has to be contained in $\left[\underline{z},\bar{z}\right]$. Without loss of generality, let us assume that $\hat{\mathcal{I}}\equiv \hat{z}$, i.e., $H(z)$ admits a unique root.
	
	A sufficient condition for \eqref{diffe} to hold is given by $H(\bar{z})>0$ and $H(\underline{z})<0$, which holds every time $\underline{z}<\hat{z}<\bar{z}$. However, not always a closed-form expression is available for $\hat{z}$. Nevertheless, we can say that the same condition is true every time that the acceptance region $\mathcal{A}$ contains the two roots $\pm z^*$, because this would imply that $\underline{z}\leq -z^*\leq \hat{z} \leq z^* \leq \underline{z}$.\\ \end{proof}

\begin{figure}[!ht]
	\centering
	\includegraphics[width=.5\linewidth]{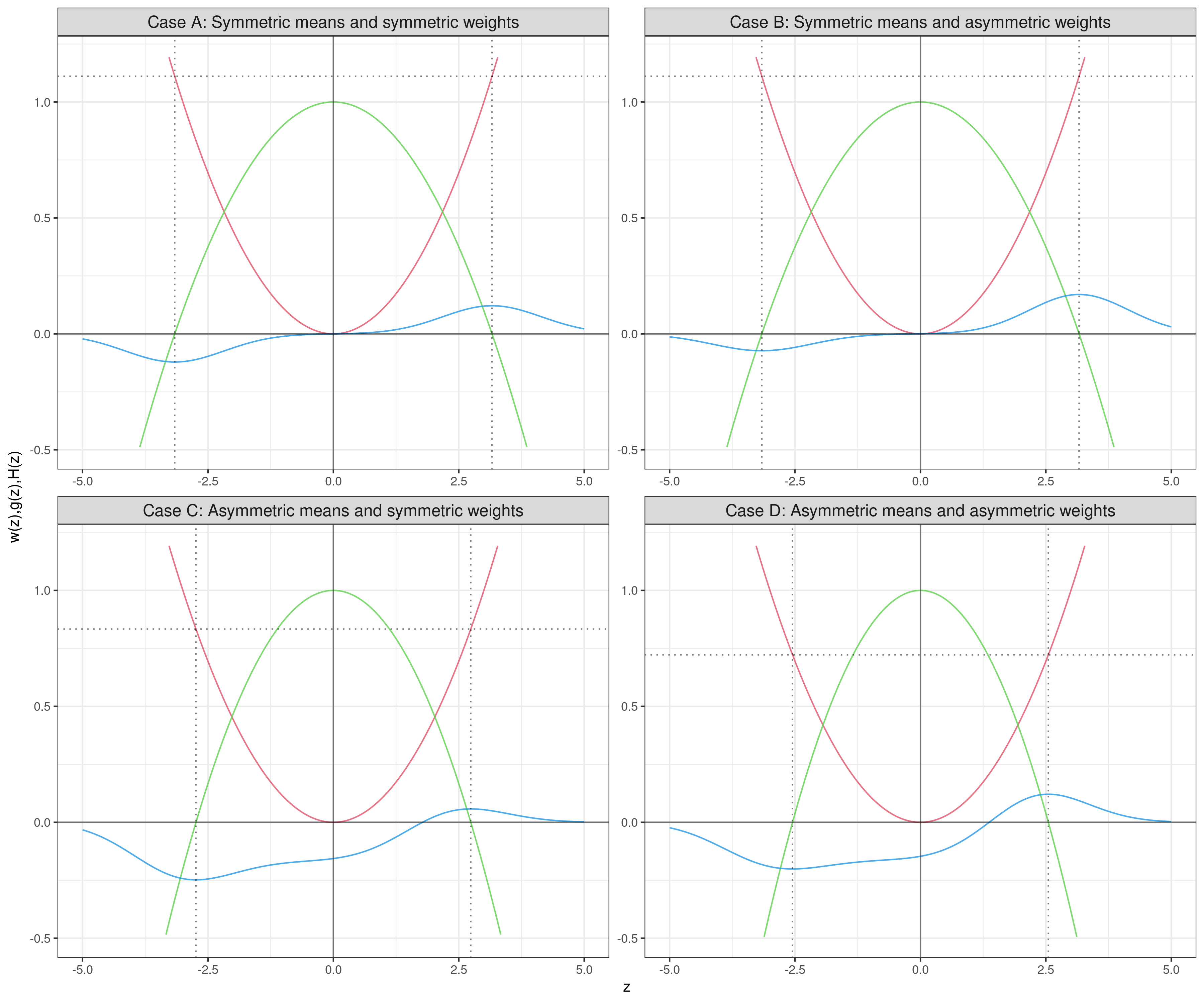}
	\caption{Different configurations of functions $w_0(z)$ (red), $g(z)$ (green) and $H(z)$ (blue) in four scenarios. The horizontal line represents the normalizing constant $\mathcal{K}$ and the vertical lines denote $\pm z^*$. }
	\label{fig:P1}
\end{figure}

\begin{figure}[!ht]
	\centering
	\includegraphics[width=.5\linewidth]{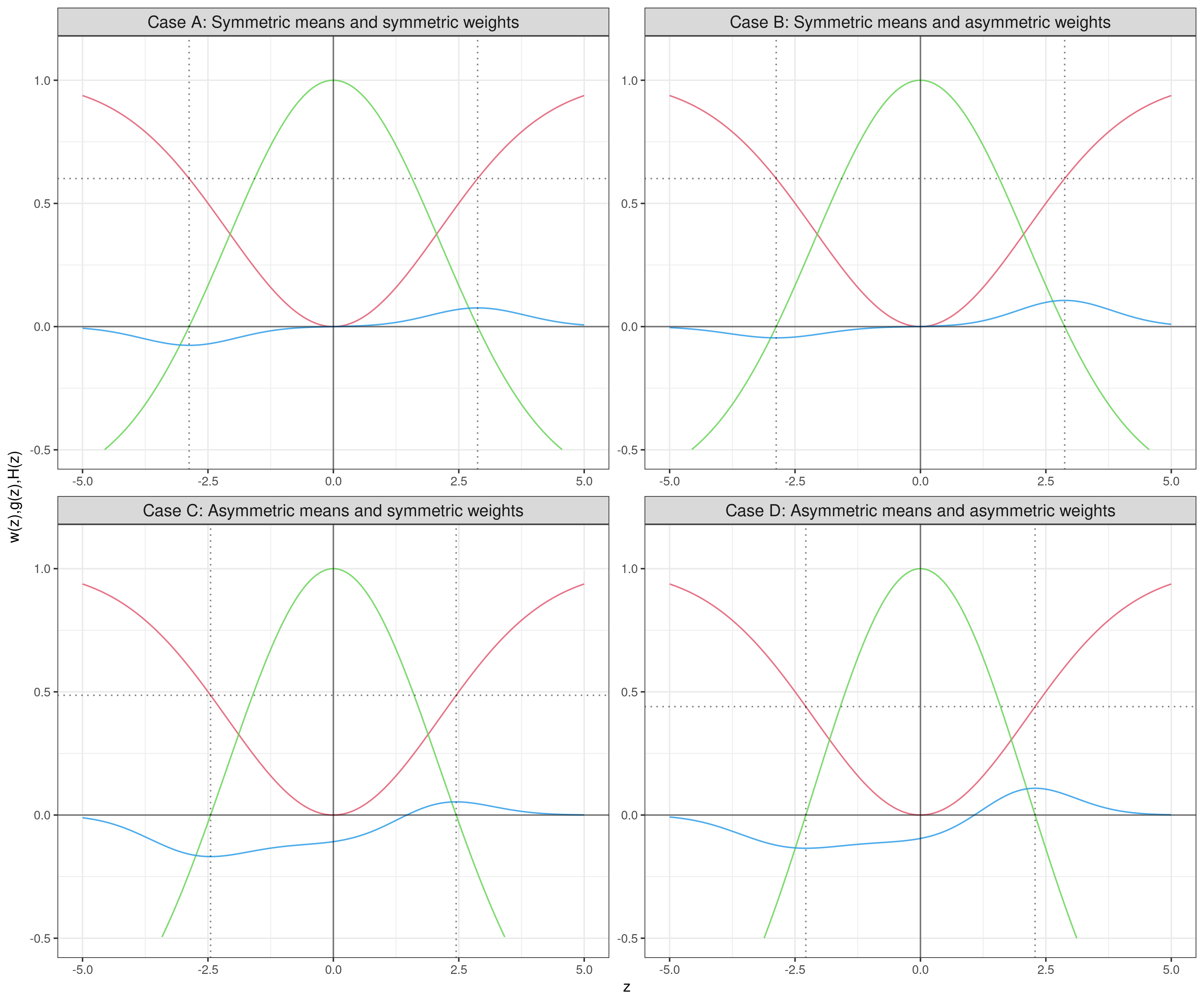}
	\caption{Different configurations of functions $w_1(z)$ (red), $g(z)$ (green) and $H(z)$ (blue) in four scenarios. The horizontal line represents the normalizing constant $\mathcal{K}$ and the vertical lines denote $\pm z^*$. }
	\label{fig:P2}
\end{figure}

\begin{figure}[!ht]
	\centering
	\includegraphics[width=.7\linewidth]{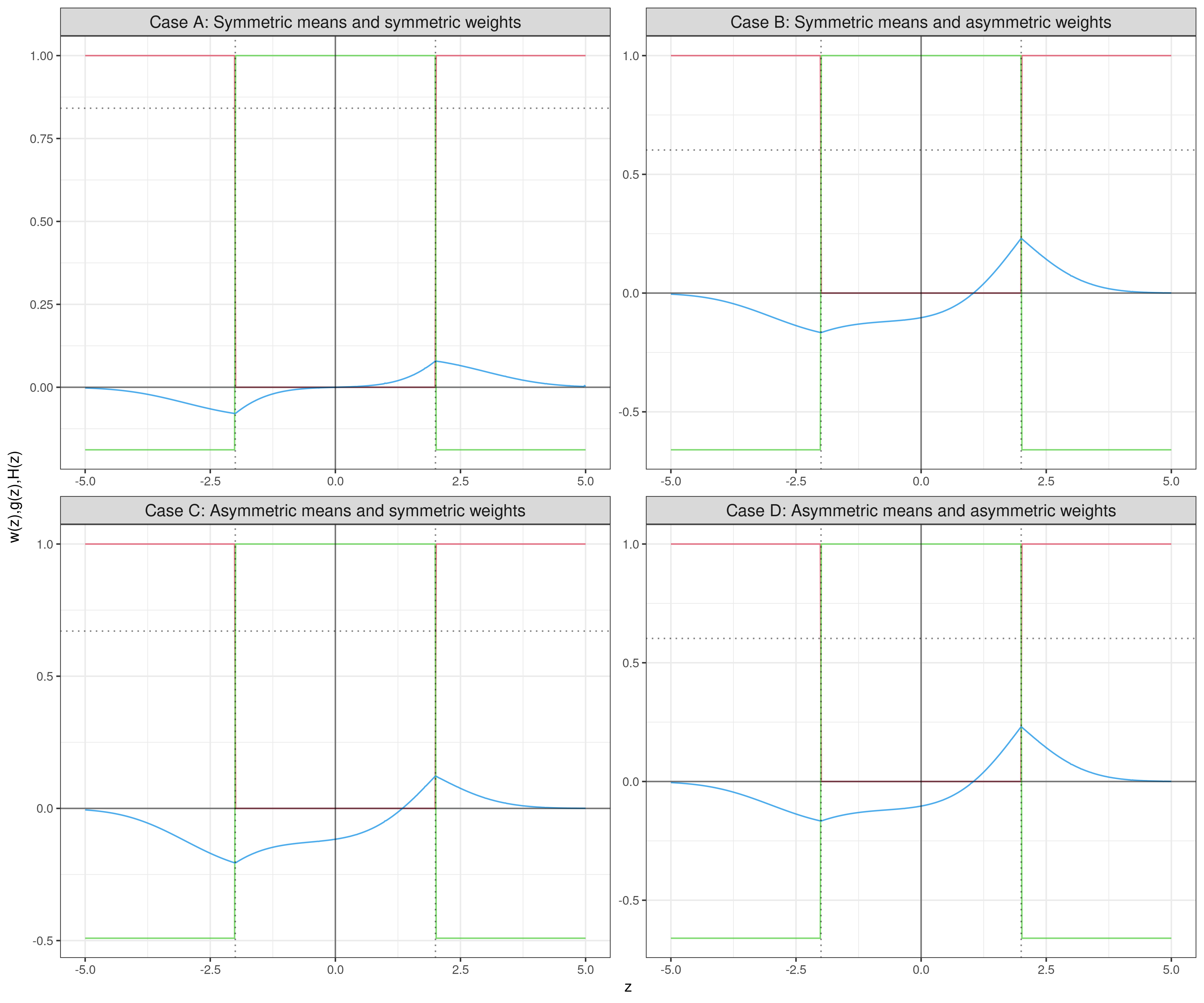}
	\caption{Different configurations of functions $w_I(z)$ (red), $g(z)$ (green) and $H(z)$ (blue) in four scenarios. The horizontal line represents the normalizing constant $\mathcal{K}$ and the vertical lines denote $\pm z^*$. }
	\label{fig:P3}
\end{figure}

\textbf{Some visual examples.} Figure \ref{fig:P1}, \ref{fig:P2}, and \ref{fig:P3} showcase a representation of the different functions we defined to prove our propositions. In Figure \ref{fig:P1} and \ref{fig:P2} the weight functions are, respectively, $w_0(z;a)= (z/a)^2$ and $w_1(z;a)= 1-\exp\{-(z/a)^2\}$, whilst in Figure \ref{fig:P3} we display the case of $w_I(z;a)= 1 -\mathds{1}_{\left[-a,a\right]}$ with $a=2$. The function $f_1$ is assumed to be a mixture of two Normals $\omega_1 \phi(m_1,s_1^2)+\omega_2 \phi(m_2,s_2^2)$. 
Each panel considers a different combinations of the mixture weights and the Normal means, to explore the generality of the proposed guidelines in case of asymmetries.


\subsection{An intuitive example}
Recall our definition of acceptance region  $\mathcal{A}$:
\begin{equation} 
	\mathcal{A}=
	\bigg\{z\in \mathbb{R}: P_1(z) = 
	\frac{\rho f_1(z)}{f(z)}\geq \nu \bigg\},
	\label{accregion}
\end{equation}
Then, consider a generic alternative density $f_1$ with c.d.f. $F_1$ and the weight function $w(z;\xi)=\mathds{1}_R$, where $R=\left(-\infty,-\delta\right)\cup \left(\delta,+\infty\right)$. This gives us $\mathcal{K}=F_1(-\delta)+1-F_1(\delta)$ and $z^*=\pm \delta$. We can derive a close expression for $H(z)$:
$$H(z)= F_1(z)-\frac{1}{\mathcal{K}}\left[ \mathds{1}_{z<-\delta}F_1(z)+\mathds{1}_{Z\in \left(-\delta,\delta\right)}F_1(-\delta) + \mathds{1}_{z>\delta}\left(F_1(z) - F_1(\delta)+F_1(-\delta) \right)   \right].$$
Knowing that the only root $\hat{z}$ can only be in $\left(-\delta,\delta\right)$, we can compute
$$\hat{z}=F_1^{-1} \left(\frac{F_1(-\delta)}{\mathcal{K}}\right).$$
In this particular example, applying the criterion showed in \eqref{accregion} we conclude that 
$\forall z \in R$ we have that $P_1^{NL}(z)=\rho f^{NL}_1(z)/f^{NL}(z)=0$. 
Thus, we are sure that, for $\nu > 0$, $\mathcal{A}$ can be either $\left(-\delta,\delta\right)$ or wider. This means that condition 
$$F_1(\bar{z})-F^{NL}_1(\bar{z})+F^{NL}_1(\underline{z})-F_1(\underline{z})  \geq 0$$
is respected and that the Bayesian $FDR$, Bayesian $FOR$ are lower in this weighted case, whilst $\beta$ and AUC are higher. 

\FloatBarrier

\section{Computational Details} \label{GIBBS}

\subsection{Gibbs sampler for the parametric model specification}
In what follows, we detail the steps of the Gibbs sampler for the Nollik model. Notice that, for the sake of brevity, we set $\phi_j = \phi(z_i;\mu_j,\sigma^2_j)$. The algorithm proceeds iteratively sampling from the following full conditionals:

\begin{enumerate}
	\item The full conditional of $\rho$ is $$\rho|\cdots \sim Beta\left(a_\rho + \sum_{i=1}^{N}\lambda_i, b_\rho +N- \sum_{i=1}^{N}\lambda_i\right),$$ thanks to conjugacy.
	\item Given their strong posterior dependence, we sample together $\left(\bm{ \Lambda },\bm{ \Gamma }\right)$ from their joint full conditional, given by: 
	\begin{equation*}
		\begin{aligned}
			\pi \left( \left(\bm{ \Lambda },\bm{ \Gamma }\right)|\cdots \right) =  &\prod_{i=1}^{N} Bern(\lambda_i;\rho) \cdot \pi\left(\gamma_i|\lambda_i,\alpha\right) \cdot  ( \left(1-\lambda_i\right) \delta_{0}\left(\gamma_i\right)\phi_0 +\\& \lambda_i
			\left[   \frac{w(z_i;\xi)}{\mathcal{K}_{\gamma_i}(\bm{\theta}_1)} \left[
			\phi_1\delta_{1}\left(\gamma_i\right)+\phi_2\delta_{2}\left(\gamma_i\right) \right]\right] ).
		\end{aligned}
	\end{equation*}
	We can update each component of $\left(\bm{ \Lambda },\bm{ \Gamma }\right)$ individually, rewriting:
	\begin{equation*}
		\begin{aligned}
			\pi \left( \left(\lambda_i,\gamma_i\right)|\cdots\right) \propto &  \rho^{\lambda_i} (1-\rho)^{1-\lambda_i}
			\pi\left(\gamma_i|\lambda_i,\alpha\right)\cdot\left( \phi_0^{1-\lambda_i} \cdot
			\left[   \frac{w(z_i;\xi)}{\mathcal{K}_{\gamma_i}(\bm{\theta}_1)}\left[\phi_1^{\delta_{0}\left(\gamma_i\right)}\cdot\phi_2^{\delta_{1}\left(\gamma_i\right)}\right]\right]^{\lambda_i}\right).
		\end{aligned}
	\end{equation*}
	In particular, the only scenarios with non-null probabilities are:
	\begin{align*}
		\pi \left( \lambda_i=0,\gamma_i=0|\cdots\right) &\propto   (1-\rho) \phi_0, \\
		\pi \left( \lambda_i=1,\gamma_i=1|\cdots\right) &\propto   \left(1-\alpha\right) \cdot \rho \left[   \frac{w(z_i;\xi)}{\mathcal{K}_1(\bm{\theta}_1)}\phi_1\right], \\
		\pi \left( \lambda_i=1,\gamma_i=2|\cdots\right) &\propto  \alpha \cdot \rho \left[   \frac{w(z_i;\xi)}{\mathcal{K}_2(\bm{\theta}_1)}\phi_2\right] .
	\end{align*}
	\item Let $n_{1,2}=\sum_{i=1}^{N}\mathbb{I}_{\{\gamma_i\neq 0\}}$. The full conditional of $\alpha$ is\\ $$Beta\left(a_\alpha + \sum_{i=1}^{N}\gamma_i, b_\alpha +n_{1,2}- \sum_{i=1}^{N}\gamma_i\right),$$ due to conjugacy.
	\item Let $n_{0} = \sum_{i=1}^{N} \mathbb{I}_{\{\lambda_i=0\}}$, $\bar{z}_{0} =  \frac{\sum_{i=1}^{N} z_i \cdot  \mathbb{I}_{\{\lambda_i=0\}}}{n_{0}}$, and $SQ_{0}^2 = \sum_{i=1}^{N}\left(z_i-\bar{z}_{0}\right)^2 \mathbb{I}_{\{\lambda_i=0\}}$.
	The full conditional for $\left(\mu_0, \sigma^2_0\right)$ is given by 
	\begin{equation*}
		\pi\left(\left(\mu_0, \sigma^2_0\right)|\cdots\right)  \sim NIG(m^*_0,\kappa^*_0,a^*_0,b^*_0)
	\end{equation*}
	where 
	$$m^*_0 = \frac{\kappa_0 m + n_{0}\bar{z}_{0}  } {\kappa+n_{0}},\quad  \kappa^*=\kappa_0+n_{0}, \quad a^*_0 = a_0+\frac{1}{2}n_{0},$$ and $$b^*_0=b_0+\frac12  SQ_{0}^2+\frac{n_{0}\kappa_0}{n_{0}+\kappa_0}\frac{(\bar{z}_{0}-m_0)^2}{2}.$$
	
	\item Let $n_{1j} = \sum_{i=1}^{N} \mathbb{I}_{\{\lambda_i=1\}}\cdot \mathbb{I}_{\{\gamma_i=j\}}$, $\bar{z}_{1j} =  \frac{\sum_{i=1}^{N} z_i \cdot  \mathbb{I}_{\{\lambda_i=1\}}\cdot \mathbb{I}_{\{\gamma_i=j\}}}{n_{1j}}$, and $SQ_{1j}^2 = \sum_{i=1}^{N}\left(z_i-\bar{z}_{1j}\right)^2 \mathbb{I}_{\{\lambda_i=1\}}\cdot \mathbb{I}_{\{\gamma_i=1\}}$.
	
	The full conditional for $\left(\mu_j, \sigma^2_j\right)$, for $j=1,2$ is given by 
	\begin{align*}
		\pi\left(\left(\mu_j, \sigma^2_j\right)|\cdots\right)  \propto & NIG(m_j,\kappa_j,a_j,b_j)\cdot \mathbb{I}_{\{(-1)^j\mu_j>0\}}\cdot \prod_{\lambda_i=1,\gamma_i=j} \frac{\phi_j}{\mathcal{K}_{\gamma_i}(\bm{\theta}_1)}\\
		\propto & NIG(m^*_j,\kappa^*_j,a^*_j,b^*_j)\cdot \mathbb{I}_{\{(-1)^j\mu_j>0\}} \cdot\frac{1}{ \mathcal{ K}_{j}(\bm{\theta}_1) ^{n_{1j} }}
	\end{align*}
	where 
	$$m^*_j = \frac{\kappa_j m + n_{1j}\bar{z}_{1j}  } {\kappa_j+n_{1j}}, \quad \kappa_j^*=\kappa_j+n_{1j}, \quad  a^*_j = a_j+\frac{1}{2}n_{1j} $$ and $$b^*_j=b_j+\frac12  SQ_{1j}^2+\frac{n_{1j}\kappa_j}{n_{1j}+\kappa_j}\frac{(\bar{z}_{1j}-m_j)^2}{2}.$$
	\item The full conditional of $\xi$ is given by:
	\begin{align*}
		\pi\left(\xi|\cdots\right)  &\propto  \pi\left(\xi\right) \prod_{\lambda_i=1}\frac{w(z_i;\xi)}{\mathcal{K}_{\gamma_i}(\bm{\theta}_1)}=
		\frac{\pi\left(\xi\right)\cdot \prod_{\lambda_i=1}w(z_i;\xi)}{\mathcal{K}_{1}(\bm{\theta}_1)^{n_{11}}\cdot\mathcal{K}_{2}(\bm{\theta}_1)^{n_{12}}}.
	\end{align*}
\end{enumerate}

Notice that Steps 2 and 5 of the algorithm are easily parallelizable over the observations and the parameters, respectively.
Steps 5 and 6 requires a Metropolis-Hastings step. 
In particular, we employ a random walk Metropolis algorithm with Gaussian proposal distribution for the means and the log-variances. To improve convergence, we adopt an adaptive strategy that allows online tuning of the variance of the proposal, as suggested in \cite{Roberts2009}.
The initial variances of the jumps or the random walk Metropolis steps are fixed equal to $ \Sigma_{\left(\mu_1,\sigma^2_1\right)}=\Sigma_{\left(\mu_2,\sigma^2_2\right)}=diag(0.5,0.5)$ and $\sigma^2_\xi=0.5$.
Every $n_{batch}=50$ iterations, these values are updated in the following way: at the $t$-th iteration, if the acceptance rate in the last examined batch is lower than the optimal rate of 0.44, the logarithm of each standard deviation is lowered by the quantity $\delta ( t ) = \min \left( 0.01, t ^ { - 1 / 2 } \right)$, otherwise it is increased of the same quantity. Notice that the adaptive term is vanishing, so the convergence to the desired target distribution is preserved
\citep{Roberts2007,Roberts2009}. \\

\subsection{Gibbs sampler for the nonparametric model specification}
To implement the sampling algorithm for the Bayesian nonparametric version of Nollik model, we use the truncated representation of \cite{Ishwaran2001a}, where the infinite sum is substituted with a sufficiently large number of mixture components $J$. The collapsed Gibbs sampler we employ mimics the finite-dimensional case with few modifications. Recall that now $\gamma_i\in \{0,1,2,\ldots\}$.
Steps 1 and 4 of the algorithm for the parametric model are unchanged, whilst the others become:
\begin{itemize}
	\item [2.] The non-null scenarios for $\left(\bm{\Lambda},\bm{\Gamma}\right)$ are
	\begin{align*}
		\pi\left(\lambda_i=0,\gamma_i=0\right)&\propto (1-\rho)\phi_0,\\
		\pi\left(\lambda_i=1,\gamma_i=j\right)&\propto \rho \pi_j\left[\frac{w(z_i;\xi)}{\mathcal{K}_{\gamma_i}}\phi(\mu_{\gamma_i},\sigma^2_{\gamma_i} )\right] \text{for } j\in\{1,2,\ldots,J\}.
	\end{align*}
	\item [3.] Let $n_{1j}=\sum_{i=1}^N \mathbb{I}_{\lambda_i=1}\mathbb{I}_{\gamma_i=j}.$ The stick-breaking weights are constructed with the auxiliary variables $u_j$, which in turn have full conditionals of the form 
	\\
	
	$u_j\sim Beta\left(1+n_{1j},a+\sum_{l<j}n_{1l} \right)$ for  $j\in\{1,2,\ldots,J\}$.
	\item [5.] With reference
	to Step 5 of the previous algorithm, we now have:
	$$
	\pi\left(\left(\mu_j, \sigma^2_j\right)|\cdots\right)  \propto  NIG(m^*_j,\kappa^*_j,a^*_j,b^*_j)\cdot  \frac{1}{ \mathcal{ K}_{\gamma_i}(\bm{\theta}_1) ^{n_{1j} }}.$$
	\item [6.] Lastly, 
	$$    \pi\left(\xi|\cdots\right)  \propto
	\frac{\pi\left(\xi\right)\cdot \prod_{\lambda_i=1}w(z_i;\xi)}{\mathcal{K}_{1}(\bm{\theta}_1)^{n_{11}}\cdots\mathcal{K}_{J}(\bm{\theta}_1)^{n_{1J}}}.$$    
	\item[7.] We can place a conjugate Gamma prior, $Gamma(\alpha_a,\beta_a)$, on the concentration parameter $a$, obtaining $a|\cdots \sim Gamma\left(\alpha_a+(J-1), \beta_a+ \sum_{j=1}^{J-1}\log(1-u_j) \right)$.
\end{itemize}

\subsection{Computational Burden}

The proposed algorithms are efficient and able to handle large datasets. To provide evidence for this claim, Table \ref{tab:times} reports the mean and the standard deviation of the running times (in seconds, where reported values are obtained by averaging over 20 Monte Carlo runs) needed to complete 1,000 iterations for different sample sizes on an i7-5500U -- 2.40GHz laptop. For comparison, we report the running times in seconds of both the parametric and nonparametric versions of the model. We truncated the infinite BNP mixture at $J=20$. The parametric model is 6 to 10 times faster than the BNP version.

\begin{table}[!ht]
	\centering
	\begin{tabular}{lcccccc}
		\toprule
		& $n=100$ & $n=500$ & $n=1,000$ & $n=5,000$ & $n=10,000$ &    $n=50,000$\\
		\midrule
		$Nollik$ &1.038 & 1.4444 & 1.8858 & 5.9271 & 10.2036 & 45.8604\\
		&(0.088) & (0.2031) & (0.1942) & (0.7047) & (0.8043) & (1.8636)\\
		\midrule
		$\text{BNP-}Nollik$ & 6.5350 & 10.4658 & 15.2201 & 50.8832 & 96.2305 & 451.3572\\
		&(0.1187) & (0.1242) & (0.1542) & (0.2324) & (0.5391) & (2.8465)\\
		\bottomrule
	\end{tabular}
	\caption{Simulation time in seconds of the Nollik parametric and nonparametric models using $w_1$ as weight function to obtain 1,000 iterations for different sample size values. 
		Results are averaged over 20 Monte Carlo runs (standard deviations are reported in parentheses).}
	\label{tab:times}
\end{table}
\FloatBarrier

\begin{figure}[ht]
	\centering
	\includegraphics[width = .75\linewidth]{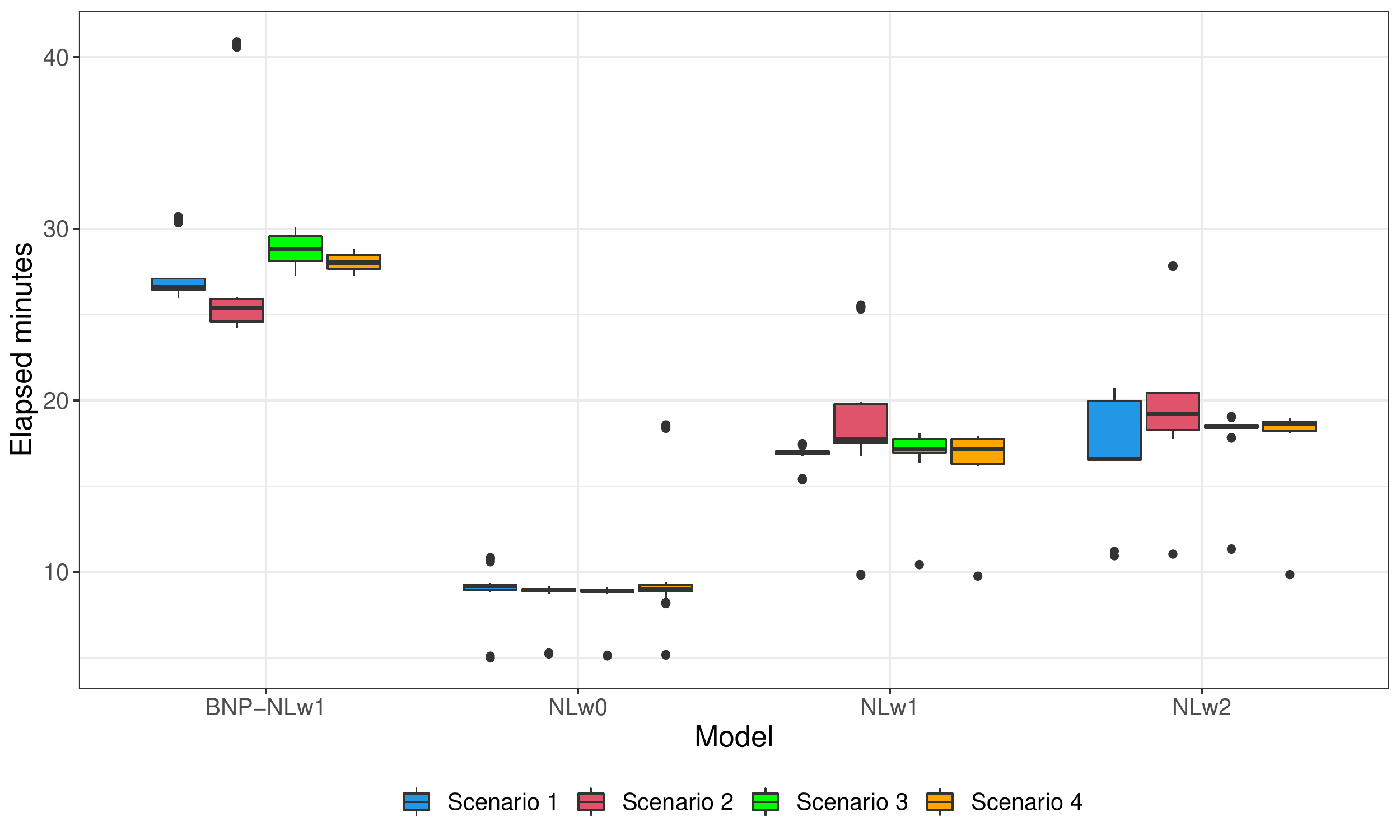}
	\caption{Boxplots stratified by the scenarios of the simulation study reported in the main text and colored according to the different Nollik specifications we considered.}
	\label{fig:time}
\end{figure}

Additionally, in Figure \ref{fig:time} we report the distributions of the computational times (in minutes) needed to fit the models in the simulation study of Section 5 of the main paper. These experiments were run on a 2.6 GHz 6-Core Intel Core i7 MacBook Pro. We see how the computational cost increases as we consider more complicated weight functions ($w_1$ and $w_2$, which implies the addition of the MH step for the parameter $\xi$). As expected, the BNP specification is the most expensive.

\subsection{Slice Sampling of Weighted Distributions} \label{SLICE}

We can take advantage of the product form of $\pi_{w}$ to introduce a general Slice Sampler that generates random variates from the weighted distribution of interest. 
\cite{Rossell2017a} showed that every non-local density can be seen as a mixture of truncated distributions, and then proposed intuitive sampling schemes (Algorithm 1 and 2 in their paper) which make the posterior simulation easier in a wide variety of cases. However, their result can be seen as a particular version of the Slice Sampler \citep{Damien1999,Neil2003,Mira2002}, exploited also in \cite{Petralia2012aaa}. We rephrase and adapt the algorithm to our framework, whenever a bounded weight function $w:\mathbb{R}\rightarrow\left[0,K\right]$ is assumed. Without loss of generality, we set $K=1$. Using the idea of data augmentation, we introduce a Uniform latent variable in the weighted density, obtaining:
\begin{equation}
	\pi_W\left(\theta,u;\xi,\eta\right) \propto  \pi\left(\theta;\eta\right)\mathbb{I}_{\{w\left(\theta;\xi\right)>u\}}.
	\label{approx}
\end{equation}
Notice that $\int_{0}^{w\left(\theta;\xi\right)}\pi_W\left(\theta,u;\xi,\eta\right) du = w\left(\theta;\xi\right)\pi\left(\theta;\eta\right)$.

Let us denote with $\left(\theta_0,u_0\right)$ the current values for the parameters of interest and let $\left(\theta_*,u_*\right)$ their updated version. The Slice Sampler algorithm for the non-local density is composed by two steps: 

\begin{enumerate}
	\item Sample $u_*$ from a $U\left(0,w\left(\theta_0;\xi\right)\right)$
	\item Sample $\theta_*$ from $\pi\left(\theta;\eta\right)\mathbb{I}_{\{A_*\}}$, i.e., sample the new value from the distribution $\pi\left(\theta;\eta\right)$ truncated on $A_*=\{ \theta:w\left(\theta;\xi\right)>u_* \}$
\end{enumerate}

This algorithm is trivial to implement every time the weight function $w$ is invertible and a sampler for a truncated version of the local density $\pi\left(\theta;\eta\right)$ is available. If a non-local density is used as a prior, as long as the local distribution is conjugate with the likelihood distribution $f\left(\bm{z};\theta\right)$, the derivation of a sampler for the posterior is immediate. In fact, we can recover the same structure of \eqref{approx} writing
\begin{equation*}
	\pi_W\left(\theta,u|\bm{z};\xi,\eta\right) \propto  \pi\left(\theta;\eta\right)\mathbb{I}_{\{w\left(\theta;\xi\right)>u\}} f\left(\bm{z};\theta\right) = \pi\left(\theta|\bm{z};\eta\right)\mathbb{I}_{\{w\left(\theta;\xi\right)>u\}}
\end{equation*}
and then applying the algorithm using $\pi\left(\theta|\bm{z};\eta\right)$
as new local distribution.\\

\begin{figure}[ht]
	\centering
	\includegraphics[scale=.5]{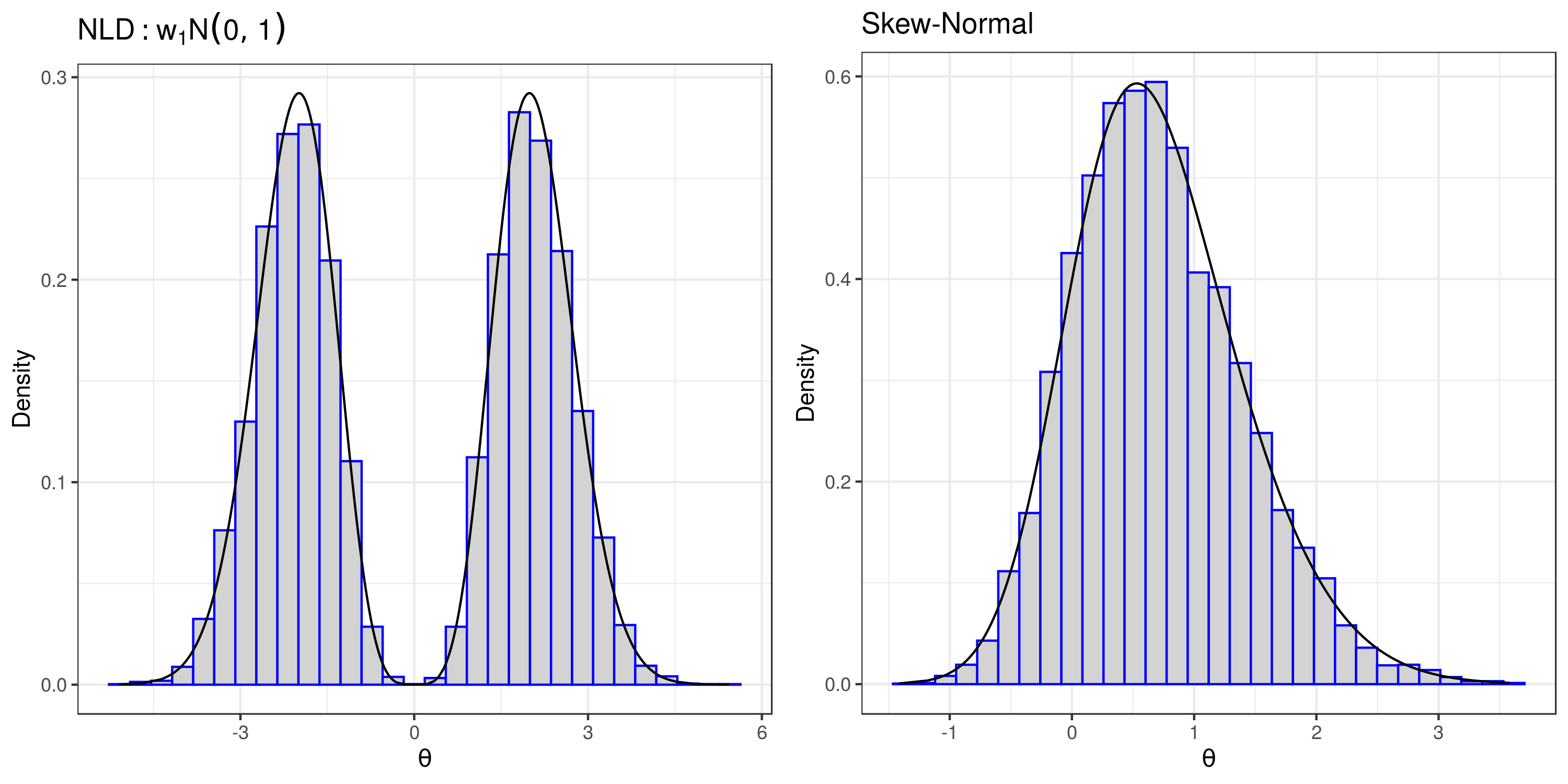}
	\caption{Histograms referring to the two distributions adopted in (S1), on the left, and (S2), on the right.}
	\label{sampler}
\end{figure}

As an example, consider these two different weighted distributions: a non-local density defined by the product of a Standard Gaussian with the weight function $w_1$ and a Skew-Normal$\left(\alpha\right)$:
\begin{equation*}
	(S1)\quad \pi_W\left(\theta\right) = w_1\left(\theta;a,k\right)N\left(\theta;0,1\right) \quad \quad \quad (S2)\quad  \pi_W\left(\theta\right) = 2\Phi\left(\alpha \theta\right)N\left(\theta;0,1\right).
\end{equation*}
To implement the algorithm, we just need to compute the set $A_*$ for both cases. Simple algebra provides the answer:
\begin{equation*}
	A_*^{(S1)} =  \left\{\theta: |\theta|> a \sqrt[2k]{-\log\left(1-u\right)}\right\}
	\quad \quad \quad A_*^{(S2)} =  \left\{\theta: \theta> \frac{1}{a} \Phi^{-1}\left(\frac{u}{2}\right)\right\}.
\end{equation*}
Both scenarios involve sampling from a Truncated Normal distribution. A recent R library, \texttt{TruncatedNormal} \citep{Botev2017} makes this operation extremely efficient. To actually simulate the values, we assumed $a=5,k=1$ and $\alpha=2$.
Figure \ref{sampler} shows the histograms of 10,000 random instances sampled with the described algorithm, where the true density has been superimposed.
\FloatBarrier

\clearpage
\section{Additional Material} 
\label{Additional}
\FloatBarrier
\subsection{Differences a priori between the weighted and unweighted two-group models}
\begin{figure}[ht!]
	\centering
	\includegraphics[width=.8\linewidth]{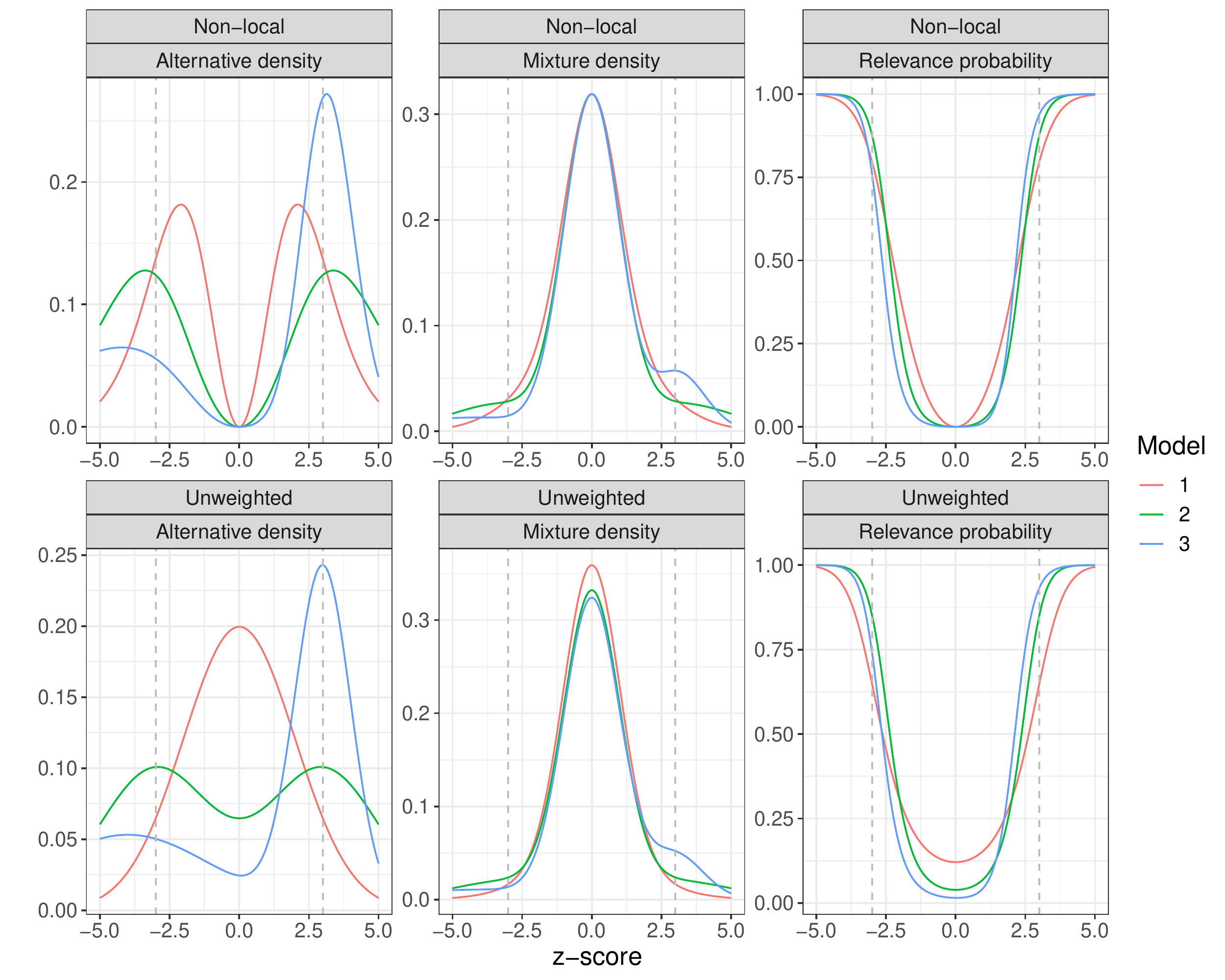}
	\caption{Comparison between the alternative densities $f_1$ (left column), the marginal densities $f$ (central column), and probabilities of relevance (right column) under the non-local (top row) and unweighted models (bottom row) for different parameters specifications summarized in Table \ref{tab:expl}, displayed in different colors. }
	\label{fig:expl}
\end{figure}

Figure \ref{fig:expl} illustrates the introduced alternative densities $f_1(z)$ (left column panels), the mixture densities $f(z)$ (central column panels) and the relevance probability functions $\rho f_1(z)/f(z)$ (right column panels), for the hyperparameter configurations in Table \ref{tab:expl} (reflected by the different colors) and the weight function $w_1$ (top row panels). The non-local alternatives are compared with their unweighted counterpart (bottom row panels). The comparison shows how the choice of weight function affects the shape of the different densities. We point out that in the multiple hypothesis framework, one typically considers a low value of the relevant proportion $\rho$; hence, the marginal mixture density $f$ is essentially unimodal in a neighborhood of the origin. From the right two panels, we appreciate that the non-local specification induces an important modification in the relevance probability. In fact, the function $\rho f_1(z)/f(z)$ under the weighted scenario is forced to assume all the values in $\left[0,1\right)$ by construction, since $\rho f_1(0)/f(0)=0$.

\begin{table}[ht!]
	\centering
	\begin{tabular}{ccccccccc}
		\toprule
		Model & $\mu_0$ & $\sigma_0$& $\mu_1$ & $\sigma_1$ & $\mu_2$ & $\sigma_2$ & $\rho$& $\alpha$ \\
		\midrule    
		1     & 0 & 1.1 & 0 & 2& 0& 2 & 0.2 & 0\\
		2     & 0 & 1   & -3& 2& 3& 2 & 0.2 & 0.5\\
		3     & 0 & 1  & -4 & 3& 3& 1 & 0.2 & 0.4\\
		\bottomrule
	\end{tabular}
	\caption{Hyperparameter specifications used to evaluate the functions depicted in Figure \ref{fig:expl}.}
	\label{tab:expl}
\end{table}

\FloatBarrier

\subsection{Simulation Study: the effect of the parameters $\xi$ and $k$ on the weight function}

\begin{figure}[ht!]
	\centering
	\includegraphics[width = \columnwidth]{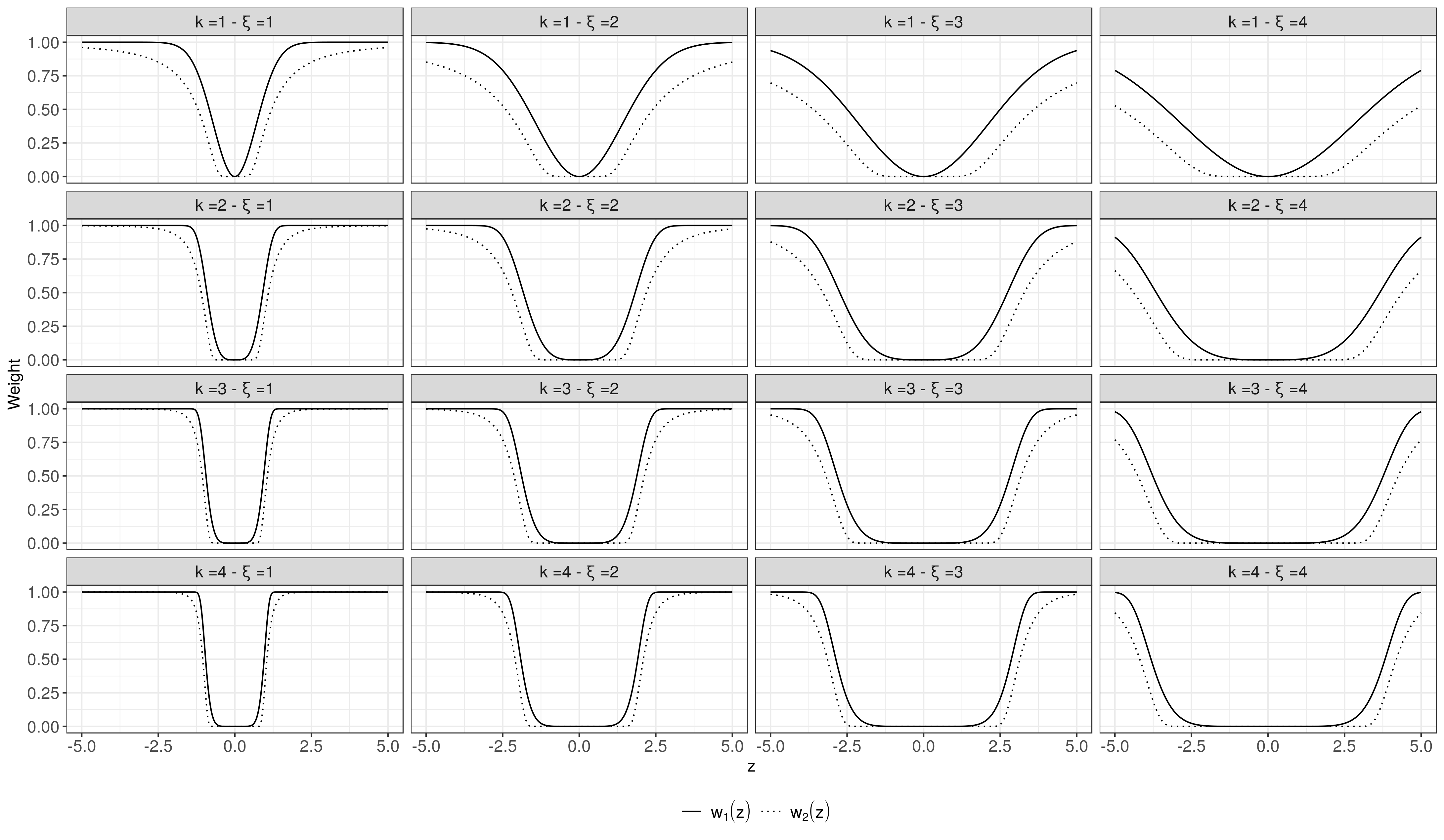}
	\caption{Different behaviors of the weight function $w_1$ (solid line) and $w_2$ (dotted line) for various combinations of $k$ and $\xi$.}
	\label{xxx}
\end{figure}

The weight functions we propose depend on one or more parameters that affect their shapes and behavior around the origin. As mentioned in the main text, Figure \ref{xxx} displays the weight functions $w_1$ and $w_2$ for different combinations of $\xi$ and $k$.

\FloatBarrier

To illustrate that our model is fairly robust to the proper choice of these hyperparameters, we also conduct a simulation study to showcase how the posterior estimates of $f_0$, $f_1$, and $P_1(z)$ vary as we consider different combinations of $\xi$ and $k$. For this study, we focus on the weight function $w_1$ and consider $k \in \{1,2,3\}$, and $\xi\in \{1,2,\hat{\xi}\}$, where $\hat{\xi}$ indicates the posterior estimate obtained in our unified Bayesian framework.

\begin{figure}[th!]
	\centering
	\includegraphics[width=.48\linewidth]{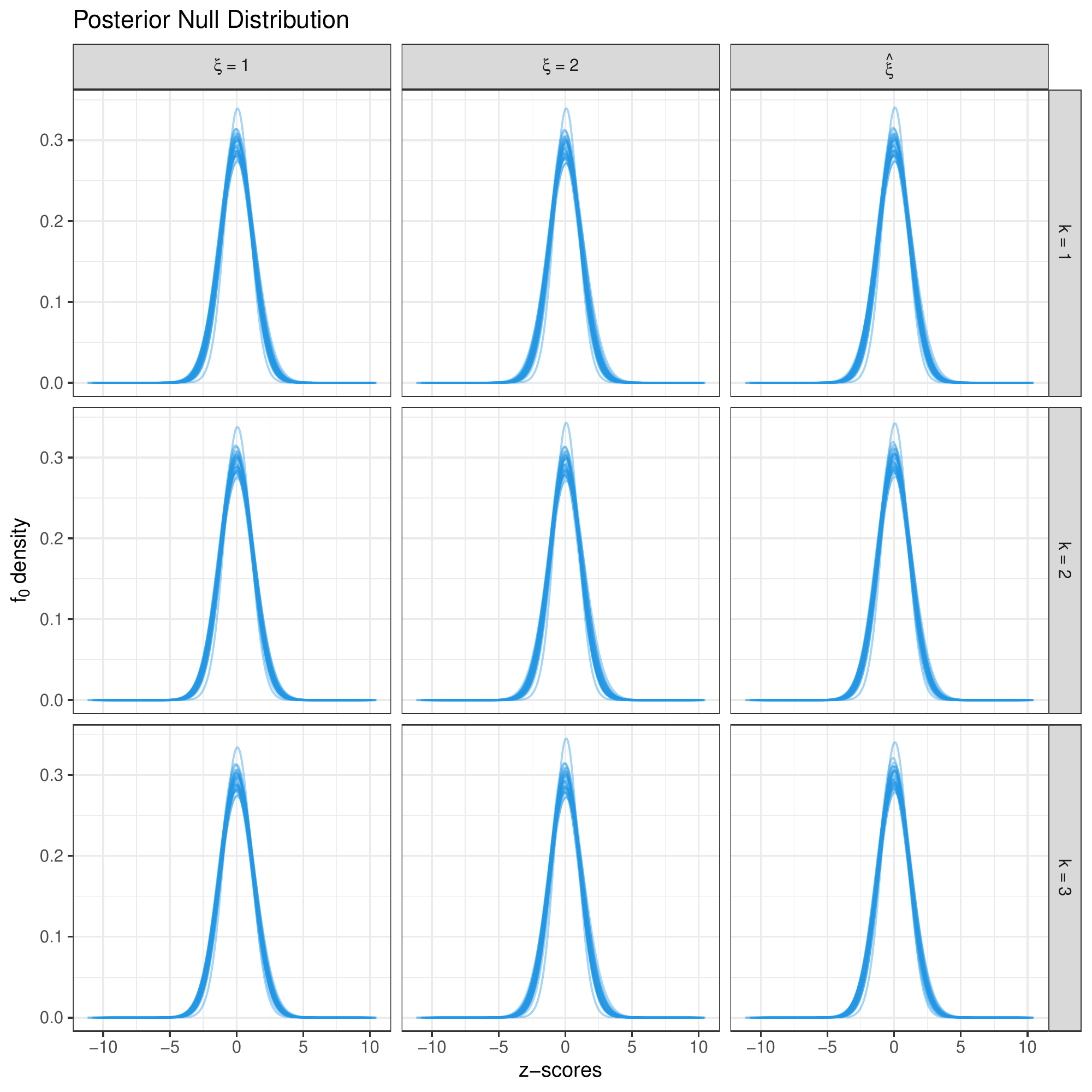}
	\includegraphics[width=.48\linewidth]{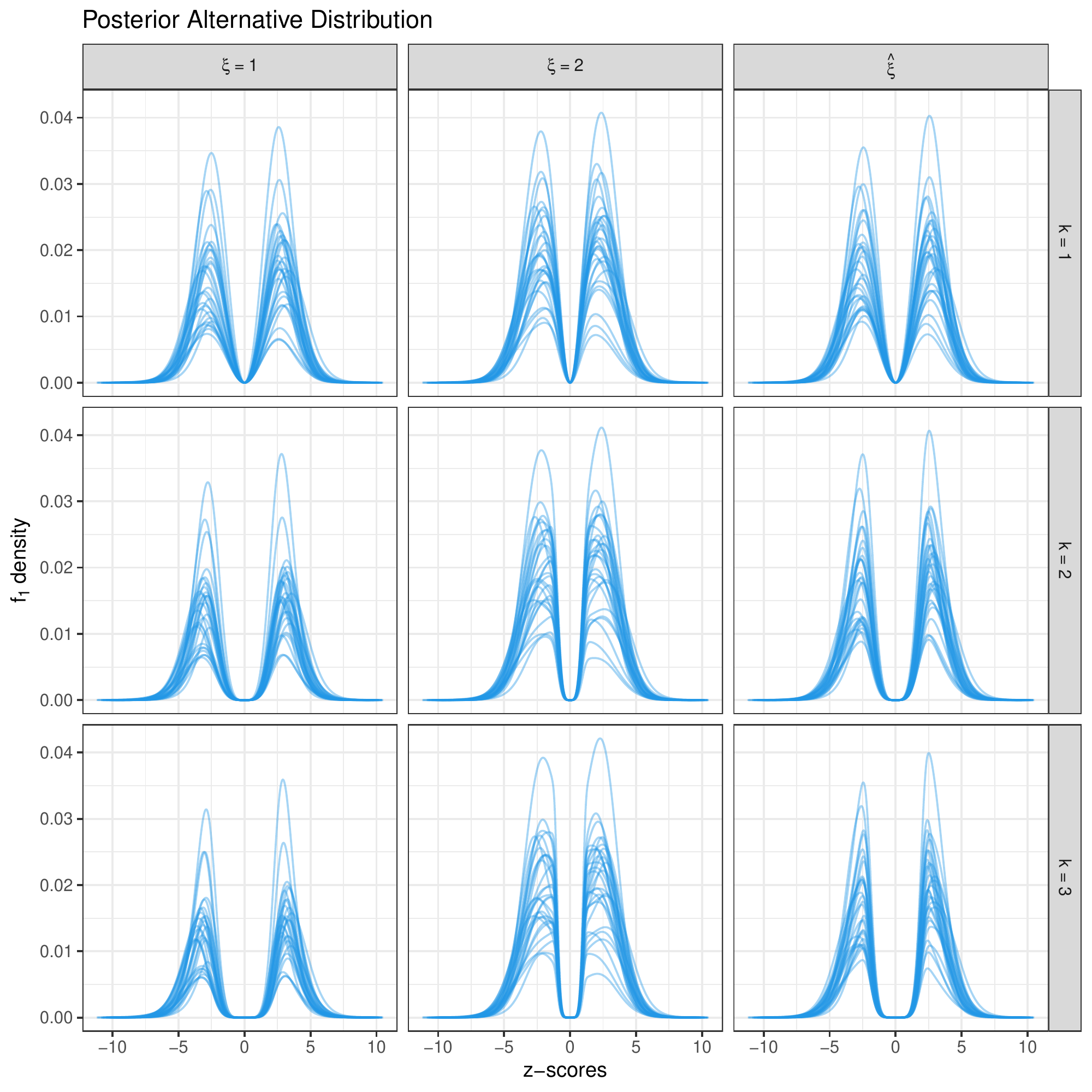}
	\includegraphics[width=.48\linewidth]{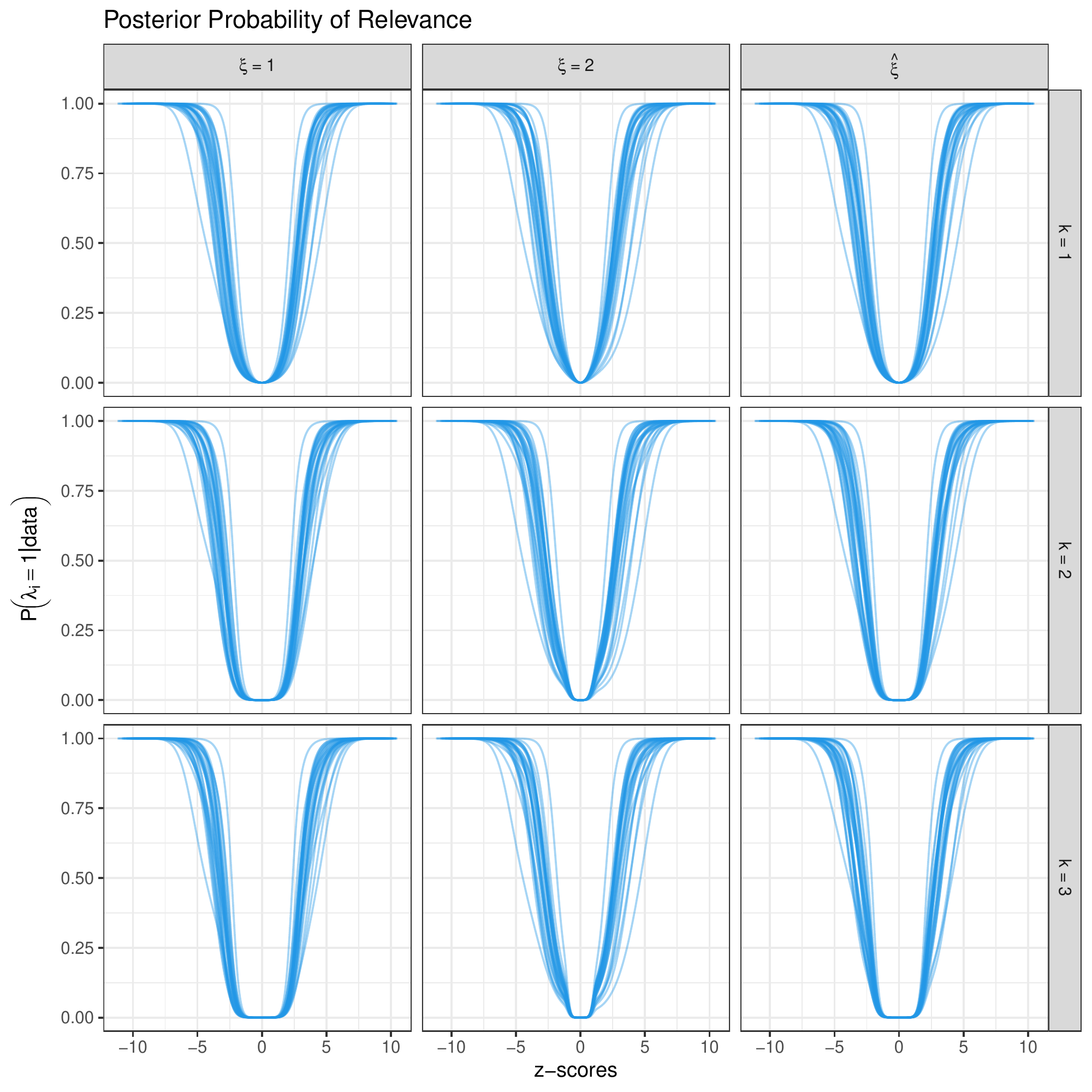}
	\caption{Every panel reports the results obtained in each of the 30 replicated datasets considering each possible pairwise combination of the values for $\xi$ and $k$. The top-left plot presents the estimated $\hat{f}_0$, the top-right displays $\hat{f}_1$, while the bottom panels show the estimated $\hat{P}_1(z)$.}
	\label{fig::postest}
\end{figure}

We generate 30 replicas of datasets comprised of 500 observations from the following mixture:
$$ z^{(k)}_i \sim \frac{9}{10}\mathcal{N}(0,1.5) + \frac{1}{20}\mathcal{N}(3,1 )+ \frac{1}{20}\mathcal{N}(-3,1 ),$$
and then we apply our model in its parametric specification for each pairwise combination of the aforementioned considered values of $k$ and $\xi$.
The results are reported in Figure \ref{fig::postest}. We appreciate that, despite some differences in the way the estimated alternative distributions approach zero (top-right panel), all the empirical null distributions (top-left panel) are very similar. The same holds for the posterior probability of relevance (bottom panels), suggesting the robustness of our method to different combinations of the weight functions' parameters. 

\FloatBarrier

\subsection{Simulation Study: Comparison between Nollik and unweighted two-group model}
\label{comparison}
As we mentioned in the main text, adopting a weighted mixture as alternative distribution in the two-group model constrains the estimated posterior local false discovery rate between 0 (always reached) and 1. Likewise, adopting a non-local weight function ensures that our model assigns negligible values to posterior probabilities of relevance $P_1(z_i)=\mathbb{P}\left(\lambda_i=1|data\right)$ to $z$-scores in a neighborhood of the origin.
This property is particularly advantageous in case of misspecification of the mixture weights $\rho$. \\

We showed in Section 3 of the main paper how Nollik always leads to better results than its unweighted counterpart for fixed mixed proportion. Here, we provide an example that highlights why to prefer the use of Nollik (weight adopted: $w_1$) over the unweighted two-group model (u2GM) when $\rho$ is stochastic. Specifically, we will compare the estimates of the posterior probability of relevance under Nollik and u2GM to investigate the robustness to misspecification of the most crucial prior of the model, the Beta distribution placed on the non-null proportion $\rho$.\\ 
We consider three simple simulated datasets, indicated by $k=1,2,3$. We sample 1,000 observations from the following mixture:
$$ z^{(k)}_i \sim \frac{9}{10}\mathcal{N}(0,1) + \frac{1}{20}\mathcal{N}(\mu_k,1 )+ \frac{1}{20}\mathcal{N}(-\mu_k,1 ), $$
with $\mu_k=k$. As $k$ increases, the overlap between the null distribution (standard Gaussian) and the alternative distribution (symmetric mixture) reduces, and the identification of the relevant test statistics becomes easier.\\ 
At the same time, we also consider three different specifications for the hyperparameters of $\rho \sim Beta(a,b)$:
\begin{itemize}
	\item[(HP1)] Classical specification: $a=1$, $b=9$, where we expect only a small fraction of $z$-scores to be generated from the alternative distribution
	\item[(HP2)] Uninformative specification: $a=1$, $b=1$, corresponding to a uniform    distribution over $\rho$
	\item[(HP3)] Strong misspecification: $a=9$, $b=1$, in the unrealistic scenario in which we expect 9 times more relevant observations than the irrelevant ones
\end{itemize}

\begin{figure}[bth!]
	\centering
	\includegraphics[width = .9\linewidth]{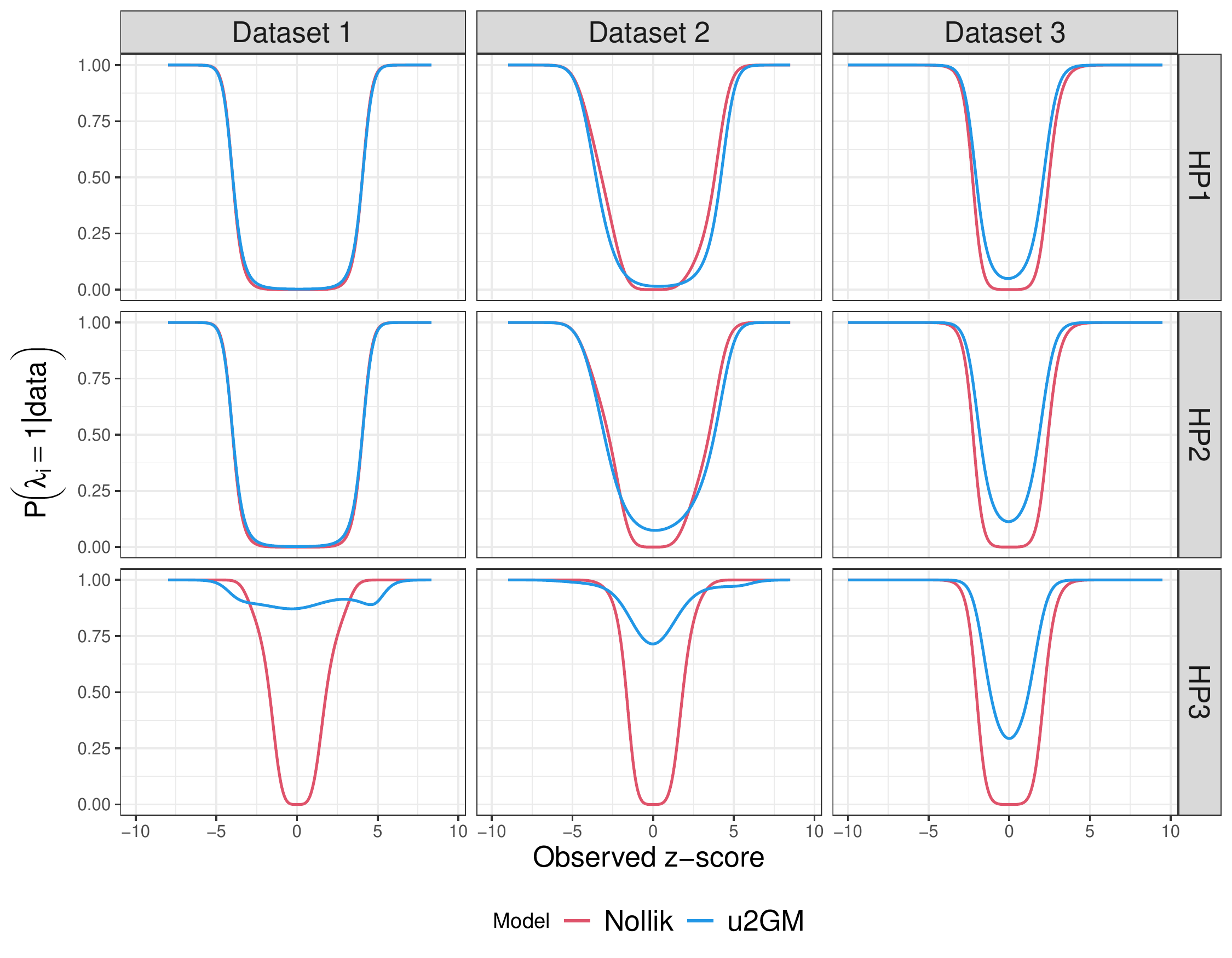}
	\caption{Posterior probability of inclusion ${P}_1(z_i)$ estimated for every $z$-score (blue: unweighted two-group model, red: Nollik). Each panel shows the results for a combination of datasets, by columns ($k=1,2,3$), and hyperparameters configurations, by rows (HP1, HP2, HP3), as detailed in Section \ref{comparison}.}
	\label{justify}
\end{figure}

Figure \ref{justify} displays the estimated posterior probabilities of inclusion ${P}_1(z_i)$ for each observation obtained with Nollik (red lines) and the u2GM (blue lines). 
We can see that, regardless of the dataset or the hyperparameter configuration, the estimates obtained from the Nollik model are close to zero in a suitable neighborhood of the origin.
Under the classical configuration HP1, the two models yield similar results across the datasets. 
The detection of the relevant statistics becomes easier as the null and alternative distributions get farther away from each other. 
The second and third rows of Figure \ref{justify} better showcase the differences between Nollik and u2GM, which become particularly evident under the HP3 configuration. 
When a vague or misspecified prior for $\rho$ is adopted, the posterior probability estimates under the unweighted model suffer from the substantial overlap between null and alternative distributions, and the estimate of ${P}_1(z_i)$ increases even for small $z$-scores, representing a large number of observations as relevant even when high thresholds are selected (bottom left panel). 
This same issue does not apply to Nollik, which is more conservative than u2GM when needed (entire bottom-row) but can also detect signal faster (central column, top and mid panel).

\FloatBarrier
\clearpage
\subsection{Additional material for the real data applications}

\subsubsection{Additional figures obtained using the weight function $w_1$}
\begin{figure}[!ht]
	\centering
	\includegraphics[scale=.35]{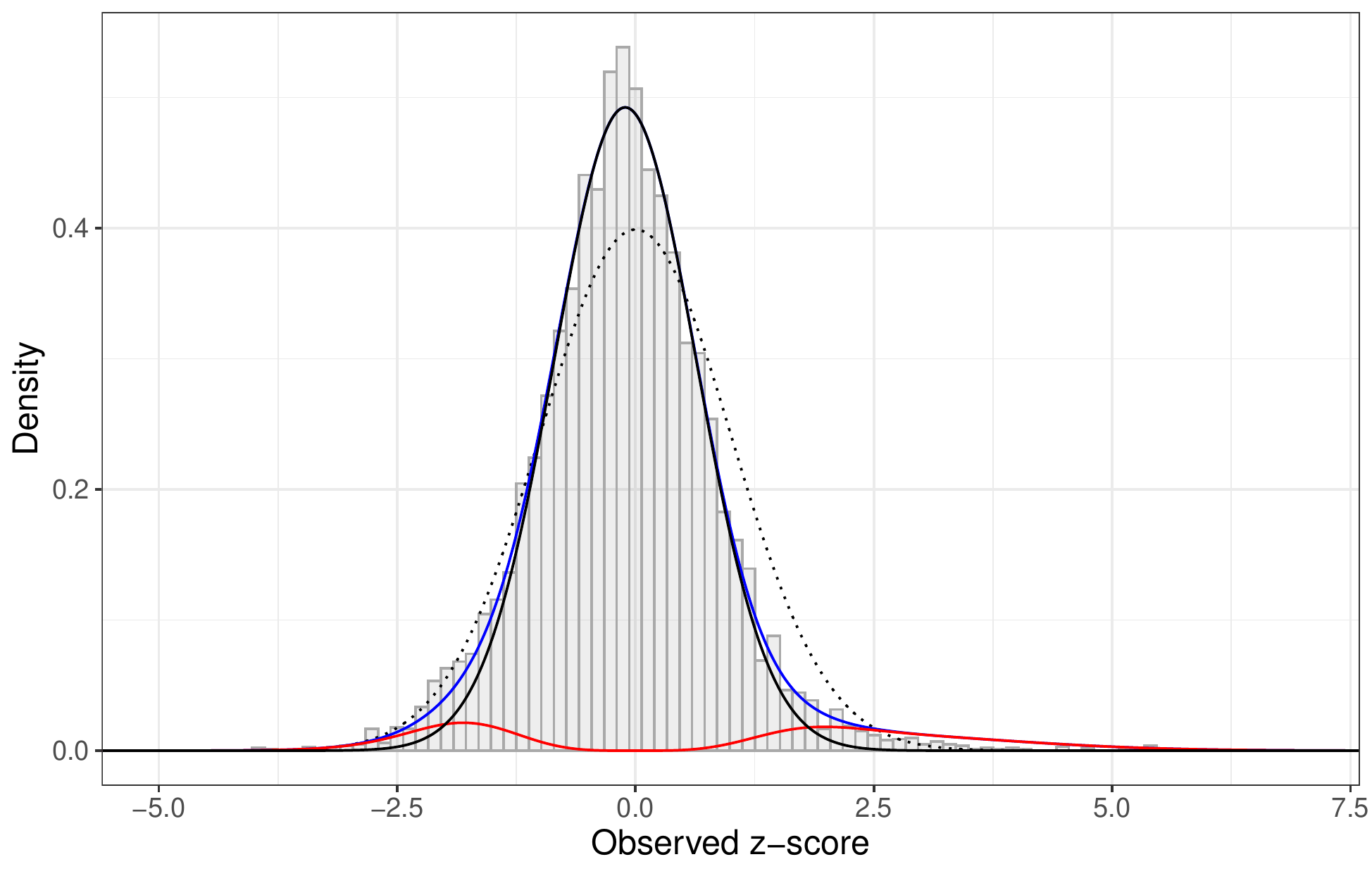}
	\includegraphics[scale=.35]{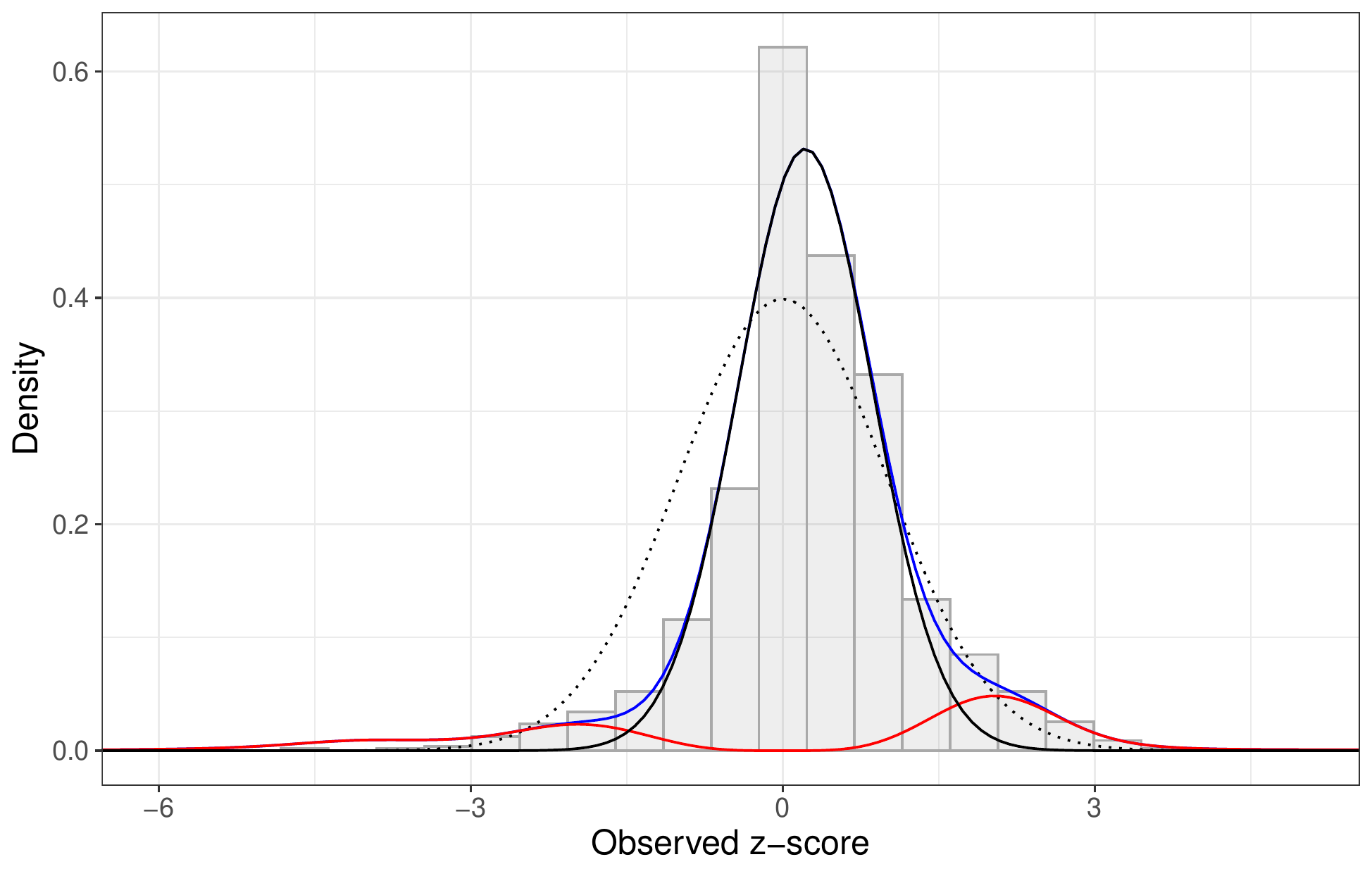} 
	\caption{Estimated null (black), alternative (red), and overall (blue) densities for the \texttt{HIV} (first panel) and \texttt{Torondel} (second panel) datasets (z-scores shown as histograms). The dotted lines denote the $\phi(0,1)$ densities. Weight function: $w_1$. The datasets are described in Section 6 of the main text.}
	\label{fig:ADDPLOTS1}
\end{figure}
\begin{figure}[!ht]
	\centering
	\includegraphics[width = \linewidth]{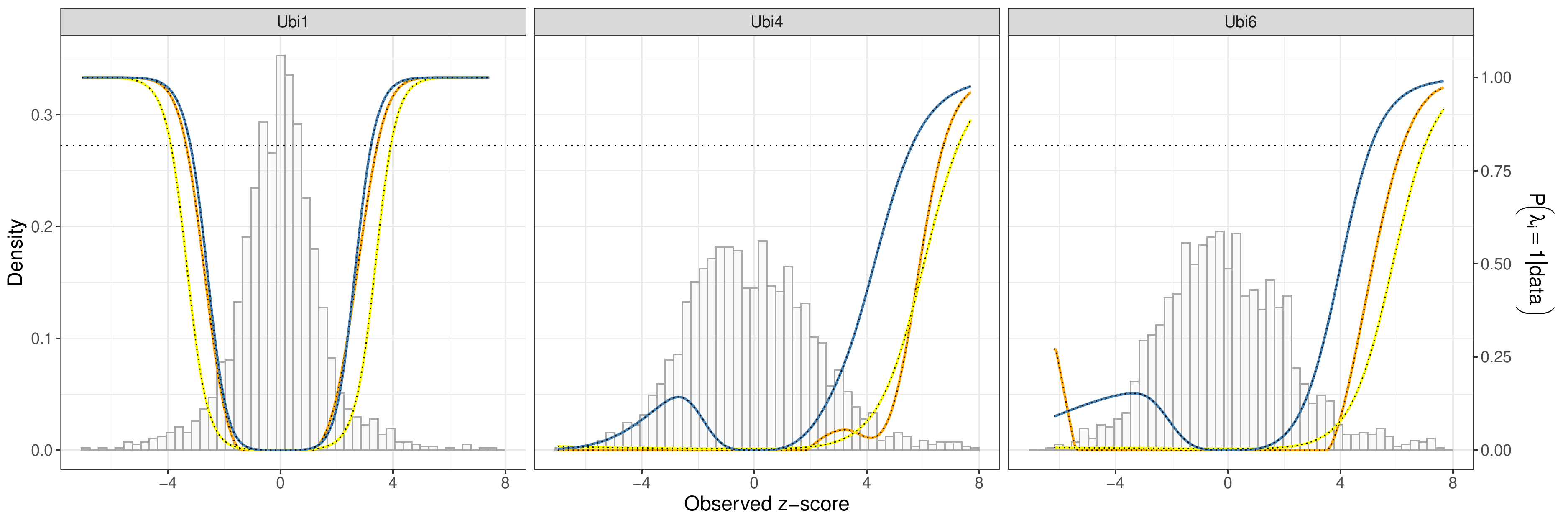}
	\caption{\texttt{Ubiquitin-proteomics} dataset. Histograms of the z-scores with function $P_1(z)$ superimposed estimated via \texttt{LocFDR} (orange), \texttt{MixFDR} (yellow), and Nollik (blue). The horizontal dotted lines represent the threshold controlling for a BFDR of 5\%. Weight function: $w_1$. The dataset is described in Section 6 of the main text.}
	\label{fig:ADDPLOTS2}
\end{figure}
\FloatBarrier
\clearpage
\subsubsection{Summary of the results obtained using the weight function $w_2$}

\begin{table}[th]
	\centering
	\begin{tabular}{lccccc}
		\toprule
		Dataset &  $\hat{\rho}$ & $\hat{\alpha}$ & $\hat{\xi}$ & 
		\text{Threshold} & \# Relevant   \\
		\midrule
		\texttt{HIV}      & $0.054\:(sd.\:0.007)$&  $0.157\:(sd.\:0.059)$&$1.816\:(sd.\:0.131)$ & $0.820$&$97$\\
		\texttt{Torondel} & $0.097\:(sd.\:0.017)$ & --- & $1.954\:(sd.\:0.237)$  & $0.862$ & $65$  \\
		\texttt{Ubiquitin} &$0.081\:(sd.\:0.012)$ & --- & --- & --- & --- \\
		\texttt{    -Ubi2} & ---                 & $ 0.469\:(sd.\: 0.260)$ & $ 2.553\:(sd.\: 0.194)$ & 0.815 & 93\\
		\texttt{    -Ubi4} & ---                 & $ 0.489\:(sd.\: 0.191)$ & $ 1.908\:(sd.\: 0.276)$ & 0.925 & 15\\
		\texttt{    -Ubi6} & ---                 & $ 0.488\:(sd.\: 0.143)$ & $ 2.115\:(sd.\: 0.347)$ & 0.881 & 28\\
		\bottomrule
	\end{tabular}
	
	\caption{Posterior estimates for $\rho$, $\alpha$, $\xi$, and threshold for the probability of relevance (thresholding the BFDR at 0.05), and number of relevant instances obtained when using the weight function $w_2$. The datasets and the models are described in Section 6 of the main text.}
	\label{tab:my_label}
\end{table}
\FloatBarrier
\subsubsection{Concordance of the results}

\begin{figure}[ht!]
	\centering
	\includegraphics[width = .49\linewidth]{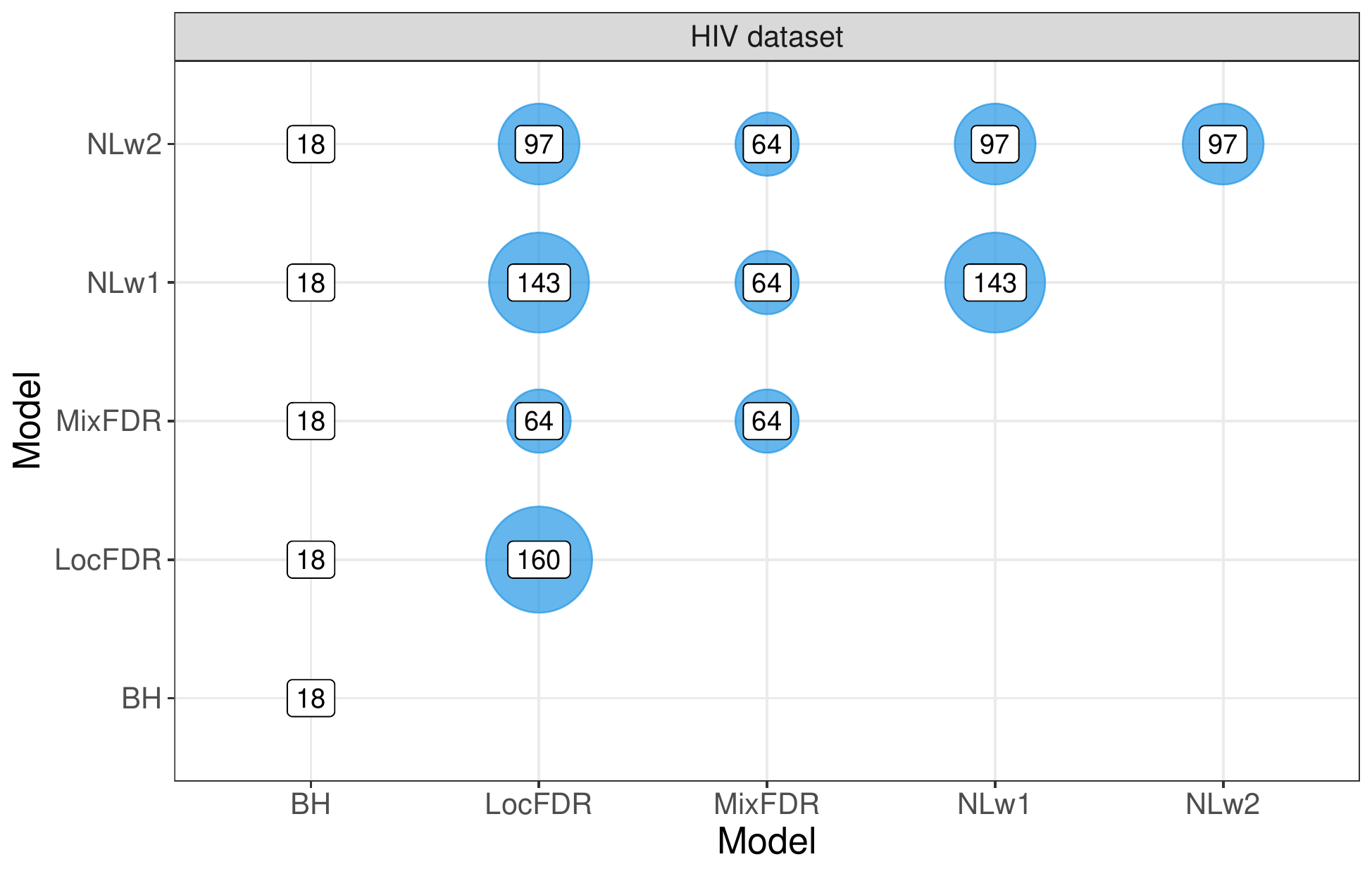}
	\includegraphics[width = .49\linewidth]{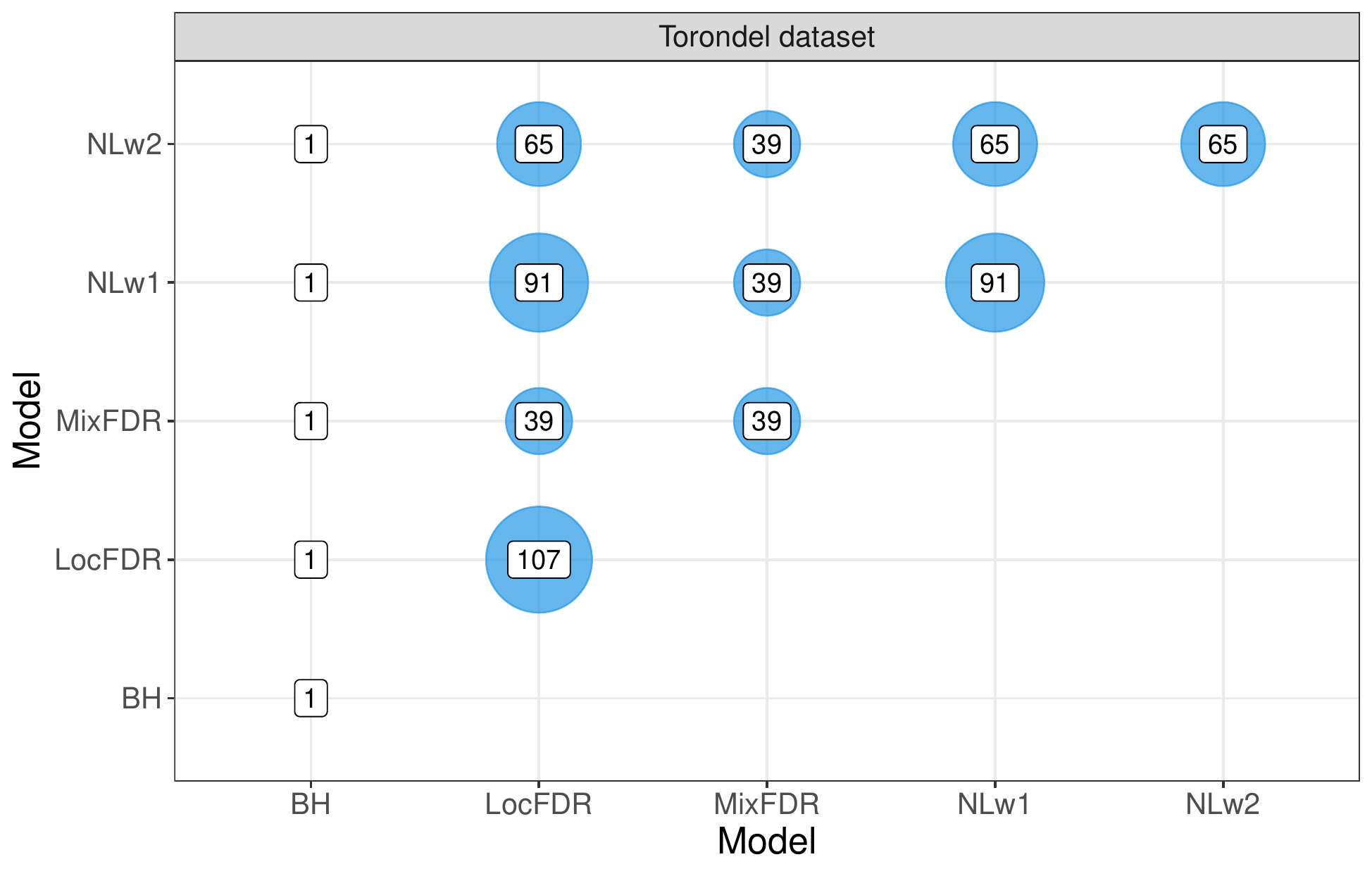}
	\includegraphics[width = \linewidth]{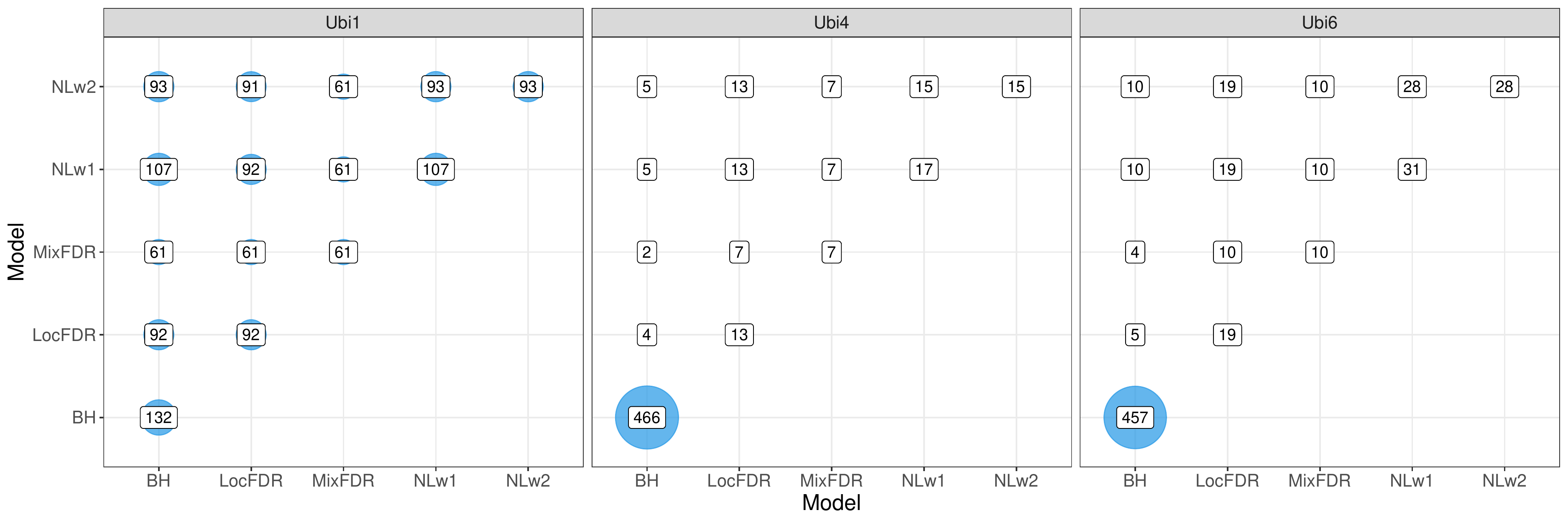}
	\caption{Visualization of the simultaneously flagged genes provided by the different methods. The plots present the results for the \texttt{HIV} dataset (top-left panel), the \texttt{Torondel} dataset (top-right), and the \texttt{Ubiquitin-protein interactors} dataset, stratified by group (bottom panels). In each panel, the main diagonals present the total number of discoveries per method. The interactions in the upper part of the plots show the number of genes simultaneously flagged as relevant by two models. }
	\label{fig:conc}
\end{figure}

Finally, we summarize in Figure \ref{fig:conc} the number of discoveries flagged by each method in all the real-world datasets we analyzed in Section 6. 
The top-left panel refers to the \texttt{HIV} dataset, the top-right refers to the \texttt{Torondel} dataset, and the bottom panels describe the results obtained from the \texttt{Ubiquitin-protein interactors} dataset, stratified by group. 
In each panel, the main diagonals present the total number of discoveries per method. The interactions in the upper part of the plots show the number of genes simultaneously flagged as relevant by two models. Note that, for each pairwise combination of methods, the frequency of the simultaneously flagged genes is always identical to the minimum number of discoveries found by one of two methods. As expected, these plots highlight that the less conservative an approach is, the higher the number of relevant genes added to a smaller set of common findings.

\bibliography{Arx_PROOF}

\begin{thebibliography}{47}
\providecommand{\natexlab}[1]{#1}
\providecommand{\url}[1]{\texttt{#1}}
\expandafter\ifx\csname urlstyle\endcsname\relax
  \providecommand{\doi}[1]{doi: #1}\else
  \providecommand{\doi}{doi: \begingroup \urlstyle{rm}\Url}\fi

\bibitem[Azzalini(1985)]{Azzalini1985}
A.~Azzalini.
\newblock {A class of distributions which includes the normal ones}.
\newblock \emph{Scandinavian journal of statistics}, 12\penalty0 (2):\penalty0
  171--178, 1985.
\newblock ISSN 0303-6898.

\bibitem[Bayarri and Berger(1998)]{Bayarri1998}
M.~J. Bayarri and J.~Berger.
\newblock {Robust Bayesian analysis of selection models}.
\newblock \emph{Annals of Statistics}, 26\penalty0 (2):\penalty0 645--659,
  1998.
\newblock ISSN 00905364.
\newblock \doi{10.1214/aos/1028144852}.

\bibitem[Benajmini and Hochberg(1995)]{Benjamini1995}
Y.~Benajmini and Y.~Hochberg.
\newblock {Controlling the False Discovery Rate : a practical and powerful
  approach to multiple testing}.
\newblock \emph{Journal of the Royal Statistical Society. Series B: Statistical
  Methodology}, 57\penalty0 (1):\penalty0 289--300, 1995.
\newblock ISSN 00359246.
\newblock \doi{10.2307/2346101}.
\newblock URL
  \url{http://www.stat.purdue.edu/{~}doerge/BIOINFORM.D/FALL06/Benjamini and Y
  FDR.pdf{\%}5Cnhttp://engr.case.edu/ray{\_}soumya/mlrg/controlling{\_}fdr{\_}benjamini95.pdf}.

\bibitem[Blei et~al.(2017)Blei, Kucukelbir, and McAuliffe]{Blei2017}
D.~M. Blei, A.~Kucukelbir, and J.~D. McAuliffe.
\newblock {Variational Inference: A Review for Statisticians}.
\newblock \emph{Journal of the American Statistical Association}, 112\penalty0
  (518):\penalty0 859--877, 2017.
\newblock ISSN 1537274X.
\newblock \doi{10.1080/01621459.2017.1285773}.

\bibitem[Botev(2017)]{Botev2017}
Z.~I. Botev.
\newblock {The normal law under linear restrictions: simulation and estimation
  via minimax tilting}.
\newblock \emph{Journal of the Royal Statistical Society. Series B: Statistical
  Methodology}, 79\penalty0 (1):\penalty0 125--148, 2017.
\newblock ISSN 14679868.
\newblock \doi{10.1111/rssb.12162}.

\bibitem[Damien et~al.(1999)Damien, Wakefield, and Walker]{Damien1999}
P.~Damien, J.~Wakefield, and S.~G. Walker.
\newblock {Gibbs sampling for Bayesian non-conjugate and hierarchical models by
  using auxiliary variables}.
\newblock \emph{Journal of the Royal Statistical Society: Series B (Statistical
  Methodology)}, 61\penalty0 (2):\penalty0 331--344, 1999.
\newblock \doi{10.1111/1467-9868.00179}.

\bibitem[Denti et~al.(2020)Denti, Guindani, Leisen, Lijoi, Wadsworth, and
  Vannucci]{Denti2020a}
F.~Denti, M.~Guindani, F.~Leisen, A.~Lijoi, W.~D. Wadsworth, and M.~Vannucci.
\newblock {Two-group Poisson-Dirichlet mixtures for multiple testing}.
\newblock \emph{Biometrics}, 2020.
\newblock ISSN 15410420.
\newblock \doi{10.1111/biom.13314}.

\bibitem[Dharmadhikari and Joag-Dev(1983)]{Dharmadhikari1983}
S.~W. Dharmadhikari and K.~Joag-Dev.
\newblock {Mean, median, mode III}.
\newblock \emph{Statistica Neerlandica}, 37\penalty0 (4):\penalty0 165--168,
  1983.

\bibitem[Do et~al.(2005)Do, Mueller, and Tang]{Do2005}
K.~A. Do, P.~Mueller, and F.~Tang.
\newblock {A nonparametric Bayesian mixture model for gene expression}.
\newblock \emph{Journal of the Royal Statistical Society, Series C},
  54:\penalty0 1--18, 2005.

\bibitem[Dunson et~al.(2007)Dunson, Pillai, and Park]{Dunson2007}
D.~B. Dunson, N.~Pillai, and J.~H. Park.
\newblock {Bayesian density regression}.
\newblock \emph{Journal of the Royal Statistical Society. Series B: Statistical
  Methodology}, 69\penalty0 (2):\penalty0 163--183, 2007.
\newblock ISSN 13697412.
\newblock \doi{10.1111/j.1467-9868.2007.00582.x}.

\bibitem[Durante(2019)]{Durante2019}
D.~Durante.
\newblock {Conjugate Bayes for probit regression via unified skew-normal
  distributions}.
\newblock \emph{Biometrika}, 2019.
\newblock ISSN 0006-3444.
\newblock \doi{10.1093/biomet/asz034}.

\bibitem[Efron(2004)]{Efron2004}
B.~Efron.
\newblock {Large-scale simultaneous hypothesis testing: The choice of a null
  hypothesis}.
\newblock \emph{Journal of the American Statistical Association}, 99\penalty0
  (465):\penalty0 96--104, 2004.
\newblock ISSN 1537274X.
\newblock \doi{10.1198/016214504000000089}.

\bibitem[Efron(2007{\natexlab{a}})]{Efron2007}
B.~Efron.
\newblock {Size, power and false discovery rates}.
\newblock \emph{Annals of Statistics}, 35\penalty0 (4):\penalty0 1351--1377,
  2007{\natexlab{a}}.
\newblock ISSN 00905364.
\newblock \doi{10.1214/009053606000001460}.

\bibitem[Efron(2007{\natexlab{b}})]{Efron2007_cor}
B.~Efron.
\newblock {Correlation and large-scale simultaneous significance testing}.
\newblock \emph{Journal of the American Statistical Association}, 102\penalty0
  (477):\penalty0 93--103, 2007{\natexlab{b}}.
\newblock ISSN 01621459.
\newblock \doi{10.1198/016214506000001211}.

\bibitem[Efron(2008)]{Efron2008}
B.~Efron.
\newblock Microarrays, empirical bayes and the two-groups model.
\newblock \emph{Statistical Science}, 23\penalty0 (1):\penalty0 45--47, 2008.
\newblock ISSN 0883-4237.
\newblock \doi{10.1214/08-sts236rej}.
\newblock URL \url{http://projecteuclid.org/euclid.ss/1215441276}.

\bibitem[Ferguson(1973)]{Ferguson1973}
T.~S. Ferguson.
\newblock {A Bayesian analysis of some nonparametric problems}.
\newblock \emph{The Annals of Statistics}, 1\penalty0 (2):\penalty0 209--230,
  1973.
\newblock ISSN 0090-5364.
\newblock \doi{10.1214/aos/1176342360}.

\bibitem[Hardcastle and Kelly(2010)]{Hardcastle2010}
T.~J. Hardcastle and K.~A. Kelly.
\newblock {BaySeq: Empirical Bayesian methods for identifying differential
  expression in sequence count data}.
\newblock \emph{BMC Bioinformatics}, 11, 2010.
\newblock ISSN 14712105.
\newblock \doi{10.1186/1471-2105-11-422}.

\bibitem[Ishwaran and James(2001)]{Ishwaran2001a}
H.~Ishwaran and L.~F. James.
\newblock {Gibbs sampling methods for stick-breaking priors}.
\newblock \emph{Journal of the American Statistical Association}, 96\penalty0
  (453):\penalty0 161--173, 2001.
\newblock ISSN 1537274X.
\newblock \doi{10.1198/016214501750332758}.

\bibitem[Johnson and Rossell(2010)]{Johnson2010}
V.~E. Johnson and D.~Rossell.
\newblock {On the use of non-local prior densities in Bayesian hypothesis
  tests}.
\newblock \emph{Journal of the Royal Statistical Society. Series B: Statistical
  Methodology}, 72\penalty0 (2):\penalty0 143--170, 2010.
\newblock ISSN 13697412.
\newblock \doi{10.1111/j.1467-9868.2009.00730.x}.

\bibitem[Love et~al.(2014)Love, Huber, and Anders]{Love2014}
M.~I. Love, W.~Huber, and S.~Anders.
\newblock {Moderated estimation of fold change and dispersion for RNA-seq data
  with DESeq2}.
\newblock \emph{Genome Biology}, 15\penalty0 (12):\penalty0 550, 2014.
\newblock ISSN 1474760X.
\newblock \doi{10.1186/s13059-014-0550-8}.
\newblock URL
  \url{http://genomebiology.biomedcentral.com/articles/10.1186/s13059-014-0550-8}.

\bibitem[MacEachern(2000)]{MacEachern2000}
S.~N. MacEachern.
\newblock {Dependent dirichlet processes}.
\newblock \emph{Technical report. Department of Statistics, The Ohio State
  Univ.}, 2000.

\bibitem[Martin and Tokdar(2012)]{Martin2012}
R.~Martin and S.~T. Tokdar.
\newblock {A nonparametric empirical Bayes framework for large-scale multiple
  testing,}.
\newblock \emph{Biostatistics}, 13\penalty0 (3):\penalty0 427--439, 2012.
\newblock ISSN 14654644.
\newblock \doi{10.1093/biostatistics/kxr039}.

\bibitem[Mira and Tierney(2002)]{Mira2002}
A.~Mira and L.~Tierney.
\newblock {Efficiency and convergence properties of slice samplers}.
\newblock \emph{Scandinavian Journal of Statistics}, 29\penalty0 (1):\penalty0
  1--12, 2002.
\newblock ISSN 03036898.
\newblock \doi{10.1111/1467-9469.00032}.

\bibitem[Muralidharan(2010)]{Muralidharan2012}
O.~Muralidharan.
\newblock {An empirical Bayes mixture method for effect size and false
  discovery rate estimation}.
\newblock \emph{Annals of Applied Statistics}, 6\penalty0 (1):\penalty0
  422--438, 2010.
\newblock ISSN 19326157.
\newblock \doi{10.1214/09-AOAS276}.

\bibitem[Neal(2003)]{Neil2003}
R.~M. Neal.
\newblock {Slice sampling (with discussion)}.
\newblock \emph{The Annals of Statistics}, 31\penalty0 (3):\penalty0 705--767,
  2003.
\newblock ISSN 00905364.
\newblock \doi{10.1214/aos/1056562461}.

\bibitem[Newton et~al.(2004)Newton, Noueiry, Sarkar, and Ahlquist]{Newton2004}
M.~A. Newton, A.~Noueiry, D.~Sarkar, and P.~Ahlquist.
\newblock {Detecting differential gene expression with a semiparametric
  hierarchical mixture method}.
\newblock \emph{Biostatistics}, 5\penalty0 (2):\penalty0 155--176, 2004.
\newblock ISSN 14654644.
\newblock \doi{10.1093/biostatistics/5.2.155}.
\newblock URL \url{http://www.ncbi.nlm.nih.gov/pubmed/15054023}.

\bibitem[O'Hagan and Leonard(1976)]{Ohagan1976}
A.~O'Hagan and T.~Leonard.
\newblock {Bayes estimation subject to uncertainty about parameter
  constraints}.
\newblock \emph{Biometrika}, 63\penalty0 (1):\penalty0 201--203, 1976.
\newblock ISSN 00063444.
\newblock \doi{10.1093/biomet/63.1.201}.

\bibitem[Petralia et~al.(2012)Petralia, Rao, and Dunson]{Petralia2012aaa}
F.~Petralia, V.~Rao, and D.~B. Dunson.
\newblock {Repulsive mixtures}.
\newblock \emph{Advances in Neural Information Processing Systems}, 3:\penalty0
  1889--1897, 2012.
\newblock ISSN 10495258.

\bibitem[Plummer et~al.(2006)Plummer, Best, Cowles, and Vines]{oro22547}
M.~Plummer, N.~Best, K.~Cowles, and K.~Vines.
\newblock {CODA: convergence diagnosis and output analysis for MCMC}.
\newblock \emph{R News}, 6\penalty0 (1):\penalty0 7--11, 2006.

\bibitem[Polson et~al.(2013)Polson, Scott, and Windle]{Polson2013}
N.~G. Polson, J.~G. Scott, and J.~Windle.
\newblock {Bayesian inference for logistic models using P{\'{o}}lya-Gamma
  latent variables}.
\newblock \emph{Journal of the American Statistical Association}, 108\penalty0
  (504):\penalty0 1339--1349, 2013.
\newblock ISSN 1537274X.
\newblock \doi{10.1080/01621459.2013.829001}.

\bibitem[Pounds and Morris(2003)]{Pounds2003}
S.~Pounds and S.~W. Morris.
\newblock {Estimating the occurrence of false positives and false negatives in
  microarray studies by approximating and partitioning the empirical
  distribution of p-values}.
\newblock \emph{Bioinformatics}, 19\penalty0 (10):\penalty0 1236--1242, 2003.
\newblock ISSN 13674803.
\newblock \doi{10.1093/bioinformatics/btg148}.

\bibitem[Rao(1985)]{Rao1985}
C.~R. Rao.
\newblock {Weighted distributions arising out of methods of ascertainment: what
  population does a sample represent?}
\newblock \emph{A Celebration of Statistics}, pages 543--569, 1985.
\newblock \doi{10.1007/978-1-4613-8560-8_24}.

\bibitem[Ritchie et~al.(2015)Ritchie, Phipson, Wu, Hu, Law, Shi, and
  Smyth]{Limma}
M.~E. Ritchie, B.~Phipson, D.~Wu, Y.~Hu, C.~W. Law, W.~Shi, and G.~K. Smyth.
\newblock {Limma powers differential expression analyses for RNA-sequencing and
  microarray studies}.
\newblock \emph{Nucleic Acids Research}, 43\penalty0 (7):\penalty0 e47, 2015.

\bibitem[Roberts and Rosenthal(2007)]{Roberts2007}
G.~O. Roberts and J.~S. Rosenthal.
\newblock {Coupling and ergodicity of adaptive MCMC}.
\newblock \emph{Journal of Applied Probability}, 44\penalty0 (2):\penalty0
  458--475, 2007.
\newblock \doi{10.1007/s11250-009-9481-x}.

\bibitem[Roberts and Rosenthal(2009)]{Roberts2009}
G.~O. Roberts and J.~S. Rosenthal.
\newblock {Examples of adaptive MCMC}.
\newblock \emph{Journal of Computational and Graphical Statistics}, 18\penalty0
  (2):\penalty0 349--367, 2009.
\newblock ISSN 10618600.
\newblock \doi{10.1198/jcgs.2009.06134}.

\bibitem[Robinson and Smyth(2007)]{Robinson2007}
M.~D. Robinson and G.~K. Smyth.
\newblock {Moderated statistical tests for assessing differences in tag
  abundance}.
\newblock \emph{Bioinformatics}, 23\penalty0 (21):\penalty0 2881--2887, 2007.
\newblock ISSN 13674803.
\newblock \doi{10.1093/bioinformatics/btm453}.

\bibitem[Rodr{\'{i}}guez et~al.(2008)Rodr{\'{i}}guez, Dunson, and
  Gelfand]{Rodriguez2008}
A.~Rodr{\'{i}}guez, D.~B. Dunson, and A.~E. Gelfand.
\newblock {The nested dirichlet process}.
\newblock \emph{Journal of the American Statistical Association}, 103\penalty0
  (483):\penalty0 1131--1144, 2008.
\newblock ISSN 01621459.
\newblock \doi{10.1198/016214508000000553}.

\bibitem[Rossell and Telesca(2017)]{Rossell2017a}
D.~Rossell and D.~Telesca.
\newblock {Nonlocal priors for high-dimensional estimation}.
\newblock \emph{Journal of the American Statistical Association}, 112\penalty0
  (517):\penalty0 254--265, 2017.
\newblock ISSN 1537274X.
\newblock \doi{10.1080/01621459.2015.1130634}.

\bibitem[Rossell et~al.(2013)Rossell, Telesca, and Johnson]{Rossell2013}
D.~Rossell, D.~Telesca, and V.~E. Johnson.
\newblock {High-dimensional bayesian classifiers using non-local priors}.
\newblock \emph{Studies in Classification, Data Analysis, and Knowledge
  Organization}, pages 305--313, 2013.
\newblock ISSN 14318814.
\newblock \doi{10.1007/978-3-319-00032-9-35}.

\bibitem[Ruggeri et~al.(2021)Ruggeri, S{\'{a}}nchez-S{\'{a}}nchez, Sordo, and
  Su{\'{a}}rez-Llorens]{Ruggeri2019}
F.~Ruggeri, M.~S{\'{a}}nchez-S{\'{a}}nchez, M.~{\'{A}}. Sordo, and
  A.~Su{\'{a}}rez-Llorens.
\newblock {On a New Class of Multivariate Prior Distributions: Theory and
  Application in Reliability}.
\newblock \emph{Bayesian Analysis}, 16\penalty0 (1), 2021.
\newblock ISSN 19316690.
\newblock \doi{10.1214/19-BA1191}.

\bibitem[Sethuraman(1994)]{Sethuraman1994}
A.~J. Sethuraman.
\newblock {A constructive definition of Dirichlet priors}.
\newblock \emph{Statistica Sinica}, 4:\penalty0 639--650, 1994.

\bibitem[Sun et~al.(2018)Sun, Kim, and Lee]{Sun2018}
P.~Sun, I.~Kim, and K.~A. Lee.
\newblock {Dual-semiparametric regression using weighted Dirichlet process
  mixture}.
\newblock \emph{Computational Statistics and Data Analysis}, 117:\penalty0
  162--181, 2018.
\newblock ISSN 01679473.
\newblock \doi{10.1016/j.csda.2017.08.005}.

\bibitem[Torondel et~al.(2016)Torondel, Ensink, Gundogdu, Ijaz, Parkhill,
  Abdelahi, Nguyen, Sudgen, Gibson, Walker, and Quince]{Torondel2016}
B.~Torondel, J.~H. Ensink, O.~Gundogdu, U.~Z. Ijaz, J.~Parkhill, F.~Abdelahi,
  V.~A. Nguyen, S.~Sudgen, W.~Gibson, A.~W. Walker, and C.~Quince.
\newblock {Assessment of the influence of intrinsic environmental and
  geographical factors on the bacterial ecology of pit latrines}.
\newblock \emph{Microbial Biotechnology}, 9\penalty0 (2):\penalty0 209--223,
  2016.
\newblock ISSN 17517915.
\newblock \doi{10.1111/1751-7915.12334}.

\bibitem[{van 't Wout} et~al.(2003){van 't Wout}, Lehrman, Mikheeva, O'Keeffe,
  Katze, Bumgarner, Geiss, and Mullins]{VantWout2003}
A.~B. {van 't Wout}, G.~K. Lehrman, S.~A. Mikheeva, G.~C. O'Keeffe, M.~G.
  Katze, R.~E. Bumgarner, G.~K. Geiss, and J.~I. Mullins.
\newblock {Cellular gene expression upon human immunodeficiency virus type 1
  infection of CD4+-T-Cell lines}.
\newblock \emph{Journal of Virology}, 77\penalty0 (2):\penalty0 1392--1402,
  2003.
\newblock ISSN 0022-538X.
\newblock \doi{10.1128/jvi.77.2.1392-1402.2003}.

\bibitem[Zellner(1986)]{Zellner1986}
A.~Zellner.
\newblock {On assessing prior distributions and Bayesian regression analysis
  with g prior distributions}.
\newblock \emph{Bayesian Inference and Decision techniques: Essays in Honor of
  Bruno de Finetti}, pages 389--399, 1986.

\bibitem[Zhang et~al.(2017)Zhang, Smits, van Tilburg, Jansen, Makowski, Ovaa,
  and Vermeulen]{Zhang2017}
X.~Zhang, A.~H. Smits, G.~B. van Tilburg, P.~W. Jansen, M.~M. Makowski,
  H.~Ovaa, and M.~Vermeulen.
\newblock {An interaction landscape of ubiquitin signaling}.
\newblock \emph{Molecular Cell}, 65\penalty0 (5):\penalty0 941--955.e8, 2017.
\newblock ISSN 10974164.
\newblock \doi{10.1016/j.molcel.2017.01.004}.

\bibitem[Zhang et~al.(2018)Zhang, Smits, {van Tilburg}, Ovaa, Huber, and
  Vermeulen]{DEP}
X.~Zhang, A.~H. Smits, G.~B. {van Tilburg}, H.~Ovaa, W.~Huber, and
  M.~Vermeulen.
\newblock Proteome-wide identification of ubiquitin interactions using ubia-ms.
\newblock \emph{Nature Protocols}, 13:\penalty0 530–550, 2018.

\end{thebibliography}

\end{document}